\documentclass[3p,times,authoryear]{elsarticle}
\usepackage{setspace}
\usepackage[colorlinks,citecolor=blue,linkcolor=blue]{hyperref}

\makeatletter
\def\ps@pprintTitle{%
   \let\@oddhead\@empty
   \let\@evenhead\@empty
   \let\@oddfoot\@empty
   \let\@evenfoot\@oddfoot
}
\makeatother

\graphicspath{ {./figures/} }

\usepackage{amsmath,amssymb,amsfonts}
\usepackage{subfig}
\usepackage{url}
\usepackage{wrapfig}

\usepackage{numprint}
\npdecimalsign{\ensuremath{.}}
\newcommand{\np}[1]{\numprint{#1}}

\usepackage{algorithm,algpseudocode,algorithmicx}

\usepackage{xcolor}
\newcommand{\rev}[1]{{#1}}
\newcommand{\revTwo}[1]{{#1}}

\usepackage{booktabs}

\newcommand{\sout}[1]{{}}

\usepackage{listings}
\newcommand{\lline}[1]{Line~\ref{#1}} 
\newcommand{\llines}[2]{Lines~\ref{#1}-\ref{#2}} 

\usepackage{xspace}
\newcommand{\mv}{MultiVeStA\xspace}
\newcommand{\mq}{MultiQuaTEx\xspace}
\newcommand{\autow}{\texttt{autoWarmup}\xspace}
\newcommand{\autobm}{\texttt{autoBM}\xspace}
\newcommand{\autord}{\texttt{autoRD}\xspace}
\newcommand{\autoW}{\autow}
\newcommand{\autoBM}{\autobm}
\newcommand{\autoRD}{\autord}
\newcommand{\manualbm}{\texttt{manualBM}\xspace}
\newcommand{\manualrd}{\texttt{manualRD}\xspace}
\newcommand{\transient}{\texttt{autoIR}\xspace}
\newcommand{\autoir}{\transient}
\newcommand{\autoIR}{\transient}

\begin{document}

\begin{frontmatter}

\title{Automated and Distributed Statistical Analysis \\ of Economic Agent-Based Models} 

\address[sssa]{Institute of Economics and EMbeDS, Sant'Anna School of Advanced Studies, Pisa, Italy.}
\address[dtu]{DTU Technical University of Denmark, Lyngby, Denmark.}
\address[pennstate]{Dept. of Statistics and Huck Institutes of the Life Sciences, Penn State University, USA}
\address[eiee]{RFF-CMCC European Institute on Economics and the Environment, Milan, Italy.}

\author[sssa,dtu]{Andrea Vandin}
\ead{a.vandin@santannapisa.it}
\author[sssa]{Daniele Giachini}
\ead{d.giachini@santannapisa.it}
\author[sssa,eiee]{Francesco Lamperti}
\ead{f.lamperti@santannapisa.it}

\author[sssa,pennstate]{Francesca Chiaromonte}
\ead{f.chiaromonte@santannapisa.it}

\begin{abstract}
	We propose a novel approach to the statistical analysis of \rev{stochastic} simulation models and, especially, agent-based models (ABMs). Our main goal is to provide fully automated, model-independent \rev{and tool-supported techniques and algorithms} 
	to inspect simulations and perform counterfactual analysis. Our approach: (i) is easy-to-use by the modeller, (ii) improves reproducibility of results, 
	(iii) optimizes running time given the modeller's machine, (iv) automatically chooses the number of required simulations and simulation steps to reach user-specified statistical confidence, and (v) automates a variety of statistical tests. In particular, our \rev{techniques are } 
	designed to distinguish the transient dynamics of the model from its steady-state behaviour (if any), estimate properties 
	in both ``phases'', and provide indications on the (non-)ergodic 
	nature of the simulated processes
	-- which, in turn, allows one to gauge the reliability of 
	a steady-state analysis. 
	Estimates are equipped with statistical guarantees, allowing for robust comparisons across computational experiments. 
	To demonstrate the effectiveness of our approach, 
	we apply it to two models from the literature: a large-scale macro-financial ABM and a small scale prediction market model. Compared to prior analyses of these models, we 
	obtain new insights and we are able  to identify and fix some erroneous conclusions. 
\end{abstract}


\end{frontmatter}


\section{Introduction}\label{sec:intro}
In this 
article we present a model-independent and fully automated approach to the statistical analysis of \rev{stochastic} simulation models and, especially, agent-based models (ABMs). \rev{In particular, our work allows one to distinguish the \textit{transient dynamics} of the model from its \textit{steady-state behaviour} (if any),
to estimate properties of the model in both ``phases'', to check whether the ergodicity assumption is reasonable, and to equip the results with statistical guarantees, allowing for robust comparison of model behaviours' across computational experiments.  

Though these tasks would improve the robustness and reliability of counterfactual analyses, especially coming from the comparison of simulated policies, we believe they are still challenging and sometimes overlooked.}\footnote{\rev{
	We focus on models that are stochastic (in the sense that they entail random draws over the simulation) and that need to be initialized. As a consequence, for every configuration of parameters and initialization, we observe a certain degree of stochasticity leading the system \rev{to exhibit properties that are either \emph{stable} or \emph{unstable} over the unfolding of simulation time. Our work focuses on the analysis of such properties.
}}}
In the last two decades, the use of
Agent-Based Models has spread across several 
fields -- including ecology \citep{grimm2013}, health care \citep{eff12}, sociology \citep{macy2002}, geography \citep{brown2005}, 
medicine \citep{An2009}, 
research in bioterrorism \citep{carley2006}, 
and military tactics \citep{ila97}. 
In economics, ABMs 
contributed to the understanding of a variety of 
micro and macro phenomena \citep{tesfatsion2006handbook}.
They provided an alternative environment for policy-testing in the aftermath of the last financial crisis, when more traditional approaches (e.g., dynamic stochastic general equilibrium models and computable general equilibrium models)
failed \citep{Fagiolo_Roventini_2012,fagiolo2017macroeconomic}. 
Moreover, they were recently used for macroeconomic forecasting with promising results \citep{gatti2020rising, poledna2020economic}.

The key advantage of ABMs lies in the flexibility they allow when modelling realistic micro-level behaviours (e.g., bounded rationality, routines, stochastic decision processes) and \rev{interactions of the agents} 
(e.g., imitation, network effects, spatial influence). \rev{These features - heterogeneity and interactions - give rise to aggregate dynamics that are qualitatively different, and cannot be deduced a priori, from those characterizing single agents. Indeed, ABMs often show the emergence of statistical equilibria \citep[e.g.,][]{gatti2005new, gatti2018agent, dosi2019more}, i.e., states of macroscopic equilibrium characterized by stable statistical distributions of variables describing aggregate phenomena accompanied by possibly unstable and evolving microscopic behaviours \citep{feller1957}. 

Depending on the applications, one might be interested in studying if - and how - the system reaches a statistical equilibrium, whether the equilibrium is unique, what properties the system displays therein, and whether they change for different parameter values and initial conditions  \citep{windrum2007empirical, grazzini2012analysis, fagiolo2019validation}. For example, the macro ABM literature has traditionally focused on characterizing long-run behaviours of macroeconomic aggregates under different policy regimes, washing away dependencies from initial conditions and transient dynamics \citep{fagiolo2017macroeconomic}; contrarily, history-friendly models rooted in evolutionary economics have often focused on microeconomic transients, interpreted as industrial paths of development \citep{malerba1999history}.}

\rev{As closed-form solutions of the distributional dynamics of ABMs are rarely available, the analysis must rely on numerical simulations.
However, we believe that little attention has been devoted to simulation protocols. While 
decisions about (i) how many steps to run, (ii) how many steps to use as  warmup (or transient) period, 
and (iii) how many runs to perform under each parameter configuration deeply influence the reliability of the analysis, they are often poorly justified.}
For example, statements like ``\emph{the results have been averaged over $n$ simulations}'' or ``\emph{we run a Monte Carlo exercise of size $n$}'', without 
a proper justification for the choice of $n$, are rather ubiquitous \citep[see, e.g.,][]{beygelzimer2012,kets2014,caiani2016agent,lamperti2018_fsc,lamperti2019public,dosi2019endogenous,fagiolo2020}. 
While irrelevant for ``thought experiments'', these aspects deserve attention when different policies are compared in counterfactual
simulation  experiments, or multiple parameter configurations are explored to 
discriminate among emerging behaviours. \cite{Secchi2017} 
conducted a study on 55 ABMs 
published 
between 2010 and 2013 in high-quality
management and organizational science journals. Their study showed that - in most 
cases - 
simulation exercises did not offer acceptable statistical quality,\footnote{The authors analyse the power of \rev{F}-tests 
in simulations on different model configurations.
} casting 
doubt on the results and their implications. The main 
cause of low statistical accuracy turned out 
being an insufficient number ($n$) of simulations performed.
Similarly, 
a poor handling of transient behaviours can 
distort results \rev{and the interpretation of steady-state behaviours} \rev{\citep[see, e.g.,][]{gal_aacute_n2005}}. 
Furthermore, 
ergodicity tests are necessary to establish whether 
performing a steady-state analysis makes sense at all.

In our opinion, these problems are \rev{partially} due to the fact that the simulation-based analysis of ABMs (i.e., the inspection of \rev{simulations of models}) is  \rev{sometimes \emph{handcrafted}, i.e., there is not \emph{one} set of well-engineered procedures and tests widely accepted by the whole community and available as robust software artifacts}, 
resulting in a time-consuming and error-prone process \citep[see also][]{lee2015complexities}. 
\rev{Notable exceptions are the R packages RNetLogo \citep{rnetlogo}, calibrar \citep{carrella2021} and freelunch \citep{carrella2021} that offer a number of statistical analysis techniques for ABMs complementary to the ones considered in this paper.}
The simulation, operations research, and computer science communities have substantially advanced the engineering of such tasks, 
developing automatic techniques equipped with statistical guarantees 
\citep[see, e.g.,][]{10.5555/554952}. 
While cross-disciplines fertilisation \rev{existed already in the previous century \citep[see, e.g.,][]{KWIATKOWSKI1992159}, and it} has \rev{further} recently 
increased \citep{dahlke2020juice}, these 
developments are often overlooked by the so-called ACE (agent-based computational economics) 
community.
In Section~\ref{sec:outputanalysis}, we use the model by \cite{grazzini2012analysis} to illustrate an example concerning the identification of transient dynamics.
%

\rev{This paper introduces a novel, automated and efficient approach to the inspection of stochastic ABMs over simulation time and across alternative configurations of parameters. 
We distinguish between: (i) \textit{transient analysis},
 which focuses on estimating properties of the model at specific points in time equipping them with a reliable measure of uncertainty; 
and 
(ii) \textit{steady-state analysis},
which focuses on the \emph{long-run} behaviour of the model and must be invariant to the transient dynamics. 
Further, we include in our steady-state analysis a methodology for 
\textit{ergodicity diagnostics}, which provides indications on whether the model behaves ergodically,
and thus on 
the reliability of the 
steady-state analysis itself. 
%
By leveraging statistical model checking~\citep{Agha18,LLTYSG19}, a successful simulation-based verification approach from computer science,  and
 \mv~\citep{SebastioV13,GilmoreRV17}, a statistical model checking tool-box that can be integrated with existing simulators to perform automated and distributed statistical analysis,
we aim to deliver a set of techniques and algorithms which:
(i) is easy-to-use by the modeller, (ii) improves 
reproducibility of the results, (iii) distributes simulations across the cores of a machine or across computer networks, (iv) automatically chooses a sufficient number of simulations and simulation steps to reach a user-specified 
statistical confidence, and (v) automatically runs a variety of statistical tests that are often overlooked by practitioners. 
\rev{While previous versions of} \mv{} have been successfully applied 
in a wide range of domains including, e.g., 
highly-configurable systems
\citep{BeekLLV20,DBLP:journals/corr/BeekLLV15,DBLP:conf/fm/VandinBLL18}, public transportation systems, 
\citep{gilmore2014analysis,ciancia2016tool},  
security risk modeling \citep{DBLP:journals/compsec/BeekLLV21}, 
 biological systems~\citep{GilmoreRV17}, 
 business process modeling \citep{DBLP:journals/jss/CorradiniFPRTV21}, collective adaptive systems \citep{DBLP:conf/wsc/GalpinGLV18}, 
robotic scenarios with planning capabilities~\citep{belzner2015onplan,belzner2014reasoning}, and crowd steering scenarios~\citep{DBLP:conf/hpcs/PianiniSV14}, it has never been employed for the analysis of the transient and steady-state properties of ABMs. 
%
\footnote{
\rev{\mv{} originally supported Java and C++,  simulators. 
We extended it in terms of  support for further environments like R and Python, and for } 
statistical tests (and their power) to compare 
different model parametrizations, and ex-novo \rev{design and} development of 
steady-state analysis. 
Furthermore, by applying it to two known ABM models, we also contribute to increasing the accessibility of ABMs and to the replicability of their results. 
\mv is maintained by one of the authors.  \rev{More information is provided in~\ref{sec:mv}.}
%
}

}

Our 
work contributes to two strands of the ACE literature. First, it complements 
many recent proposals for the validation of simulated models \citep[see the surveys in][]{fagiolo2019validation, lux2018empirical}. The method proposed by \cite{guerini2017method} for macro ABMs requires the model to be in a steady state and
to remove the transient period from the analysis;   
calibration approaches based on simulated moments \citep{winker2007, franke2012structural, grazzini2015estimation}, 
as well as recent Bayesian techniques \citep[see, e.g.,][]{Grazzini2017}, 
\rev{apply} to ergodic models.\footnote{\rev{In non-ergodic settings, a possible calibration procedure might rely upon techniques like those in, e.g.,~\cite{SERI202162}.}}
Further, 
our transient analysis 
can be employed to evaluate probabilities of observing certain patterns, 
which would support the use of validation metrics recently proposed in the literature \citep[see, e.g.,][]{barde2016direct, lamperti2018empirical, lamperti2018information}. 
Second, we contribute to the analysis of the complexities of ABM output \citep{lee2015complexities, mandes2017complexity, kuka2020} 
by providing fast and
practical tools  
to inspect 
models 
with statistical guarantees \citep{Secchi2017}, and by complementing the proposals in \cite{seri2017} 
to determine the adequate number of simulation runs. 
Finally, we offer an automated environment to carry out tests across \rev{some of  the} experiments are typical in the macro ABM literature \cite[see, e.g.,][]{Dosi15, lee2015complexities}.
 

\rev{We proceed as follows. Section \ref{sec:outputanalysis} introduces the framework of analysis; Sections~\ref{sec:algorithms} and ~\ref{sec:methodologyergodicity} introduce our algorithms and methods, respectively. Next, we showcase the proposed approach on two established ABMs from the literature.}\footnote{\rev{All} materials \rev{and instructions} for replicating the experiments presented in this paper are available at \url{https://github.com/andrea-vandin/MultiVeStA/wiki}. \rev{Furthermore, \mv is freely available from the website together with information on how to integrate new simulators.}}
%
Section~\ref{sec:macro} 
replicates and enriches the transient analysis from~\cite{caiani2016agent} on a large-scale benchmark 
stock-flow consistent macro ABM.
We optimize the number of simulations 
to reach a given (user-defined) level of statistical precision for each time point of interest and \rev{use such information 
to establish, in a statistically sound manner, differences across model configurations.
We show that our approach  
facilitates the interpretation of counterfactual experiments. In particular, we focus on a behavioural and a policy experiment, which highlight that business cycles are sensitive to changes in risk aversion (yet just in the short run), while relatively small income tax variations tend to produce significant and enduring effects on aggregate dynamics.}
Section~\ref{sec:predmarket} performs a steady-state analysis of the prediction market model of \cite{kets2014}. 
This model has been chosen because of its analytical tractability, which provides an effective ground truth against which we test our \rev{techniques}.\footnote{The model has been analytically studied in~\cite{bottazzigiachini2019b}, proving asymptotic results about agents' wealth and market price.} We show that an erroneous identification of the transient period 
led to misleading qualitative and quantitative results in the original simulation-based analysis by~\cite{kets2014}. Indeed, we first show that the agents' relative 
wealths and the 
relationships between market price and other model parameters were 
incorrectly characterized and, second, we provide a numerical solution matching the analytical results of \cite{bottazzigiachini2019b}.
%
Finally, Section~\ref{sec:nonergodic} applies our methodology for ergodicity analysis to (non-ergodic) variants of 
this prediction market model, showing how it can be used to further increase the reliability of a steady-state analysis, while Section~\ref{sec:conclusions} concludes the paper.
\rev{The paper contains an appendix where: \ref{appendix:MHT} discusses the potential multiple hypothesis problem related to our techniques; \ref{appendix:macro} presents further details on the transient analysis performed on the considered macro ABM; \ref{appendix:ketsetal} and \ref{appendix:market} provide further details on the market ABM model and on the steady-state analysis performed on it, respectively; \ref{sec:mv} provides details on   statistical model checking and on the tool-support for our techniques, while \ref{sec:parallelizationstudy} demonstrates the run-time gains offered by offered parallelization capabilities.}

\section{
Analysis of simulation output
}\label{sec:outputanalysis}
%
%
\rev{The analysis of ABMs} 
typically employs
stochastic simulations, relying on Monte Carlo methods, \rev{e.g.,}
to derive reliable estimates of the true model characteristics \citep{richiardi2006common, lee2015complexities, fagiolo2019validation}. 

Without loss of generality, one can represent an ABM as a mapping $\textit{map}: I \rightarrow O$ from a set of input parameters $I$ into an output set $O$. \rev{Usually, }
$I$ is 
a multidimensional space spanned by the support of each parameter\rev{, while} 
$O$ is typically larger and more complex,
as it 
comprises time-series realizations of a very large number of micro- and macro-level variables. 
In most cases we can think of the output of an ABM as a discrete-time stochastic process $(\mathbf{Y}_t)_{t>0}$ describing the longitudinal evolution of a 
vector of variables of interest (e.g., the wealth of an agent, 
the GDP of a country, etc.).
For 
simplicity, here we focus on the case in which $(\mathbf{Y}_t)_{t>0}$ contains only one 
time series of interest $(Y_t)_{t>0}$. 
However, 
our framework straightforwardly
covers the concurrent analysis of multiple time series.

\begin{figure}[t]
\centering
\subfloat[Simulations]{\includegraphics[scale=0.5]{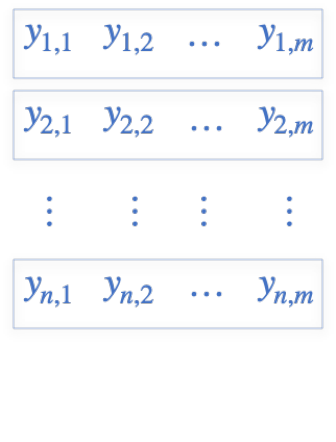}}
\hfill
\subfloat[Transient analysis]{\includegraphics[scale=0.5]{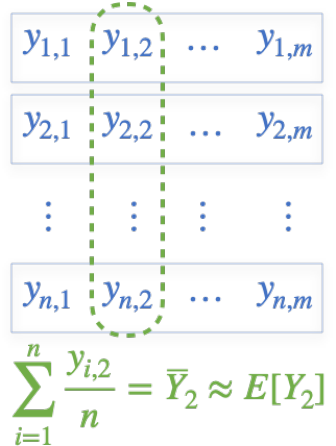}}
\hfill
\subfloat[Steady-state analysis by \emph{Replication and Deletion} (RD)]{\includegraphics[scale=0.5]{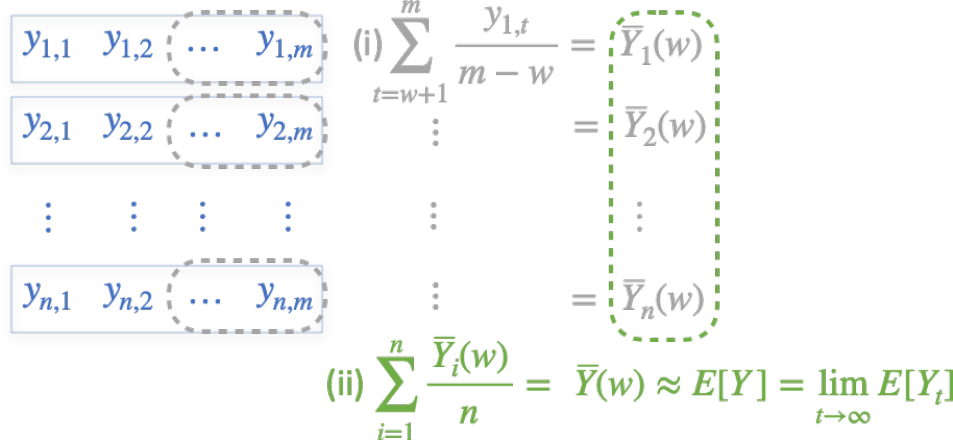}}
\caption{Transient and steady-state analysis using \emph{n} simulations of \emph{m} steps each}
\label{fig:nmsimsAnalysis}
\end{figure}

Figure~\ref{fig:nmsimsAnalysis}(a)
depicts 
$n$ independent simulations of 
$Y_t$ (one per row) each comprising 
$t=1,\ldots, m$ steps (one per column).
\footnote{By {\em independent simulations} we 
mean runs obtained from different random seeds that have been used for each replication, with the simulator status reset to an initial configuration at the beginning of each replication.}
The outcome of a simulation $i$ 
is therefore a sequence $\{y_{i,1},\ldots,y_{i,m}\}$
denoting a realization of length $m$.
%
Clearly, the observations within the same row $i$ are not independent, while those in the same column $t$ are independent and identically distributed (IID). 
%
%
Here we focus on two typical classes of properties:
\begin{itemize}
\item \emph{Transient properties} concerning $E[Y_t]$; what is the expected value of a property of a model at a given 
time $t$ (or 
within a time range, or at the
occurrence of a specific event)? 
\item \emph{Steady-state properties} concerning $E[Y]\!=\!\lim_{t\to\infty} E[Y_t]$; what is the expected value of a property of a model at steady state (
i.e., when the system has reached a statistical equilibrium)?~\footnote{\rev{
		What we mean by steady state for a stochastic simulation model consists in reaching a statistical equilibrium, that is - to repeat -  a state of macroscopic equilibrium maintained by a large number of transitions in opposite directions \citep{feller1957}.}}
\end{itemize}
%
An example of transient property
is given in Section~\ref{sec:macro}:
\emph{what is the expected 
unemployment rate in each of the first 400 quarters of a macro ABM?}
In this example, a transient analysis is
particularly important because the model has been designed to study fluctuations in the quarters following a given initial condition. 
In contrast,
an example of steady-state property
is given in Sections~\ref{sec:predmarket} and~\ref{sec:nonergodic}:
\emph{given a market with repeated sessions, what is the expected wealth of each agent at steady state?}
In this 
example, a steady-state analysis is
particularly important because the model has been designed to study problems of market selection and informative efficiency.
Obviously, a steady-state analysis is meaningful only ``around'' a statistical equilibrium.
This requires that $\lim_{t\to\infty} E[Y_t]$ exists and is finite. 
We 
first present two complementary techniques for steady-state analysis that 
rely on such assumption, 
and then (in Section~\ref{sec:methodologyergodicity}) we combine them into a methodology for ergodicity diagnostics; that is, \rev{a methodology} for assessing whether the assumption is reasonable or clearly violated. 
\footnote{We leave to future work 
extensions of our framework 
that would allow us to detect the number and nature of the statistical equilibria of a simulation model.}
%


Figures ~\ref{fig:nmsimsAnalysis}(b) and (c) depict how to compute statistical estimates for $E[Y_t]$ and $E[Y]$\rev{, respectively}. 
Such estimates can 
and should be accompanied by 
appropriate measures of uncertainty, e.g., 
computing 
``$\alpha$-$\delta$ confidence intervals'' (CI) around them. 
Given two user-specified parameters $\alpha \in (0,1)$ and $\delta \in \mathbb{R}^{+}$, we will show how to guarantee with statistical confidence $(1-\alpha)\cdot 100\%$ that the actual expected value belongs to the interval of width $\delta$ centred 
at its estimate, and how to optimize the number of runs needed to obtain such 
guarantee. These steps,
which are sometimes overlooked in the ABM community, 
can make the statistical analysis of 
\rev{ABMs} sounder and more informative for policy analysis.
We now provide more details on our proposals for transient and steady-state analyses.


\paragraph{Transient analysis}
Procedures for 
transient analysis are well-established and relatively simple. 
As shown in Figure~\ref{fig:nmsimsAnalysis}(b), for a given 
time point $t$ (a column) we obtain an 
unbiased estimator for $E[Y_t]$ by computing the \emph{vertical} mean $\overline{Y}_t$ of the observations at 
$t$ (across the rows). 
Since these observations are IID, we can use 
standard statistical techniques
based on the law of large numbers to build CIs as follows (see Chapter 9 of~\citealp{10.5555/554952}):
\begin{equation}\label{eq:ci}
\overline{Y}_t \pm \mathbf{t}_{n-1,1-\frac{\alpha}{2}}\cdot\sqrt{\frac{s^2_t}{n}}\text{,}
\end{equation}
where $n$ is the number of 
simulations, $s^2_t$ is the sample variance of $Y_t$, and the multiplier
$\mathbf{t}_{n-1,1-\frac{\alpha}{2}}$
is obtained from the tabulation of the Student's t distribution with $n-1$ degrees of freedom
(the area under the density function integrated from minus infinity to $\mathbf{t}_{n-1,1-\frac{\alpha}{2}}$ is equal $1-\frac{\alpha}{2}$).~\footnote{\rev{The Student’s t distribution is exact if the data are normally distributed with the same variance; it is an approximation, widely accepted by the researchers (see~\citealp{10.5555/554952}), if they are not. In particular, it is a  penultimate distribution \citep[see, e.g.,][]{penultimate}.}}
%
For any fixed confidence level $\alpha$,
the width of 
the CI decreases as $n$ increases. Therefore, in an automated procedure for computing an $\alpha$-$\delta$ CI, we can 
continue performing new simulations 
until the width 
becomes smaller than the desired $\delta$ 
(the target width can also be expressed as a fraction of the  mean value; $\delta$\% of $\overline{Y}_t$ ). Note that the CI width shrinks slowly, 
at the rate of the square root of $n$. Therefore, it is important to perform the \emph{correct} number of simulations to 
guarantee the target width without performing unnecessary computations. 
%
%
\rev{Our proposed techniques offer} 
an automated procedure for doing so. Furthermore, 
since in many cases different times $t$ might have different variances $s^2_t$, we account for the fact that different number of simulations 
 might be required at each
$t$ to get CIs of homogeneous width
across times.
As we will see in Section~\ref{sec:counterfactual}, this is particularly important for  counterfactual analysis.

\revTwo{We remark that the parameter $\delta$ provides a sort of ‘precision’ on the performed estimations. Equation~\eqref{eq:ci} shows an estimated mean ($\overline{Y}_t$) with its confidence interval (the right-hand-side part starting with $\pm$). The parameter  $\delta$ is used to impose a maximal width to all such CIs: the algorithms that we will present here are designed so to perform enough simulations (and simulation steps) to guarantee that all the computed CIs have a width of at most $\delta$.}
\rev{Clearly, for a given statistical confidence $\alpha$, the smaller is $\delta$, the tighter will be the computed CIs, therefore the more \emph{precise} will be the estimations. At the same time, smaller values of $\delta$ require more  simulations. Therefore, when choosing the value of $\delta$ for the analysis at hand, one has to keep into account such a trade-off.}

A common exercise that builds upon transient analysis is to compare estimates obtained for 
different model configurations 
(typically 
corresponding to different sets of input parameter values) -- 
as to assess whether 
the configurations differ significantly in terms of the output variables under consideration.
Given the outcomes of the transient analyses for the two 
configurations and a user-defined significance level $a_w$, our \rev{set of techniques} 
performs 
a  \emph{Welch's t-test} of means' equality~\citep{welch1947} for every
$t$ of interest. 
\rev{The proposed techniques also allow for the computation of} 
power of such test~\citep{chow2002} 
in detecting a difference of at least a given precision $\varepsilon$ (see  Section \ref{sssec:eqtest_power}).\footnote{
In Section~\ref{sec:counterfactual} we show that a reasonable choice is to set $\varepsilon=\delta$.
\rev{Our techniques also support so-called \emph{Wilcoxon-Mann-Whitney} test, or just \emph{u-test}~\citep{10.1214/aoms/1177730491}. This is an alternative to Welch's t-test with 
	milder assumptions, for which, however, to the best of our knowledge no closed-formula exists for power estimation. This is presented in \ref{sec:utest}.}
}

%
%

\paragraph{Steady-state analysis}
Figure~\ref{fig:nmsimsAnalysis}(c) depicts how 
a steady-state analysis can be performed similarly to 
a transient 
\rev{one} by adding a pre-processing \rev{phase}. We first compute the \emph{horizontal} mean $\overline{Y}_i(w)$ within each simulation $i$, ignoring a given number $w$ of initial observations. Since all 
these means are IID, 
we can compute their \emph{vertical} mean 
$\overline{Y}(w)$ 
and build a CI around it as in Equation~\eqref{eq:ci}. 
%
%
Unfortunately, this approach has 
intricacies 
that 
hinder its automatic implementation and 
can lead to relevant analysis errors. 
Depending on the chosen number $w$ of initial observations to discard,  
the estimator $\overline{Y}(w)$ of $E[Y]$ might carry a bias due to the transient behaviour of the system,
and not give us reliable information on 
its steady state 
(see Section~\ref{manual} for a notable example from the literature). 
In order to avoid this issue, we need to identify the \emph{correct} 
$w$ 
the system needs to exit 
its transient (or \emph{warmup}) period, and discard 
the initial $w$ 
observations from each simulation. Such procedure is known as \emph{Replication and Deletion}~\citep[RD, ][]{10.5555/554952}. 
Effectively identifying the length of the warmup period is a difficult problem.  The most popular approaches in the ABM community are rooted in the Welch's 
method~\citep{welch1983statistical}: 
\begin{enumerate}
\item Perform $n$ simulations of 
given 
length $m$ and compute averages $\overline{Y}_t$, $t=1,\ldots,m$ as in Figure~\ref{fig:nmsimsAnalysis}(b); 
\label{point1}
\item Plot $\overline{Y}_t$, $t=1,\ldots,m$
\label{pointplot}; \footnote{In Point~(\ref{pointplot}) 
one might smooth the plot, e.g., 
employing moving-windows averages, 
where one is in effect further averaging each $\overline{Y}_t$ with a few neighbouring steps.}
\item Choose the 
time $w$ after which the plot \emph{seems to converge}. If no such
time exists, iterate the procedure from point~(\ref{point1}), 
performing a new batch of $n$
simulations of length $m$, and computing averages over all 
simulations. 
\end{enumerate}
%
%
Being only semi-automated and
based on a visual assessment, this procedure is 
time consuming, 
error-prone, and
not backed by a strong statistical justification. It also critically depends on choosing a large enough ``time horizon'' $m$ -- of course progressively larger $m$ can be tried, adding to the computational burden.

More recently, \cite{grazzini2012analysis} presented an alternative approach where 
a single simulation of length $m$ is performed and divided into 
 \emph{windows} of length $\mathit{wi}$
 ($m$ and $\mathit{wi}$ are arbitrarily chosen). If the distribution of the means computed within each window passes a randomness test~\citep[in particular the Runs Test by][]{doi:10.1002/bimj.4710280806,10.2307/2235872}, 
then the author concludes that the system is in steady state. 
The use of statistical tests rather than 
visual assessments makes the approach more reliable, fostering its
use in the ABM literature~\citep[e.g.,][]{guerini2017method, lamperti2020climate}. 
However, the approach is still not fully automated -- and relies 
on the arbitrary choice of $m$ and $\mathit{wi}$: quoting from the author ``\emph{with appropriate settings the tests can detect non-stationarity}''~\citep{grazzini2012analysis}. In the next \rev{section} we introduce a fully automated statistical procedure for estimating the end of the warmup period. 

\section{Automated simulation-based analysis with statistical guarantees}\label{sec:algorithms}

Our approach to transient 
and steady-state 
analysis 
is fully automated, in that all parameters are computed automatically or have default values.  
The user specifies the properties to be studied, the 
$\alpha$ and $\delta$ parameters to be employed in the CI construction, and an optional maximum number of allowed simulations (if this number is reached before satisfying the CI constraints, the analysis terminates with the currently computed CIs). 
%
As in Section~\ref{sec:outputanalysis}, we focus the description on 
a single 
variable $Y_t$, 
but our treatment applies straightforwardly to the
analysis of 
multiple model characteristics 
%
(indeed, 
\rev{our actually implemented techniques support}
multi-variable analyses). 

\rev{We have to notice here that our techniques make extensive use of statistical testing and the probability of observing a supposedly significant difference under the null hypothesis may be larger than the nominal value $\alpha$. This problem goes under the \emph{multiple hypothesis testing problem}. We critically discuss such an issue in \ref{appendix:MHT}.}

\subsection{Transient analysis}\label{subsec:transient}
Section~\ref{sec:mean_ci} 
describes how to estimate transient properties 
expressed as expected values, $E[Y_t]$, and how to build CIs around them. After this, 
Section~\ref{sssec:eqtest_power} 
describes how to statistically compare estimates 
from different experiments. 

\subsubsection{
Mean estimation and CI computation}\label{sec:mean_ci}
Algorithm~\ref{algorithm:autotransient} 
illustrates \transient, a simple automated algorithm for transient analysis 
that takes in input $\mathit{bl}$ (discussed later), a 
time of interest $t$, and 
 $\alpha$ and $\delta$, 
and produces in output  
an estimate of $E[Y_t]$ and a corresponding CI. The algorithm 
determines automatically the number $n$ of simulations required 
to guarantee that the  $(1-\alpha)
\times 100\%$ CI centred 
at the estimate 
has width at most $\delta$.

\llines{tr1}{tr3} 
set $t$ as time horizon $m$, 
and initialize the counter $n$ of computed simulations and the list $\mu$ to store the observations at step $t$ from each simulation (the $y_{i,t}$ in Figure~\ref{fig:nmsimsAnalysis}(b)). 
%
\llines{for1}{forend} perform a \emph{block} of $\mathit{bl}$ simulations\rev{, by default $20$}~\citep{10.5555/554952}, 
populating $\mu$. 
In \lline{draw}, $y$ is a list of size $m$  containing a value $y_{i,t}$ for each time point $t$ from $1$ to $m$ for the current simulation $i$, but only the value for $t=m$ is used, adding it to $\mu$. 
After performing $\mathit{bl}$ simulations, \autoir computes the mean $\overline{\mu}$ and variance $s^2$ of $\mu$, used to compute the width $d$ of the current CI. 
If $d$ is greater than $\delta$,
\footnote{For the sake of presentation, all algorithms in the paper consider $\delta$ given as absolute values. The case of $\delta$ given in percentage terms relatively to the studied means is trivially obtained by changing the comparisons $d>\delta$ in  ${d}/{\overline{\mu}}>\delta$.}
\autoir performs another block of $\textit{bl}$ simulations, otherwise it 
returns the 
current CI. 
The 
\rev{actual} 
implementation of \autoIR{} \rev{ used in later sections} 
allows one to concurrently estimate $E[Y_t]$ for different time points $t$ (e.g., average bankruptcies 
in each $t$ from $1$ to $400$ in Section~\ref{sec:macro}). This is 
done by computing, at each iteration, 
mean, variance and CI only for the elements 
of $y$ 
(\lline{draw}) 
that correspond to time points whose current 
CI width is still above $\delta$. 
At each iteration of 
a block of $\mathit{bl}$ simulations, the time horizon $m$ is updated with the largest $t$ still to be processed.

\subsubsection{Test for equality of means and power computation}
\label{sssec:eqtest_power}
\rev{Our proposed set of techniques} 
allows 
one to compare, 
in a statistically meaningful and reliable way, expected values corresponding to different settings or parametrizations of a model. 
Given that the compared means might come from experiments 
with different 
sample sizes and variances, we 
use the Welch's t-test~\citep{welch1947}, whose \emph{power} can be computed as in  \cite{chow2002}.

\begin{figure}[t]
\centering
\scalebox{0.83}{
\begin{minipage}{.53\linewidth}
\begin{algorithm}[H]
	\caption{\label{algorithm:autotransient} \transient: \emph{Transient analysis}} 
	\begin{algorithmic}[1]
\Require {\small $\mathit{bl}$, $\alpha$, $\delta$, $t$ (default: 20, 0.05, 0.1, NA)}
		\State //\emph{Set $t$ as time horizon $m$, and initialize data structures}\label{tr0}
		\State $\mathit{m} \leftarrow \mathit{t}$\label{tr1}
		\State $n \leftarrow 0$\label{tr2}
		\State $\mu \leftarrow \text{empty list}$\label{tr3}
		\Repeat \label{repeat1}
			\For{$i \in \{1,\ldots, bl\}$}\label{for1}
				\State $y \leftarrow \texttt{drawIndependentSimulation}(m)$\label{draw}
				\State //\emph{Add $y_{i,t}$ to $\mu$}
				\State $\mu.\texttt{add}(y[m])$\label{trym}
				\State $n \leftarrow n+1$
			\EndFor\label{forend}
		\State $(\overline{\mu},s^2) \leftarrow \texttt{computeMeanAndVariance}(\mu)$\label{mean}
		\State $d \leftarrow \texttt{computeCIWidth}(\overline{\mu},s^2,n,\alpha)$\label{ci}
		\Until{$d > \delta$}\label{repeatend}
		\State \Return $(1-\alpha)\cdot100\%$ CI $[\overline{\mu} - \frac{d}{2},\overline{\mu} + \frac{d}{2}]$ of width at most $\delta$ 
	\end{algorithmic}
\end{algorithm}
\end{minipage}
}
\hfill
\scalebox{0.83}{
\begin{minipage}{.615\linewidth}
\begin{algorithm}[H]
	\caption{\label{algorithm:autord} \autord: \emph{Steady-state analysis by Replication and Deletion}} 
	\begin{algorithmic}[1]
\Require {\small $B$, $b$, $\mathit{bs}$, $\mathit{minVar}$, $\mathit{bl}$, $\alpha$, $\delta$ (default: 128, 4, 16, 1E-7, 20, 0.05, 0.1)}
		\State $\mathit{w} \leftarrow \autow(B,b,\mathit{bs},\mathit{minVar})$ \label{autord1}
		\State $\mathit{m} \leftarrow \mathit{w}\cdot 2$\label{autord2}
		\State $n \leftarrow 0$
		\State $\mu \leftarrow \text{empty list}$\label{autord3}
		\Repeat 
			\For{$i \in \{1,\ldots, bl\}$}
				\State $y \leftarrow \texttt{drawIndependentSimulation}(m)$
				\State $y' \leftarrow (y_{w+1},\ldots, y_{m})$ \label{rdw}
				\State $\mu.\texttt{add}(\texttt{computeMean}(y'))$\label{rdmean}
				\State $n \leftarrow n+1$
			\EndFor
		\State $(\overline{\mu},s^2) \leftarrow \texttt{computeMeanAndVariance}(\mu)$
		\State $d \leftarrow \texttt{computeCIWidth}(\overline{\mu},s^2,n)$
		\Until{$d > \delta$}
		\State \Return $(1-\alpha)\cdot100\%$ CI $[\overline{\mu} - \frac{d}{2},\overline{\mu} + \frac{d}{2}]$ of width at most $\delta$ 
	\end{algorithmic}
\end{algorithm}
\end{minipage}
}
\end{figure}

\paragraph*{Welch's t-test}
Given 
estimates 
from two transient analyses for a 
set of time points $T$, 
our \rev{approach allows one to perform} 
a test for 
equality of means 
for each $t\in T$ using 
\cite{welch1947}. 
In symbols, given two experiments $\{j,k\}$, define the set of triplets $\mathcal{D}=\{(\overline{Y}_{i,t},s^2_{i,t},n_{i,t})\mid  i\in \{j,k\}, t\in T\}$, 
each containing 
the mean, the sample variance, and number of simulations for time $t$ in experiment $i$. \rev{We take} 
$\mathcal{D}$ as input and, for each $t$, compute\\
 
%
\begin{equation}
	\tau_t=\dfrac{\overline{Y}_{j,t}-\overline{Y}_{k,t}}{\sqrt{f_{j,t}+f_{k,t}}} \ \ ,
	\label{eq:Welchtt}
\end{equation}
where $f_{i,t}={s^2_{i,t}}/{n_{i,t}}$, $i \in \{j,k\}$. Following \cite{welch1947}, under the null hypothesis that the difference between the two means 
is zero, each $\tau_t$ 
\rev{is asymptotically normal, and it is usually approximated by a Student’s t distribution -- which can be considered a penultimate distribution --}
with degrees of freedom approximated as in \cite{satterthwaite1946}:

\begin{equation*}
\nu_t\approx\dfrac{(f_{j,t}+f_{k,t})^2}
                  {{f_{j,t}^2}/{(n_{j,t}-1)} + {f_{k,t}^2}/{(n_{k,t}-1)}}
                  \ \ .
\end{equation*}
Therefore, given a 
statistical significance $a_w$, 
\rev{we use} 
$\tau_t$ to perform the test of no difference between the two means producing $1$ if $\tau_t\in[-\mathbf{t}_{\nu_t,1-\frac{
a_w}{2}},\mathbf{t}_{\nu_t,1-\frac{
a_w}{2}}]$ (the null hypothesis of equal means is not rejected) and $0$ otherwise. The significance $a_w$ is user-specified, and can be set to be equal to the $\alpha$ used for the transient analysis.

\paragraph*{Power of the test}
Following \cite{chow2002}, \rev{we estimate} 
the power $1-\beta_t$ 
of Welch's t-test 
in  detecting 
a difference of at least a given precision $\varepsilon$ between the two means at 
time $t$. 
This is 
\begin{equation}\label{eq:power}
	\beta_t=\mathcal{T}_{\nu_t}\left(\mathbf{t}_{\nu_t,1-\frac{a_w}{2}}\Bigg{|}\frac{|\varepsilon|}{\sqrt{f_{j,t}+f_{k,t}}}\right) \ \ ,
\end{equation} 
where $\mathcal{T}_{\nu_t}(x \,|\, \theta)$ is the cumulative distribution function of a non-central t-distribution with $\nu_t$ degrees of freedom and non-centrality parameter $\theta$, evaluated at point $x$.
%
Calculating the power of 
Welch's t-test 
requires specifying 
the minimum 
difference $\varepsilon$ 
\citep{chow2002}. As a rule of thumb, we suggest setting $\varepsilon\geq\delta$, 
the parameter used in the transient analysis,
which expresses a precision for the estimated mean. In Section~\ref{sec:macro}, 
setting $\varepsilon=\delta$ leads to 
very good power for the considered macro ABM.


\subsection{Steady-state analysis}\label{subsec:steadystate}
\begin{figure}[t] \centering 
	\includegraphics[scale=0.5]
	{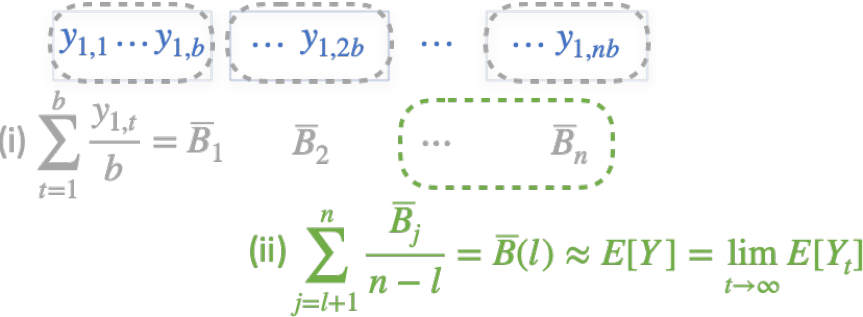}
	\caption{\small  
Steady-state analysis by \emph{Batch Means} (BM) using one long simulation: 
	(i) We split the simulation into batches (consecutive steps) of size $b$, and we compute the mean within each batch (the batch means $\overline{B}_i$); 
(ii) We compute the mean of such means, the \emph{grand mean}, ignoring the first $l$ batches where it is assumed to terminate the warmup. 
	We obtain  $\overline{B}(l)$, an estimator for $E[Y]$. 
}
	\label{fig:steadyBM}
\end{figure}

%
A statistically sound analysis of steady-state properties 
poses 
challenges that have been thoroughly investigated by the simulation community -- at the boundary of
computer science and operations research.
%
Two main 
approaches have emerged~\citep{alexopoulos2004batch,whitt1991efficiency,10.5555/554952}: those based on 
{\em Replication and Deletion} (RD, see Section~\ref{sec:outputanalysis}),
and those based on 
\emph{batch means} 
\citep[BM,][]{Conway1963,Alexopoulos1996,steiger2005asap3}.
%
Unlike RD,
which computes \emph{many short} simulations, 
BM computes \emph{one long} run which is evenly divided into adjacent non-overlapping subsamples labelled as \emph{batches}. Intuitively, if certain statistical properties hold, each batch can be used \rev{similarly to a} a simulation in RD -- as depicted
in Figure~\ref{fig:steadyBM}. 
This can be seen as a generalized version of the proposal by \cite{grazzini2012analysis}, which
allows one to estimate the end of the warmup period rather than to 
check whether a given time 
is subsequent to such end.\footnote{More precisely, \cite{grazzini2012analysis} 
appears to employ a non-automated version of BM.
Yet the first automated version of BM 
was published in 1979~\citep{doi:10.1287/opre.27.5.1011}. This is a clear signal of the potential (and often overlooked) complementaries 
between the simulation community and the 
ABM community in economics.
}

There is no \emph{best} approach 
between RD and BM~\citep{alexopoulos2004batch,whitt1991efficiency,DBLP:journals/ior/KeltonL84}. 
%
They are complementary, and therefore have complementary (dis)advantages. RD, 
which uses many short simulations, suffers from biases 
due to initial conditions.  BM, 
which uses many short batches from one long simulation, is less affected by initialisation bias but \rev{has to adopt corrections to deal with} 
correlations among batch means. 
%
While some  automated BM-based procedures have been proposed \citep[e.g.,][]{steiger2005asap3,tafazzoli2011performance,GilmoreRV17}, to the best of our knowledge, little attention has been paid to RD. Interestingly, \cite{lada2013ard} 
tried to combine the two approaches exploiting their respective strengths: they use BM 
for warmup analysis, 
and automate 
RD by discarding the estimated transient behaviour from each simulation. 
Following a similar approach, 
we extract and condense the warmup analysis capabilities inspired by BM 
into a simple self-standing procedure for warmup estimation 
(\autow), and we 
introduce automated RD- and BM-based algorithms (\autord and \autobm, respectively)  which use \autow. 
In all algorithms (see Figure \ref{fig:algorithms}) we 
favour simplicity and accessibility. 

\subsubsection{Steady-state analysis by replication and deletion}\label{sec:steadystaterd}
Algorithm~\ref{algorithm:autord} 
illustrates \autord. 
The \emph{difficult part} in automating 
RD 
is the warmup analysis. 
However, in our setting we can easily do this 
by invoking \autow (\lline{autord1}; see Section~\ref{sec:steadystatewu} below). For now 
it is sufficient to know that $w$ is the last step of the estimated warmup period.
Once $w$ 
has been determined, we have to set a 
\emph{substantially larger} time horizon~\citep{10.5555/554952}. We can do this 
using a (small) multiplier. In \lline{autord2} 
the default multiplier for $w$ is $2$.
%
%
The code of \autord presents also a second modification with respect to that of \autoIR: we replaced \lline{trym} of Algorithm~\ref{algorithm:autotransient} with \llines{rdw}{rdmean} of Algorithm~\ref{algorithm:autord} to discard the first $w$ observations from $y$, and add the mean of the remaining values of $y$ (the horizontal mean in Figure~\ref{fig:nmsimsAnalysis}(b)) to $\mu$. 

%

\subsubsection{Warmup estimation 
}\label{sec:steadystatewu}
Algorithm~\ref{algorithm:autow} provides pseudo-code for our automatic warmup estimation, inspired by existing BM-based approaches for steady-state analysis~\citep{steiger2005asap3,GilmoreRV17,tafazzoli2011performance}. Indeed, 
such algorithms include a form of warmup analysis that we extract and refine into a simple self-standing procedure. 
%
\llines{w1}{w2} perform a simulation of $m=B
\times bs$ steps (by default, $B=128$ and $\mathit{bs}=16$). The simulation is divided in $B$ adjacent non-overlapping \emph{batches}, each containing $bs$ steps. After this, 
the array $\mu$ stores the mean of each batch (therefore the name \emph{batch means}): each entry $\mu[i]$ stores the corresponding $\overline{B}_i$ (see Figure~\ref{fig:steadyBM}).
The algorithm then proceeds iteratively by performing statistical tests to check whether $m$ is large enough to cover the warmup period, doubling the number of performed steps while keeping 
the number of batches fixed (doubling the steps $\mathit{bs}$ in each batch) until all tests 
are passed. The key point is that if the process satisfies 
properties required for steady-state analysis (\rev{see the discussion on the \emph{mixing} property in Section~\ref{sec:methodologyergodicity}}),
then such iterative procedure will lead to \emph{approximately} IID normally distributed batch means $\mu$ for a sufficiently large value of $\mathit{bs}$. 

\begin{figure}[t]
\centering
\scalebox{0.83}{
\begin{minipage}{.47\linewidth}
\begin{algorithm}[H]
	\caption{\label{algorithm:autow} \autow: \emph{Warmup estimation}}
	\begin{algorithmic}[1]
		\Require {\small $B$, $b$, $\mathit{bs}$, $\mathit{minVar}$, (default: 128, 4, 16, 1E-7)}
		\State //\emph{Draw the first $B\cdot \mathit{bs}$ steps}\label{w1}
		\State $\mu \leftarrow \texttt{array}(\mathit{B})$
		\For{$i \in \{1,\ldots, B\}$}\label{wf1}
			 \State $\mu[i] \leftarrow \texttt{drawBatchAndComputeMean}(\mathit{bs})$
		\EndFor \label{w2}
		\State $(a, \rho) \leftarrow \texttt{goodnessOfFitTests}(\mu,b,\textit{minVar})$\label{wt1}
		\State //\emph{Keep doubling $\mathit{bs}$ and time horizon until tests pass}
		\While{$a > a^* \texttt{ or } \rho > \rho^*$} \label{wwhile1}
			\For{$i \in \{1,\ldots, B/2\}$}\label{wsqueeze1}
				\State $\mu[i] \leftarrow (\mu[2\cdot i] + \mu[2\cdot i+1]) / 2$
			\EndFor\label{wsqueeze2}
			\State $\mathit{bs} \leftarrow 2\cdot \mathit{bs}$
			\For{$i \in \{B/2+1,\ldots, B\}$}\label{wsecondhalf1}
			 \State $\mu[i] \leftarrow \texttt{drawBatchAndComputeMean}(\mathit{bs})$
			\EndFor\label{wsecondhalf2}
		\State $(a, \rho) \leftarrow \texttt{goodnessOfFitTests}(\mu,b,\textit{minVar})$
		\EndWhile\label{wwhile2}
		\State \Return Warmup period estimated to terminated after $\mathit{B}\cdot\mathit{bs}$ steps
	\end{algorithmic}
\end{algorithm}
\begin{algorithm}[H]
	\caption{\label{algorithm:goodnessOfFitTests}\texttt{goodnessOfFitTests}} 
	\begin{algorithmic}[1]
		\Require {\small $\mu$, $b$, $\emph{minVar}$}
		\State $\mu' \leftarrow (\mu_{b+1},\ldots, \mu_B)$\label{wmp}
		\State $(\overline{\mu},s^2) \leftarrow \texttt{computeMeanAndVariance}(\mu')$\label{wvar}
		\State $(\alpha,\rho) \leftarrow (0,0)$
		\If{$s^2 > \mathit{minVar}$}\label{wvarc}
			\State $a \leftarrow \texttt{AndersonDarlingNormalityTest}(\mu',\overline{\mu},s^2)$\label{wnorm}
			\State $\rho \leftarrow \texttt{lag1Autocorrelation}(\mu',\overline{\mu},s^2)$\label{wautocorr}
		\EndIf
		\State \Return $(a, \rho)$;
	\end{algorithmic}
\end{algorithm}
\vspace{0.0cm}
\end{minipage}
}
\hfill
\scalebox{0.83}{
\begin{minipage}{.65\linewidth}
\vspace{-0.05cm}
\begin{algorithm}[H]
	\caption{\label{algorithm:autobm} \autobm: \emph{Steady-state analysis by Batch Means}} 
	\begin{algorithmic}[1]
		\Require {\small$B$, $b$, $\mathit{bs}$, $\mathit{minVar}$, $\alpha$, $\delta$ (default: 128, 4, 16, 1E-7, 0.05, 0.1)}
		\State $\autow(B,b,\mathit{bs},\mathit{minVar})\ $ //\emph{Fast-forward simulation  after warmup}\label{alg:autobm1}
		\State $\mu \leftarrow \texttt{array}(\mathit{B})$\label{alg:autobm3}
		
		\For{$i \in \{1,\ldots, B\}$}
			 \State $\mu[i] \leftarrow \texttt{drawBatchAndComputeMean}(\mathit{bs})$
		\EndFor
		\State $(a, \rho, \overline{\mu},d) \leftarrow \texttt{goodnessOfFitTestsAndCI}(\mu,b,\textit{minVar},\alpha)$
		\State //\emph{Keep doubling $\mathit{bs}$ and time horizon until tests pass}
		\While{$a > a^* \texttt{ or } \rho > \rho^* \texttt{ or } d > \delta$} 
			\For{$i \in \{1,\ldots, B/2\}$}
				\State $\mu[i] \leftarrow (\mu[2\cdot i] + \mu[2\cdot i+1]) / 2$
			\EndFor
			\State $\mathit{bs} \leftarrow 2\cdot \mathit{bs}$
			\For{$i \in \{B/2+1,\ldots, B\}$}
			 \State $\mu[i] \leftarrow \texttt{drawBatchAndComputeMean}(\mathit{bs})$
			\EndFor
		\State $(a, \rho, \overline{\mu},d) \leftarrow \texttt{goodnessOfFitTestsAndCI}(\mu,b,\textit{minVar},\alpha)$
		\EndWhile
		\State \Return 
		$(1-\alpha)\cdot100\%$ confidence interval $[\overline{\mu} - \frac{d}{2},\overline{\mu} + \frac{d}{2}]$ of width at most $\delta$, adjusted for keeping into account residual correlation
	\end{algorithmic}
\end{algorithm}
\begin{algorithm}[H]
	\caption{\label{algorithm:goodnessOfFitTestsAndCI}\texttt{goodnessOfFitTestsAndCI}} 
	\begin{algorithmic}[1]
		\Require {\small $\mu$, $b$, $\emph{minVar}$, $\alpha$}
		\State $\mu' \leftarrow (\mu_{b+1},\ldots, \mu_B)$
		\State $(\overline{\mu},s^2) \leftarrow \texttt{computeMeanAndVariance}(\mu')$
		\State $(\alpha,\rho,d,n) \leftarrow (0,0,0,B-b)$
		\If{$s^2 > \mathit{minVar}$}
			\State $a \leftarrow \texttt{AndersonDarlingNormalityTest}(\mu',\overline{\mu},s^2)$
			\State $\rho \leftarrow \texttt{lag1Autocorrelation}(\mu',\overline{\mu},s^2)$
			\State $d \leftarrow \texttt{computeCIWidth}(\overline{\mu},s^2,n,\alpha,\rho)$
		\EndIf
		\State \Return $(a, \rho, \overline{\mu},d)$;
	\end{algorithmic}
\end{algorithm}
\end{minipage}
}
\caption{BM-based algorithms for estimating the initial warmup period (left), and for studying steady-state properties (right).}
\label{fig:algorithms}
\end{figure}

BM-based approaches perform different statistical tests on $\mu$ to check whether 
$m$ is large enough for completing the warmup period:
\cite{tafazzoli2011performance} use the \cite{neumann} randomness test,  
while \cite{steiger2005asap3} use a test for stationary multivariate normality on groups of 4 consecutive batches followed by a check for low correlation among consecutive batch means (i.e., the lag-1 autocorrelation of $\mu$). 
\cite{GilmoreRV17} apply the Anderson-Darling test for normality on $\mu$, followed by a check for low lag-1 autocorrelation of $\mu$. In all cases, a few (typically 4) initial batches are ignored as they are likely the most affected ones by the initial transient. 
We follow the latter approach, as specified in the subprocedure \texttt{goodnessOfFitTests} 
of Algorithm~\ref{algorithm:goodnessOfFitTests}: \lline{wmp} skips $b$ (by default 4) initial batches, obtaining $\mu'$, 
%
then \lline{wvar} computes the variance of the batch means, used in \lline{wvarc} to decide whether the statistical tests are necessary or pass by default. The rationale is that if the variance among the batch means is below a minimum threshold (parameter $\mathit{minVar}$ with default value 1E-7), then the process is likely 
converging to a deterministic fixed point, therefore we can safely assume that the initial warmup period has terminated. 
%
Concerning normality, \lline{wnorm} uses 
the Anderson-Darling test implemented in
the SSJ library~\citep{iLEC16j,sLEC02a} to check whether it is statistically plausible that $\mu'$ has been sampled from a normal distribution specified by
its mean and variance,
and 
obtain 
a p-value $a$.\footnote{\rev{The Anderson-Darling test may overweight the tails of the distribution, hence we provide the Cramer-Von Mises normality test as an alternative. See \ref{appendix:market} for a discussion and a comparative exercise.}} 
\lline{wautocorr} 
stores $\rho$, the lag1-autocorrelation of $\mu'$. 
The subprocedure thus returns 
$a$ and $\rho$, which are used in \lline{wwhile1} of 
\autow to decide whether the tests are passed --
using minimum thresholds $a^*$ and $\rho^*$ 
based on prior publications~\citep{GilmoreRV17,steiger2005asap3}.\footnote{In particular, the significance level for  the normality test is set to 
$a^*=1\%$
while the lag-1 autocorrelation threshold is set to $\rho^*= \sin(0.927-\frac{q}{\sqrt{\mathit{size}(\mu)}})$, where $q$ is the 99\% quantile of the 
standard normal distribution. \rev{See~\cite{steiger2005asap3}} for 
more details.}
If any of the tests fail, then an iteration of the {\em while} loop in \llines{wwhile1}{wwhile2} is performed to double the number of 
steps $m$ by doubling the current batch size $\mathit{bs}$. We note that the current $B$ 
batch means 
are \emph{squeezed} in the first half of $\mu$, (\llines{wsqueeze1}{wsqueeze2}), and $m$ new steps are performed to create the new batch means in the second half of $\mu$ (\llines{wsecondhalf1}{wsecondhalf2}). The statistical tests are performed on the new batch means, and new iterations of the loop are performed until both statistical tests 
are passes. The algorithm terminates returning the final value of $m=B
\times \mathit{bs}$ as the estimated end of the warmup period.

\subsubsection{Steady-state analysis by batch means}\label{sec:steadystatebm}
Algorithm~\ref{algorithm:autobm} 
illustrates our automatic BM-based procedure for steady-state analysis. 
%
\lline{alg:autobm1} invokes \autow, which \emph{moves} the simulator 
to the end of the estimated warmup period. 
%
No other information from 
\autow is used. 
%
The algorithm then proceeds similarly to \autow, the only difference being that we add a third statistical test: we also compute the width $d$ of the CI according to the current batch means. This is obtained by invoking \texttt{goodnessOfFitTestsAndCI} from Algorithm~\ref{algorithm:goodnessOfFitTestsAndCI} rather than 
\texttt{goodnessOfFitTests}. 
Since the tests for normality and 
absence of correlation 
already passed during \autow, 
one might expect that they are no longer necessary. 
We note however that the tests passed for the last
value of $\mathit{bs}$ used in \autow, 
so they could potentially fail for initial small values of $\mathit{bs}$. 
When all statistical tests are 
passed, we return the computed $(1-\alpha)100\%$ CI of width at most $\delta$, adjusted by the computed residual correlation among the batch means. This is done similarly to \cite{steiger2005asap3}, using an inverse Cornish-Fisher expansion~\citep{stuart} based on a standard normal density.~\footnote{See \rev{Section~2 of~\cite{steiger2005asap3}} for the 
exact formula.} 
%

\subsubsection{
Some remarks on \emph{\autobm} and \emph{\autord}}\label{sec:bmvsrd}
We note that both \autobm and \autord proceed by iterations, during which new samples are drawn and new statistical tests are performed. In 
\autobm, new simulation steps are added onto the same long simulation 
(the number of simulations $n$ is constant, 
the time horizon $m$ grows).
In \autord, new simulations of fixed length
are added ($n$ grows, $m$ is constant). 
In some sense, the computational burden of \autord
is higher, as $w$ steps from each newly performed simulation are ignored. However, the 
simulations performed in
each iteration of \autord can be trivially parallelized 
-- so the additional computation can be efficiently handled. Rather, which approach to prefer 
depends on the model at hand and on the available hardware: 
\begin{itemize}
\item The 
longer the initial warmup period, the more 
advantageous is \autobm 
relative to \autord. 
\item The \rev{higher is the} 
parallelism 
supported by the hardware, the more 
advantageous is \autord 
relative to \autobm. 
\end{itemize}

The ABM community favours the RD approach due to its simplicity, but its \emph{trivially parallelisable} nature is not always exploited due to limited computer engineering skills. Notably, 
some studies have shown that 
the BM approach might provide more accurate results in specific cases~\citep{whitt1991efficiency,alexopoulos2004batch}.
%
\rev{Our tool-supported algorithms} 
enable modellers to freely choose 
between the two \rev{approaches} and to exploit the distributed nature of the tool-box to parallelize simulations.
The next section shows how \rev{\autoRD{} and \autoBM{}} 
can be combined to obtain a methodology for ergodicity diagnosis.


\section{
Ergodicity diagnostics 
using \autord and \autobm}
\label{sec:methodologyergodicity}

\begin{figure}[t] \centering 
	\includegraphics[width=1.0\linewidth]{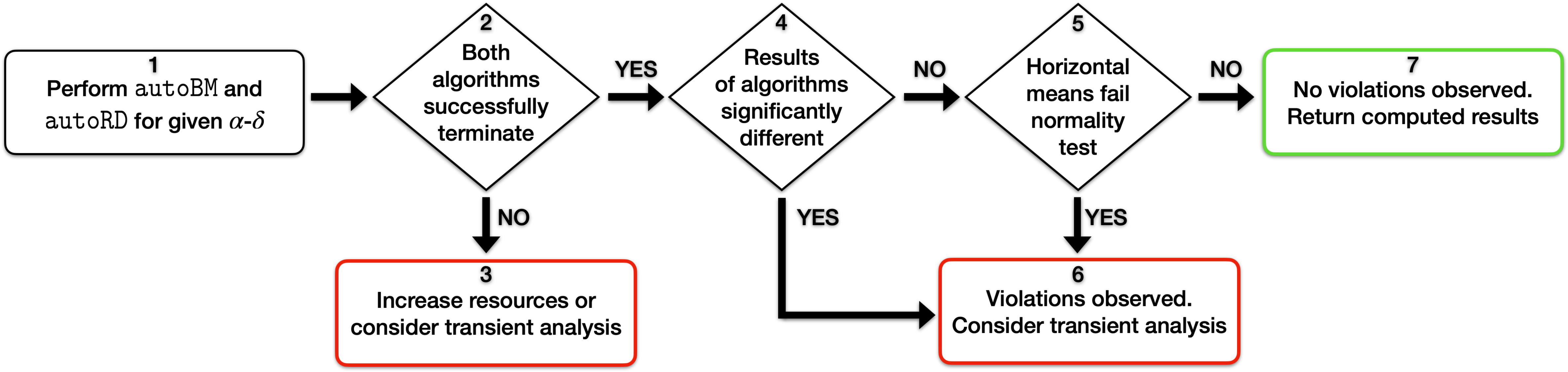}
	\caption{
	Procedure for ergodicity 
	diagnostics based on \autord and \autobm to assess the reliability of a steady-state analysis}

	\label{fig:ergo}
\end{figure}

\rev{As discussed, not all ABMs can be studied at steady state. We hereby provide a methodology to help the modeller understand whether she should focus instead on transient analysis. As a matter of fact,}
consistency and unbiasedness of the estimates produced by \autoBM and \autoRD rely on 
the underlying process 
possessing the {\em 
strong mixing} property.~\footnote{The 
strong mixing property 
guarantees that two sufficiently distant observations 
in $(\mathbf{Y}_t)_{t>0}$
are approximately independent\rev{, e.g., any correlation among them is negligible}. There are various definitions of the property; we utilize the 
$\phi$-mixing definition
provided in \cite{doi:10.1287/ijoc.13.4.277.9737} 
} 
Indeed, once normality of batch means in \autoW is
well approximated and autocorrelation is low, we can be confident that future observations will not have initialization bias \citep{steiger2005asap3}.
If at a certain point in time 
the batch means resemble a sample of IID observations from the same Gaussian population, then the 
effects of initial conditions must have 
disappeared.
Further, the 
strong mixing assumption ensures that such  a  point in time can eventually be reached by increasing the batch size 
\citep{doi:10.1287/opre.27.5.1011,doi:10.1287/ijoc.13.4.277.9737}. 
%
\rev{Figure~\ref{fig:ergo} depicts a procedure}
that combines \autoRD and \autoBM to assess 
whether this assumption is met. 
\rev{We have fully implemented this procedure, allowing for its validation} 
in Section~\ref{sec:nonergodic} on variants of a prediction market model.

\rev{The procedure is as follows.}
We start 
performing both \autoBM and \autoRD for given $\alpha$ and $\delta$ (step 1). If any of the two fails to converge in due time (step 2), we have 
evidence that the process is eventually non-stationary
(or fails to reach its stationary phase 
within the allotted computational time/resources). 
In such cases, performing any steady-state analysis
could be misleading and should be avoided (step 3). 
If both \autoBM and \autoRD successfully terminate, 
we can be confident that the process possesses \emph{ergodic properties}~\citep{gray2009,billingsley1995} -- meaning, intuitively, that the horizontal means of its realisations 
(i.e., the means across 
simulations as in Figure~\ref{fig:nmsimsAnalysis}(c)) indeed converge asymptotically to a finite number 
\rev{ \citep[see also][]{9526649}}.
However, there could be potentially 
different limits for different simulations.
In these cases, a natural 
check for ergodicity is to compare the results of \autoBM and \autoRD (step 4). This is in line with previous approaches to ergodicity analysis from the literature
\citep[e.g.,][]{grazzini2012analysis}, where BM-like means across one long simulation are compared with RD-like means across several simulations. The difference is that our 
BM and RD results are 
obtained using automated algorithms (\autobm and \autord), rather then from arbitrarily parametrized experiments. 
%
Our test, performed in step 4, 
checks whether the difference between BM and RD estimates is larger in absolute value than the $\delta$ used for CI implementation. 
%
If this is the case, 
we have evidence of non-ergodicity and therefore of violation of the strong mixing assumption (step 6). For example, this could be due to the presence of multiple stationary points in the process: \autoBM would end up exploring only one of such stationary points, while \autoRD would provide averaged information on the possible realizations. 
If the difference between BM and RD estimates is small, 
we have no evidence that the assumption is violated and we proceed 
with a second 
test on \rev{the horizontal means of} \autord (step 5). 
Indeed, under the null hypothesis that the process is strongly mixing and that the initial warmup phase has been effectively discarded, the central limit theorem for weakly correlated variables states that the 
horizontal means should be approximately normally distributed. In particular, we perform an Anderson-Darling normality test 
(with significance level 1\%) on the sample of horizontal means. If the null hypothesis is not rejected 
we again have no evidence 
of violation of the strong mixing assumption, and we therefore return the values computed by either of the two algorithms (step 7). 
\section{Transient analysis of a large macro ABM}\label{sec:macro}

\subsection{The macro ABM of \cite{caiani2016agent}
}
The model has been developed to bridge the stock-flow consistent approach (SFC; \cite{godley2006}) with the macroeconomic agent based literature \citep[see, e.g.,][]{gatti2005new, cincotti2010credit, dosi2010, dawid2012eurace, popoyan2020winter}.\footnote{A rather detailed overview of the macro ABM literature can be found in \cite{fagiolo2017macroeconomic} and \cite{dosi2019more}.} 
It depicts an economy composed of households selling their labour to firms in exchange for wages, consuming, and saving into deposits at (commercial) banks. Households own shares of firms and banks in proportion to their wealth, and receive a share of \rev{profits of the firms and banks}  as dividends; they also pay taxes as set by the Government, which runs fiscal policy. There are two categories of firms. Consumption firms produce a homogeneous good using labour and the capital goods manufactured by the other class of firms: capital firms. Firms may apply for loans in order to finance production and investment. Retained profits enter the financial system as \rev{deposits of the banks.} 
Banks provide credit to firms, buy bonds issued by the Government and need to satisfy mandatory capital and liquidity ratios. Finally, a Central Bank holds  reserve accounts \rev{of the banks} and the government account, accommodates \rev{the demand of banks} for cash advances at a fixed discount rate, and possibly buys government bonds that have not been purchased by banks.

Here we focus on two key indicators of economic activity: the unemployment rate, and the bankruptcy rate of business firms. They have been chosen because of: 
(i) the relative large fluctuations they exhibit during the transient dynamics \citep[see Figure 2 in][]{caiani2016agent}, which we aim at reproducing and testing and (ii) their well known role as proxies of \rev{the health of an economy} at macro and micro level, respectively. We sketch how these two quantities are modelled by Caiani and co-authors while leaving the additional details about the model to the original paper.\footnote{Of course, all variables present in the original model could be analysed using the very same procedure; we selected two for illustrative purposes.} 

The labour market is composed of workers, firms, and the public sector. 
Firms in the capital good sector (indexed by $k$) demand workers based on their desired level of production $y^D_{xt}$ and the productivity of labour ($\mu_N$), which is assumed to be constant and exogenous:
\begin{equation}
N^{D}_{kt}=\frac{y^D_{xt}}{\mu_N}.
\end{equation}
Differently, the request of workers by consumption good firms (indexed by $c$) is given by
\begin{equation}
N^D_{ct}=u_{ct}^D\frac{\kappa_{ct}}{l_{k}},
\end{equation} 
where $\kappa_{ct}$ is the capital stock, $l_{k}$ is a constant expressing the capital-to-labour ratio and $u_{ct}$ is the utilization capacity 
needed to obtain the desired production. Workers can be fired under two circumstances: workers in excess of production needs are randomly sampled from the pool of firm employees and fired, and workers can \rev{lose} their job because of an exogenous positive employee turnover (a fixed share of workers is fired in every period). Finally, a constant share of households are employed by the public sector and public servants are also subject to an exogenous turnover.

After having planned production, firms and the government interact with unemployed households on the labour market. Workers follow an adaptive heuristic to set the wage they ask for: if over the year (i.e., four periods), they have been unemployed for more than two quarters, they lower the asked wage by a stochastic amount. In the opposite case, they increase their asked wage. The share of workers that is not employed at the end of each session of interaction in the labour market represents the prevailing unemployment rate.

After production, firms sell their products and need to compensate for the inputs they received. Firms may default when they run out of liquidity to pay wages or to honour the debt service. Defaulted firms are bailed-in by households (who are the owners of firms and banks and receive dividends) and depositors, as the authors seek to maintain the number of firms constant. Hence, the bankruptcy rate emerges as the ratio between defaulted firms before the bailing-in event and the total number of firms in the economy. As the defaulted firms create non-performing loans that might trigger vicious cycles and - ultimately - a financial crisis, they offer key information on the turbulence and riskiness of the business cycle.

\subsection{Transient analysis with \emph{\autoIR}: automatic computation of confidence intervals}

The model is run in its baseline configuration 
considered in Section 5.1 of \cite{caiani2016agent}. The artificial time series show the model first experiences a sequence of expansionary and recession regimes, then converging, in most cases, to a relatively stable behaviour where aggregate variables (including the unemployment and the bankruptcy rates) fluctuate around particular values, and nominal aggregates grow at similar rates. Our focus is centred on the first part of 
\rev{this} process. 

\begin{figure}[t] \centering 
	\includegraphics[scale=0.35]{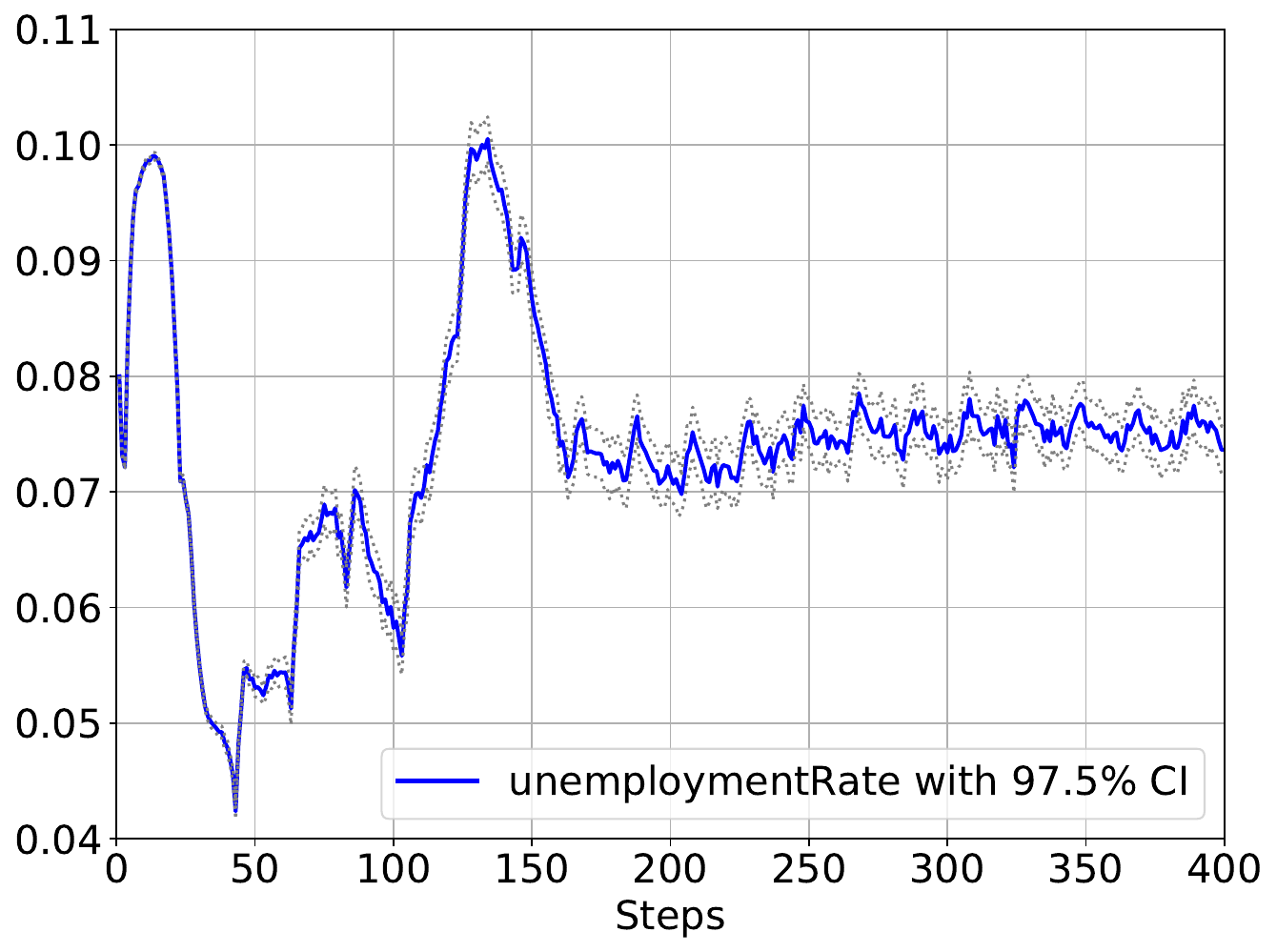}
\hfill
	\includegraphics[scale=0.35]{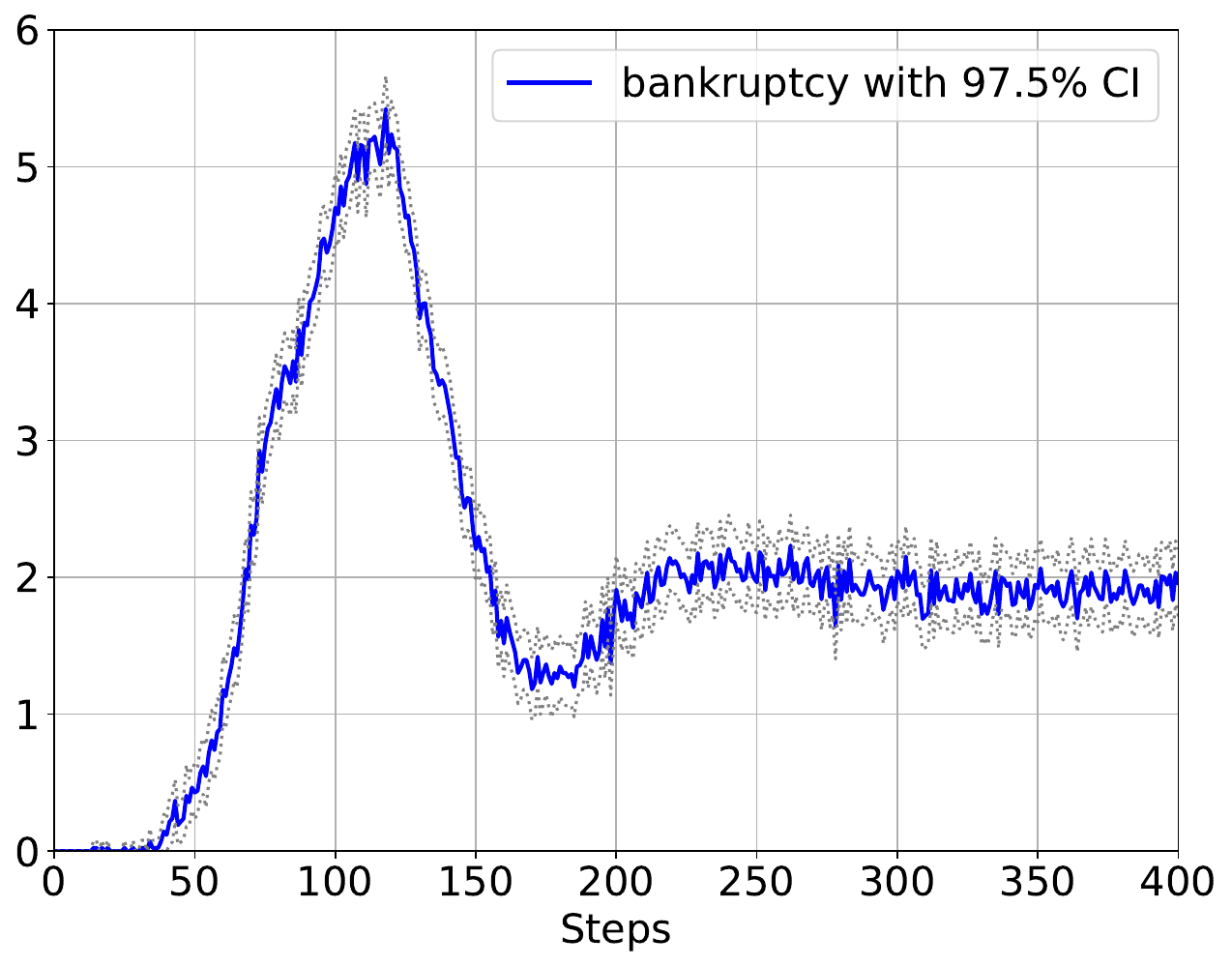}
	\caption{Unemployment rate and bankruptcy over time. The dashed lines are the computed 97.5\% CI of size $\delta$ at most $0.005$ and $0.5$, respectively.}
	\label{fig:analysisMacroABM}
\end{figure}

As a first exercise, we reproduce (Figure \ref{fig:analysisMacroABM}) the behaviour of the economic indicators we 
selected in the first 400 steps of the simulation, \rev{as done in the original paper,} and construct CIs 
around their mean according to Equation \eqref{eq:ci}. In particular, we choose $\alpha=0.025$ and set $\delta$ at the maximal allowed width of the confidence intervals around our central estimates (i.e., $\delta_U=0.005$ for the unemployment rates and $\delta_B=0.5$ for the average bankruptcies) and let \autoir 
automatically decide the number of simulations needed to obtain the desired confidence intervals. 
\rev{We deem $\delta_U=0.005$ and $\delta_B=0.5$ to offer an \emph{adequate precision} because, as shown in Figure~\ref{fig:analysisMacroABM}, the unemployment rate goes  from 0.04 to 0.10 while the average bankruptcies go from 0 to about 5 (we remark that, alternatively, one could have used a percentage value for $\delta$ as discussed in Section~\ref{sec:mean_ci}).}

\begin{figure}[t] 
\includegraphics[scale=0.275]{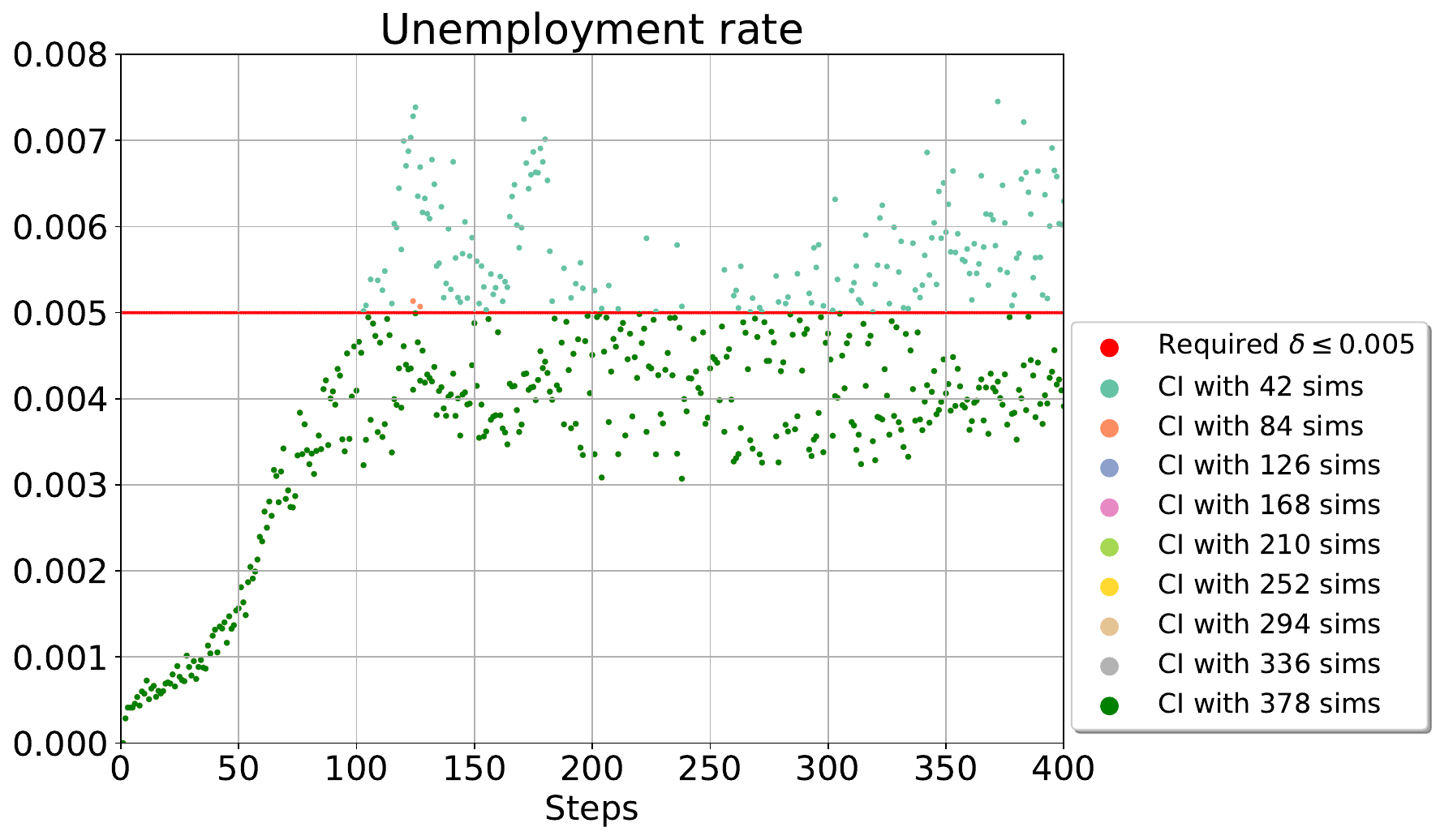}
\hfill
\includegraphics[width=0.485\linewidth]{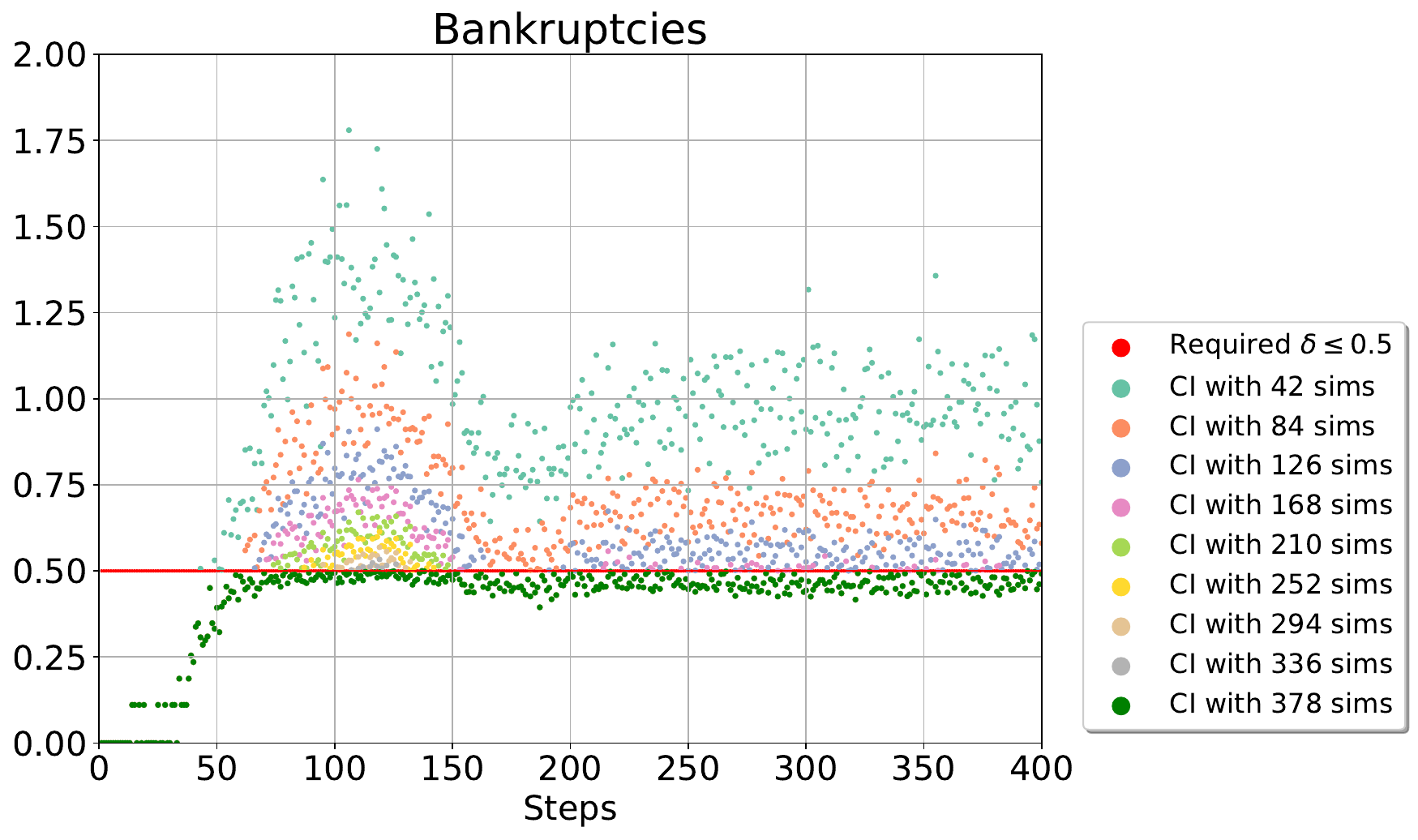}
\\
\includegraphics[scale=0.275]{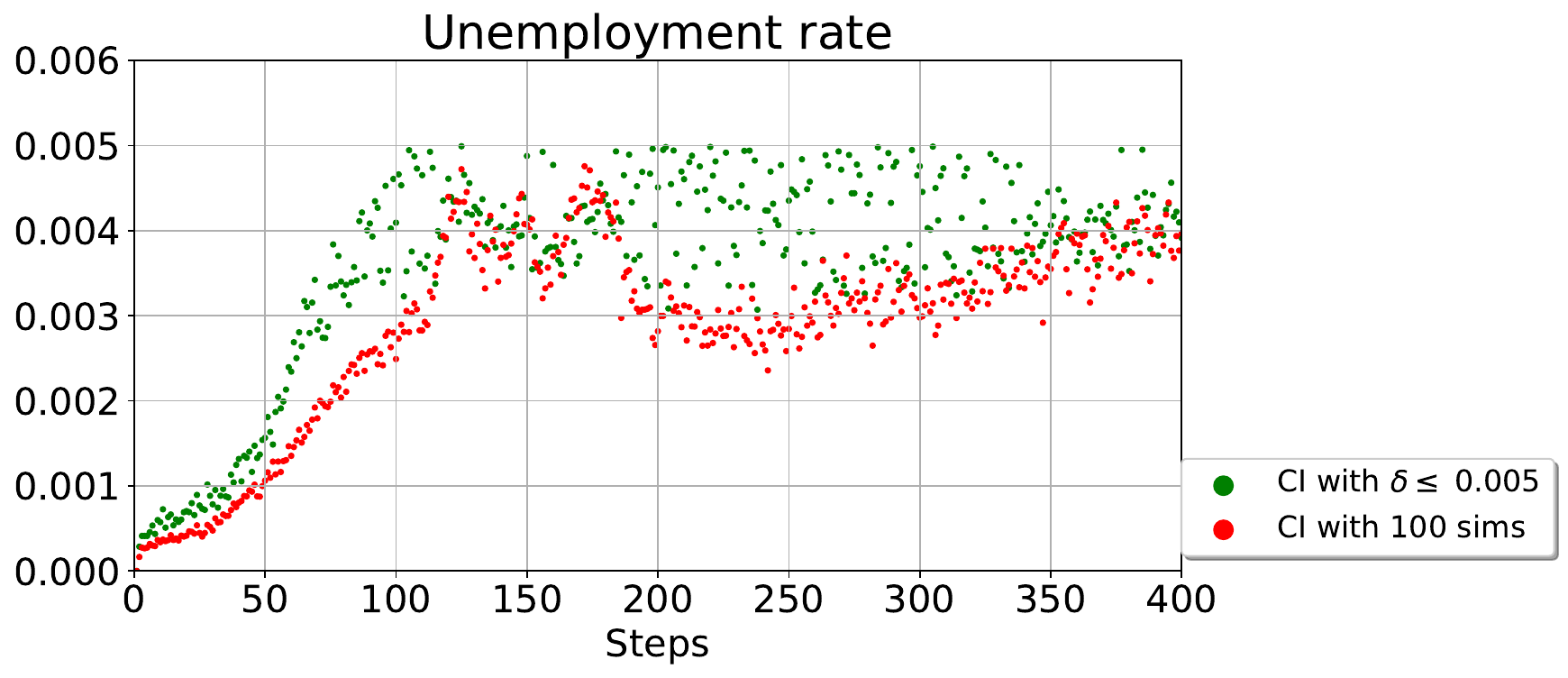}
\hspace{0.216cm} %
\includegraphics[width=0.475\linewidth]{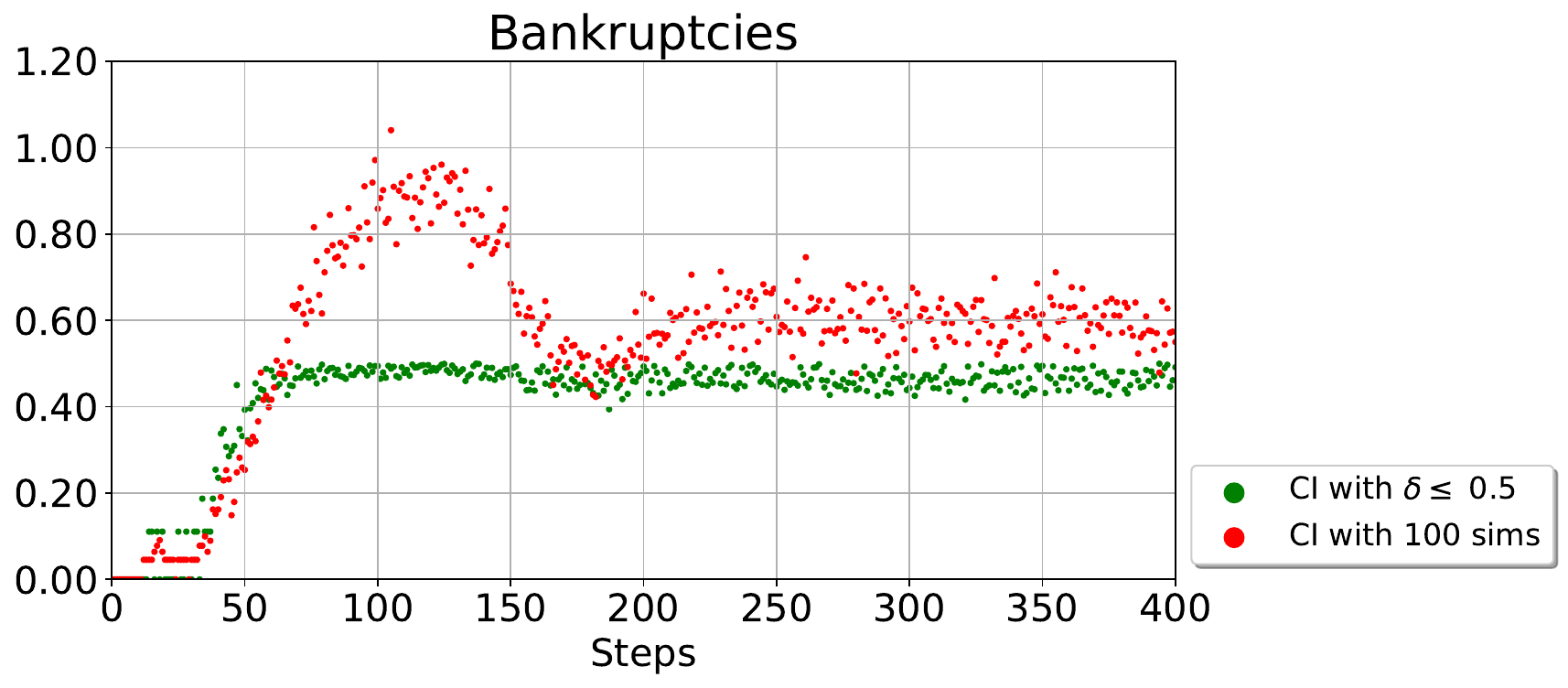}
\caption{
Top: Intermediate \rev{absolute} widths of the 97.5\% CI for unemployment rate and bankruptcy computed by \mv{} (top); 
Bottom: Comparison among \rev{absolute} CI widths computed by \mv{} and by setting 100 simulations for all properties and $t$ as in \cite{caiani2016agent} (obtained using \autoIR and setting both $bl$ and the maximum number of simulations performed to \np{100}).
\rev{In each plot, the dots are drawn iteratively from the top one in the legend downwards. E.g., in the top-left plot we see steps where the CI got below the red line after: (i) 42 simulations, e.g., all steps smaller than 100, where we see only one dot; 
(ii) 84 simulations, e.g., step 400, where we see two dots: a higher cyan one (42 sims), and a lower dark green one; (iii) 126 simulations, the two steps close to 125 where we see three dots, including a middle orange one (84 sims).
}
}
\label{fig:analysisMacroABM_CI}
\end{figure}

We stress that \rev{our algorithms} 
automatically determine the number of runs 
required to obtain the desired CI for each point in time and for whatever variable of interest. 
%
As shown in the top of Figure~\ref{fig:analysisMacroABM_CI}, this required at most 378 simulations for both properties. 

 As a concept-proof of our approach, the inspection of Figure \ref{fig:analysisMacroABM} confirms that our 
 algorithms 
 do not modify the 
 model and deliver the same dynamics (see Figure 2 of \citealp{caiani2016agent}).\footnote{We highlight that the artificial time series we generate are somehow comparable to Figure 2 of \cite{caiani2016agent}; however, Caiani et al. apply a bandpass filter over their series and just show the emerging trend component. Contrarily, we show the ``raw'' series that the model generates. We notice that the latter is the prevailing practice in the literature \citep[see for example the models reviewed in][]{fagiolo2017macroeconomic}.} However, \cite{caiani2016agent} performed 100 simulations for all the 400 time steps, without providing information on the obtained confidence intervals.

The ability to specify the precision of the confidence intervals comes with a number of advantages. First, it is a flexible requirement that can be expressed either in absolute or relative terms (see Section \ref{sec:algorithms}), leaving the chance to statistically compare the expected behaviour of the model to a certain target (say, an employment rate not higher than $5\%$ or an inflation rate of $2\%$) or to its mean (e.g., allowing one to compute for each period the probability to observe bankruptcy rates $10\%$ higher than the average). Second, and more relevantly, it allows evaluating the robustness (and the uncertainty) of the dynamics simulated by the model. 
In particular, Figure~\ref{fig:analysisMacroABM_CI} 
shows how the \rev{absolute} width of the confidence intervals vary, for each time point and property, across the simulation span for various number of simulations. 
The top row shows the intermediate CI widths obtained after every iteration of the blocks of simulations performed by \rev{our approach} (see the discussion in Section~\ref{sec:mean_ci} - we use $bl=42$). We note that the widths decrease at every iteration, and that some time points (from 100 to 200) require more simulations than the others to get the desired CI width. 
\rev{The two top figures also show that the unemployment rate required only a few iterations of simulations to obtain the required CI widths for all time points, while the bankruptcies required up to 9 iterations.}
Instead, the bottom of the figure compares the CIs obtained by \rev{our approach} (in green) against those obtained using the setting of the original paper (i.e., 100 simulations for all time steps, in red). We note that, apart for the first time points which present very low variance, our CIs computed tend to be homogeneous and close to the required $\delta$, demonstrating that the minimum number of simulations are computed for the given $\delta$. Instead, the setting used by \cite{caiani2016agent} 
might lead to CIs of different widths which follow the trend of the computed means. This is particularly evident for the case of bankruptcies \rev{of firms}, cf. Figure \ref{fig:analysisMacroABM_CI} (bottom-right), while the same does not happen for unemployment rates (bottom-left of the figure).
The figure suggests that each property and time point should be studied using its \emph{own best} number of simulations, confirming that a trade-off between an insufficient and an excessively large number of simulations exists. When this is too low the across-runs variability might not be adequately washed-out and the representation of (stochastic) uncertainty could depend on the level of the relevant variable; conversely, when  the number of simulations is too large, simulations are redundant and the same representation of uncertainty can be effectively offered saving computational time.
Finally, in line with \cite{Secchi2017}, the right-hand panels of both figures confirm that the arbitrary choice of $n=100$, which is common in the literature (see the discussion in Sections \ref{sec:intro} and \ref{sec:outputanalysis}), is unjustified by the properties of the model itself.

\subsection{Experiment comparison and statistical testing: A \rev{behavioural experiment}}\label{sec:counterfactual}
The second exercise we perform uses the confidence intervals 
(and the means, variances and number of samples 
 \rev{computed by our algorithms}) 
to automatize a series of tests that identify statistical differences across model configurations discussed in Section \ref{sssec:eqtest_power}. Indeed, one of the most common approaches in the macro ABM literature is to focus on key parameters or mechanisms of interest and test how the 
dynamics of the model respond to changes. The difference across experiments is tested comparing the value of some statistic of interest (e.g., the growth rate of output) - usually averaged over the entire time span - by means of t-tests \citep[e.g., in][]{Dosi15, popoyan2020winter}. Obviously, the ability of the test to discern across experiments and to validate the counterfactual policy intervention is affected by the choice of $n$, as an insufficient number or runs is likely to make model configurations (i.e., experiments) difficult to distinguish. 
Even further, it is not infrequent that statistical tests about differences across experiments are completely missing \citep[see, e.g.,][]{cincotti2010credit, caiani2019does}, which weakens the potential of the paper and the eventual policy recommendations.

\begin{figure}[h!]
\centering
\subfloat[\rev{CIs width for $\alpha=0.025$ and $N=100$ simulations. T-tests ``\emph{are means point-wise equal?}'' not rejected for significance $a_w\!=\!0.025$}]{\includegraphics[width=0.47\linewidth]{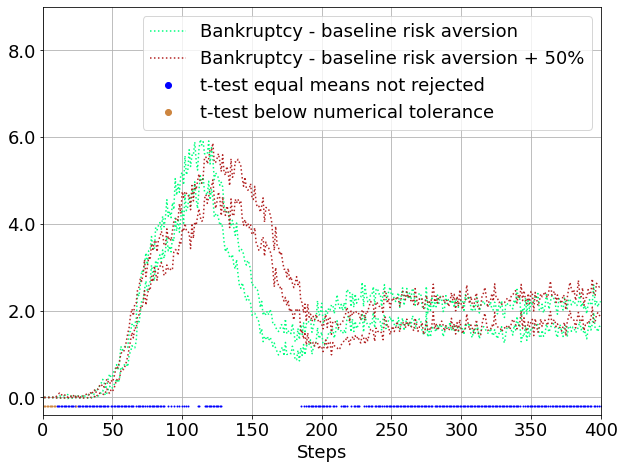}
 }
\hfill
\subfloat[\rev{CIs width for $\alpha=0.025$ and $\delta=0.5$. T-tests ``\emph{are means point-wise equal?}'' not rejected for significance $a_w\!=\!0.025$}]{\includegraphics[width=0.47\linewidth]{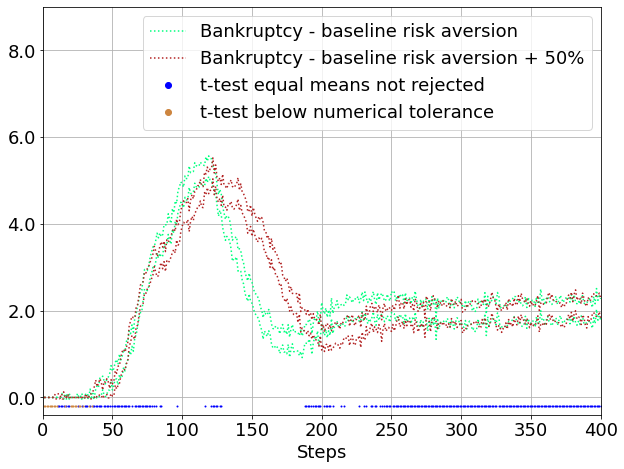}
}
\\
\subfloat[Power of t-test in (a) for difference $\varepsilon=0.5$]{\includegraphics[width=0.47\linewidth]{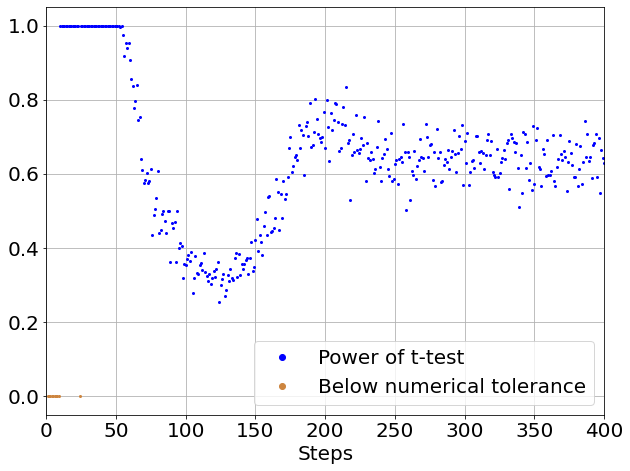}}
\hfill
\subfloat[Power of t-test in (b) for difference $\varepsilon=0.5$]{\includegraphics[width=0.47\linewidth]{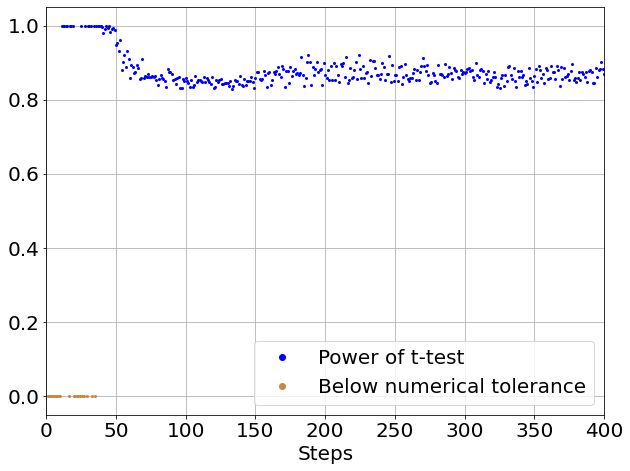}}
\caption{
Evolution of bankruptcies for \rev{two} different risk aversions $C$ for consumption firms: are they point-wise equal? \rev{The left-column considers an analysis setup involving $n=100$ simulations for each time point, while in the right-column we let our algorithms find automatically the correct $n$ for each time point. The top-row provides the confidence intervals of the estimations computed for the two values of $C$. The same row also 
provides results of the t-tests on the results obtained for the two values of $C$ (where yellow} dots denote initial steps with variances so small to get intermediate results below the numerical tolerance of our implementation of the test, 1E-15). \rev{The bottom row shows the power of the computed t-tests.}
}
\label{fig:behaviouralWithPower}
\end{figure}

Our \rev{techniques} 
provide an automatic series of t-tests across experiments,\footnote{\rev{These experiments are repeated in \ref{sec:utest} using the u-test rather than the t-test.}} where the expected value of any variable of interest in any pre-determined set of experiments is tested against a baseline configuration for each step of the transient period. As discussed in Section~\ref{sssec:eqtest_power}, tests are run post-mortem and consist in Welch's t-tests (Equation~\eqref{eq:Welchtt}),
whose power can be computed with respect to a minimum distance $\varepsilon$ between the means that the test is expected to detect (Equation~\eqref{eq:power}). 
%
Figure \ref{fig:behaviouralWithPower} shows the results in our testbed macro ABM\rev{, evaluating} 
the effects that changes in the degree of risk aversion \rev{of agents} ($C$) produce on the number of bankruptcies. 
Caiani and co-authors show that when the risk aversion of the agents increases, the economy tends to completely avoid the recession phase experienced in the baseline configuration. While they did not offer a statistical analysis of these differences, our approach automatically embeds it. In particular, we contrast model behaviours across the baseline value of $C$ and a 50\%  increase of the latter. \rev{The two top-plots of} Figure \ref{fig:behaviouralWithPower} show the results of our tests comparing the set-up of the original analysis (i.e., with $n=100$ for all properties and time points, left column) to our approach ($n$ automatically determined for each property and time point, right column). \rev{In each plot we compare the estimated average bankruptcies obtained for the two values of $C$ (we show the CIs rather than the actual estimations, which are the centre of the CIs) and show the results of the t-tests.}
While increasing risk aversion delays the peak of bankruptcy rates, we find that\rev{, independently on the analysis set-up (the choice of $n$),} no statistical difference between the two experiments is found but for the central part of the simulation, that is when the economy first experiences a deep crisis and then recovers \citep[see Figure 2 in][]{caiani2016agent}. This 
suggests that doubling risk-aversion modifies the shape of the crisis (smoother surge of bankruptcies and slower decline) but not its existence nor duration, partially contradicting the original results.
\footnote{We also highlight that our approach identifies a statistically significant difference between the two experiments for the slight increase in business insolvencies between period 200 and 250.} 

Though using $n=100$ or our approach \rev{(left- and right-column of Figure~\ref{fig:behaviouralWithPower}, respectively)} makes little difference in terms of type 1 errors, a key advantage is evident when comparing powers \rev{of t-tests provided in the bottom-plots of Figure~\ref{fig:behaviouralWithPower}}. Indeed, our setup \rev{(right-column)} guarantees a much higher power of the tests, thereby reducing dramatically the chance of not rejecting the null hypothesis of equality across experiments when it is actually false \citep[see also][]{Secchi2017}. \footnote{\rev{Figure~\ref{fig:behaviouralWithPower_unemplRate} provides a similar study for the unemployment rate which confirms the same results, though the discrepancy is less marked.}}
Further, our approach delivers - for given significance $a_w$ 
and setting $\varepsilon$ equal to the $\delta$ used for the transient analysis - a \emph{good} and stable power across the simulation horizon, i.e., above $0.8$, which is usually considered an acceptable threshold in the applied statistics literature \citep[see, e.g.,][]{Secchi2017,cohen1992,lehr1992}.
This comes by the fact that for each property and time point, we run the correct number of samples to obtain a constant width of the CI embedded in the choice of $\delta$ (and by the assumption that the minimum difference we want to detect -- $\varepsilon$ -- is equal to $\delta$). Indeed, it is interesting to note how the t-test for bankruptcies obtained for the setting with 100 simulations (Figure~\ref{fig:behaviouralWithPower} bottom-left) has a low power which appears to decrease specularly to how the corresponding CI width increases in Figure~\ref{fig:analysisMacroABM_CI} (bottom-right).
Hence, we can derive a rule of thumb to support the modeller's choice of the two free parameters in a set-up that compares different experiments: first of all, $a_w$ can be set equal to the $\alpha$ used for the transient analysis, from $5\%$ to $1\%$, whose extrema are the most diffused levels of statistical significance in the social sciences. 
Then, by setting $\delta$ in the transient analysis equal to the $\varepsilon$ of interest, we expect to obtain t-tests with good power.
If this is not the case, one can perform the transient analysis for smaller values of $\delta$ while keeping constant $\varepsilon$.
In case the maximum budget of simulations that have been originally chosen does not allow to meet such conditions, a trade-off exists in accepting an higher chance of type II error (not detection of false negatives) and the computational resources at disposal of the modeller. However, we stress that while increasing the size of the simulation exercise might come at the expenses of computational time, the proposed tool automatically parallelise model's runs to speed-up the analysis. \rev{\ref{sec:parallelizationstudy} shows the efficiency of our approach in parallelizing tasks.}
\footnote{We remark that when using the original code and simulation environment (JMAB) of the Caiani et al. paper alone, it is not possible to perform any form of statistical analysis automatically, requiring to process CSV files created by the framework. Our integration of  \mv{} provides JMAB with analysis capabilities described so far, encompassing both the transient and the steady-state analysis, while leaving the simulation environment unaltered.} 

\subsection{\rev{Experiment comparison and statistical testing: A policy experiment}}\label{sec:counterfactual_tax}
\begin{figure}[h!]
\centering
\subfloat[\rev{CIs for $\alpha=0.025$ and $\delta=0.5$. T-tests ``\emph{are means point-wise equal to the baseline?}'' not rejected for significance $a_w\!=\!0.025$}]{\includegraphics[height=0.356\linewidth]{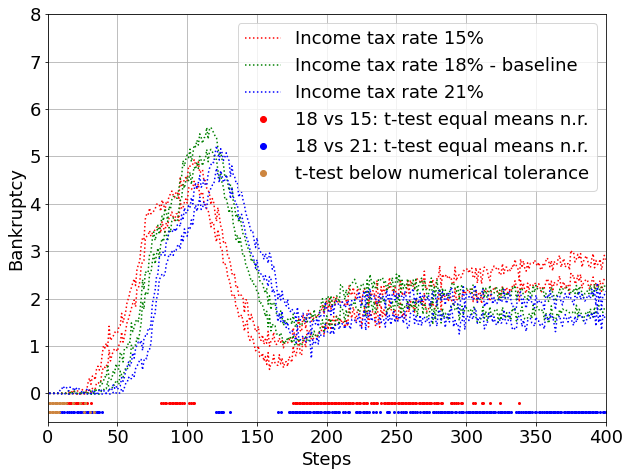}
 }
\hfill
\subfloat[\rev{CIs for $\alpha=0.025$ and $\delta=0.01$. T-tests ``\emph{are means point-wise equal to the baseline?}'' not rejected for significance $a_w\!=\!0.025$}]{\includegraphics[height=0.35\linewidth]{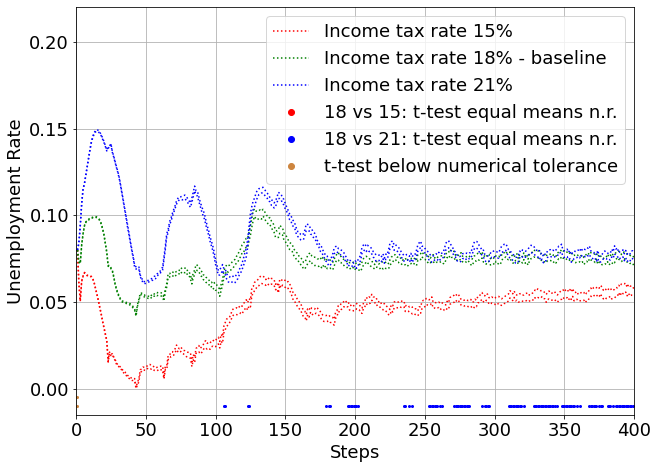}
}
\\
\subfloat[\rev{CIs for $\alpha=0.025$ and $\delta=500$. T-tests ``\emph{are means point-wise equal to the baseline?}'' not rejected for significance $a_w\!=\!0.025$}]{\includegraphics[height=0.35\linewidth]{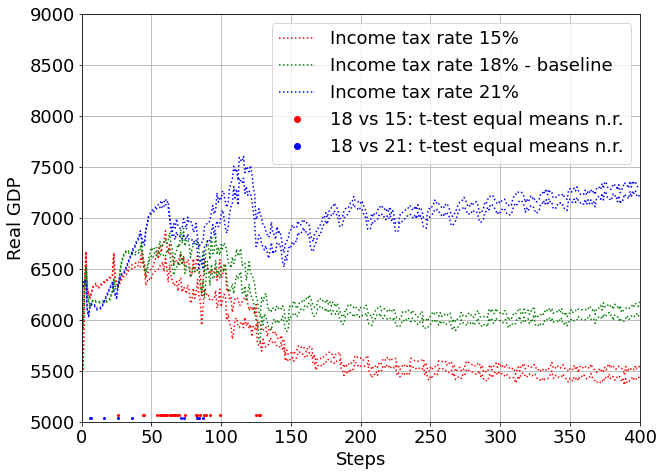}}
\hfill
\subfloat[\rev{CIs for $\alpha=0.025$ and $\delta=300$. T-tests ``\emph{are means point-wise equal to the baseline?}'' not rejected for significance $a_w\!=\!0.025$}]{\includegraphics[height=0.35\linewidth]{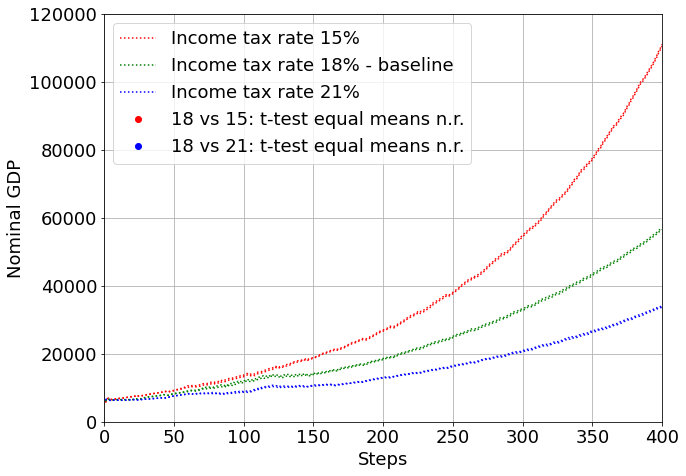}}
\caption{
\rev{
Evolution of bankruptcies (a), unemployment rate (b), real GDP (c) and nominal GDP (d) for three different income tax rate: the baseline 18\%, and variations 15\% and 21\%. Are the two variations point-wise equal to the baseline? For each parametrization we provide the CIs computed by \mv for the given $\alpha$ and $\delta$. At the bottom of each plot we provide the non-rejected t-tests for equal means of the variations against the baseline 18\%. We note that yellow dots in (a) and (b) denote initial steps with variances so small to get intermediate results below the numerical tolerance of our implementation of the test, 1E-15. Moreover, the t-tests are always rejected in (d), and always rejected for case \emph{18 vs 15} in (b).}
}
\label{fig:policyTaxRate}
\end{figure}

\rev{
%

Macro ABM are ubiquitously employed to perform ex-ante policy experiments. Just to make few examples, \cite{Dosi15} compare a series of rules for monetary and fiscal policy, \cite{lamperti2020climate} explore feed-in tariffs and R\&D subsidies, and \cite{caiani2019does} extend the model analysed in this section 
to study various progressive tax schemes and their effects on growth and inequality. Here we focus on a fiscal policy exercise where we let vary the tax rate that the government charges on  gross incomes \rev{of households} and study the economy-wide effects of such alternative policies, leaving government spending unaltered. Similarly to the previous section, we employ the means, variances, confidence intervals and number of runs from each experiment to provide pairwise t-tests comparing the baseline to alternative tax regimes. Results are shown in Figure~\ref{fig:policyTaxRate}.\footnote{\rev{As for the t-tests presented in Figure~\ref{fig:behaviouralWithPower}, all t-tests have high power as reported in Figure~\ref{fig:policyTaxRate_power}.
}}
First, we notice that cutting and raising the income tax rate induce asymmetric effects. Higher taxes with respect to the baseline (i.e. 18\% vs. 21\%) smooth inflation and boost real output in the short run (with a significant difference with respect to the baseline), delaying the rise in bankruptcies that leads to the recession observed in the middle of the simulation; finally, they allow the economy stabilizing on a regime characterized by higher output while statistically indifferent bankruptcy and unemployment rates. Conversely, lower income taxes (i.e. 18\% vs. 15\%) spur inflation by raising demand, which leads to over-investment episodes that anticipate the increase in bankruptcies bringing about a longer recession than in the baseline, finally forcing the model to stabilize on a lower output regime with higher bankruptcies.

These experiments let us conjecture that the long-run dynamics of the model are influenced by its transient behaviour, though heterogeneously across ``state'' variables. This reinforces the urgency of adequate tools to rigorously inspect \rev{behaviours of models} across the phases of their simulation.
}

\section{Steady-state analysis in a model of market selection}\label{sec:predmarket}
Here we use \rev{our proposed techniques and algorithms} 
to perform a statistical analysis of the steady-state expected value of wealth shares and market price in a simple repeated prediction market model. 
\rev{We consider an accepted ABM model }
extensively studied in the literature \citep{beygelzimer2012,kets2014,bottazzigiachini2017,bottazzigiachini2019b}.
\rev{We believe that it offers a good testbed for our procedures for automated steady-state analysis because} its steady-state properties have been investigated by \cite{kets2014} \rev{via simulation-based analysis} and, later on, studied analytically in \cite{bottazzigiachini2019b} showing that the numerical results of \cite{kets2014} were inaccurate both qualitatively and quantitatively. \rev{In addition, we can use the analytical solutions from \cite{bottazzigiachini2019b} as an oracle to compare with the results obtained using our algorithms. }
As briefly reported by \cite{bottazzigiachini2019b} and as we shall see, the source of inaccuracy can be traced back to the strong autocorrelation and initial condition bias that process possesses. 
The \emph{post-mortem}\rev{, or \emph{offline},} nature of the numerical analysis carried on by \cite{kets2014}\rev{, where the number of steps and of performed simulations is decided a priori while the analysis is performed afterwards,} is unable to properly deal with those issues.
Our approach, instead, uses statistical tests and procedures able to manage both autocorrelation and the initial condition bias in an automated way.
Thus, in what follows, we first introduce the model, then we repeat the numerical analyses of \cite{kets2014} showing how and why the inaccuracies emerge, and finally we use \autord and \autobm to accurately perform the steady-state analyses. In fact, we match the correct analytical results from~\cite{bottazzigiachini2019b}. 

\subsection{The prediction market model by \cite{kets2014}
}

\rev{The model consists in a pure exchange economy in discrete time where $N$ agents repeatedly bet on the occurrence of a binary event. 
The probability of observing the event is $\pi^*$ in each period.
Such a probability is unknown to agents:  each agent $i\in\{1,2,\ldots,N\}$ assigns a subjective probability $\pi^i$ to the realization of the event. 
Agent $i$ has initial wealth equal to $w^i_0$ and at the end of every betting round it evolves in $w^i_t$ depending on the results of her betting. 
Every agent $i$ bets on the occurrence of the event at time $t$ a fraction $\alpha^i_t$ of her wealth $w^i_{t-1}$, while $1-\alpha^i_t$ is the fraction bet against the occurrence.  
As in \cite{kets2014,bottazzigiachini2017,bottazzigiachini2019b}, we focus on the so-called fractional Kelly rule, that is $\forall i,t$
\begin{equation}
	\alpha^i_t=c\pi^i+(1-c)p_t\,,
	\label{eq:alpha}
\end{equation}
with $c\in(0,1]$ and $p_t$ the price at time $t$ of the security paying 1 if the event occurs. Security prices are fixed in every period according to market clearing conditions or, equivalently, to a parimutuel procedure.\footnote{\rev{See \ref{appendix:ketsetal} for the details of the model.} Notice that the mathematical specification we use may appear different from the one presented in \cite{kets2014}. However, as explained in \cite{bottazzigiachini2019b}, the two specifications are indeed equivalent.} }
In this setting,  \cite{kets2014} want to explore the selection dynamics of the model and are particularly interested in 
the asymptotic (steady-state) value of expected wealth shares and price for $c\to0$.
Indeed, they conjecture that in such a limit the steady-state expectation of $p_t$ matches $\pi^*$ and use the case $c=0.01$ as a proxy. 

In Section~\ref{manual}, we 
 replicate exactly the analysis of \cite{kets2014}, reproducing their Figures 3(c) and 3(d). 
Thus, we follow the procedure proposed in~\cite{kets2014} to estimate steady-state expected wealth shares and price for several values of $\pi^*$ under the parametrization in Table~\ref{tab:params}\rev{, where we stress that we have $N=3$ agents.}
In doing that, we highlight some issues related to initial condition bias and strong autocorrelation. 
Indeed, the warmup period appears not correctly determined, and the autocorrelation within the observations of each performed simulation not correctly accounted for. This is due to the procedure for computing CIs used in~\cite{kets2014} and has the result of producing extremely wide CIs.
\begin{table}[t]
\center
%
\begin{tabular}{c c ccc ccc} 
\toprule
&
&\multicolumn{3}{c}{\emph{Beliefs}} & \multicolumn{3}{c}{\emph{Wealth}}
\\
\cmidrule(r){3-5}  \cmidrule(r){6-8} 
$N$ & $c$ & \ $\pi^1$ & \ $\pi^2$ & \ $\pi^3$ & $w^1_0$ & $w^2_0$ & $w^3_0$\\
$3$ & $0.01$ & $0.3$ & $0.5$ & $0.8$ & $0.33$ & $0.33$ & $0.34$
\\
\bottomrule
\end{tabular}
\caption{Parameters used for the prediction market model. 
}
\label{tab:params}
\end{table}

After this, in Section~\ref{sec:automatic}, we perform the steady-state analyses using our approach. 
We show that, thanks to the provided automatic procedures, our estimates (and the corresponding confidence intervals) of steady-state expected wealth shares and price are correctly determined. 
Our conclusions differ not just quantitatively, but also qualitatively from those in~\cite{kets2014}, and match those from~\cite{bottazzigiachini2019b}. 
Overall, our analysis shows the importance of using an automated procedure provided with statistical guarantees. 

\subsection{Steady-state analysis with \emph{\manualrd} using original wrong warmup estimation}\label{manual}
\rev{In order to replicate exactly the erroneous analysis in Figures 3(c) and 3(d) of  \cite{kets2014}, we implemented also a \emph{manual} version of \autoRD{} named \manualrd  \footnote{\rev{See also \ref{sec:queries}.}} which allows one to manually set an \emph{a-priori} estimate of the warmup period, and to fix the maximum number of simulations used in an analysis based on independent replication.}
In general it is always advisable to do not fix such parameters \emph{a-priori}, but to use the offered automated procedures so to avoid bias in the estimates and excessively large CIs. In this section we exemplify these issues by fixing a priori the erroneous warmup estimate and number of simulations used in~\cite{kets2014}, and discuss the problems this introduced in the obtained results. 

\cite{kets2014} performed an RD-based steady-state analysis of the \rev{wealth of the agents}   $(w^1_t,w^2_t,w^3_t)$ and market price $p_t$ using the parametrization of Table~\ref{tab:params}. The authors arbitrarily estimated the end of the warmup period after \np{90000} steps, and fixed  the time horizon of each simulation to \np{100000} steps and the number of performed simulations to \np{1000}. This means that estimates were computed averaging the last \np{10000} observations in each simulation (the horizontal means of Figure~\ref{fig:nmsimsAnalysis}(c)) and then further averaging the so-computed means from each simulation (the vertical means). We shall see how this led to estimates highly biased by the initial conditions. 

Regarding computations of confidence intervals, \cite{kets2014} did not follow the standard approach relying on the central limit theorem  used by \rev{our approach}. 
Rather, \cite{kets2014} considered how the above discussed \np{1000} averages built for each simulation distribute. 
In particular, the 5-th and 95-th percentiles of such distribution are taken as the bounds of confidence intervals with 10\% statistical significance. 
The problem with this approach is that, differently from \rev{ours,} 
it is based on the assumption that each of the considered \np{1000} averages has the same distribution of 
an average across independent replications computed at a time $t$ large enough to have reached steady state. This is not correct because, as well as the initial condition bias, the process is characterized by strong autocorrelation. We shall discuss how this led to erroneous interpretation of the results.

\begin{figure}[t] \centering 
	\includegraphics[width=0.49\linewidth]{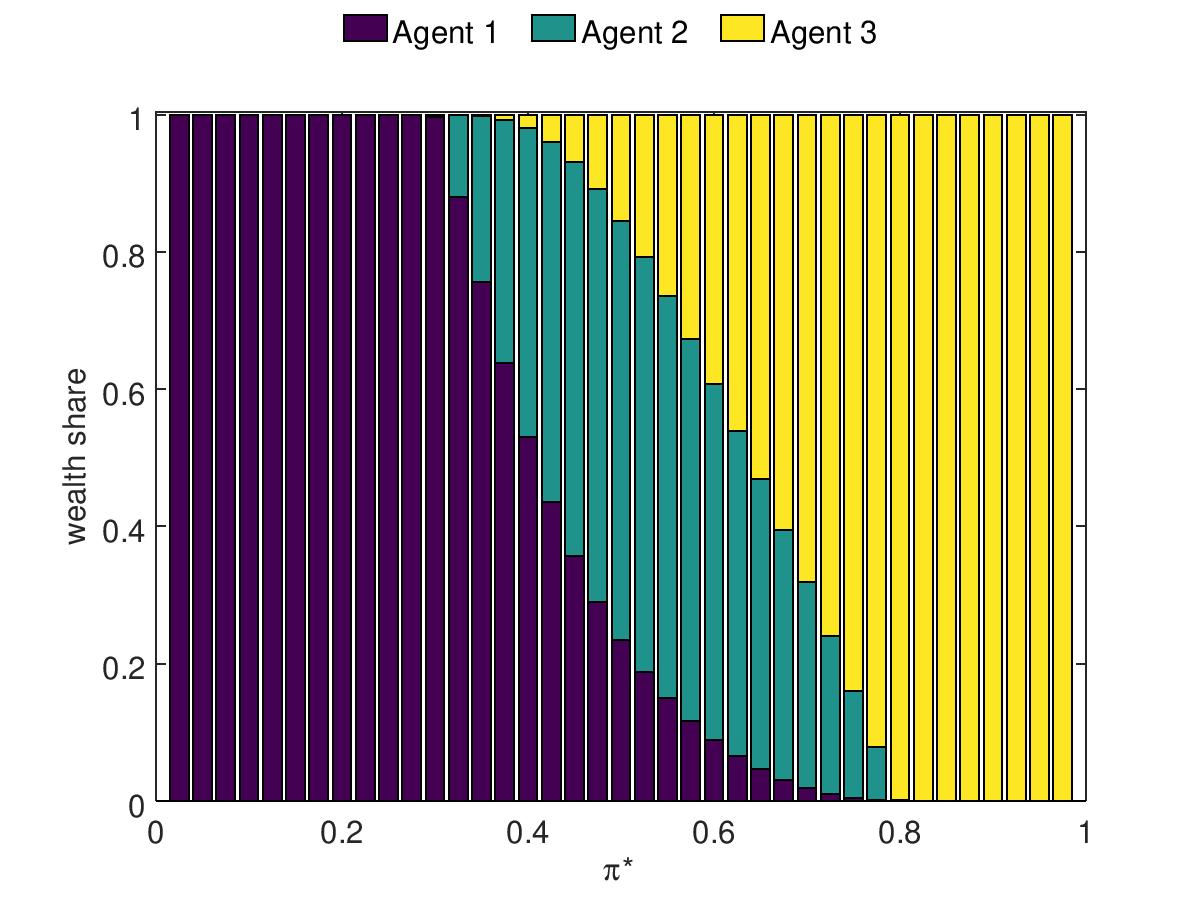}~
	\includegraphics[width=0.49\linewidth]{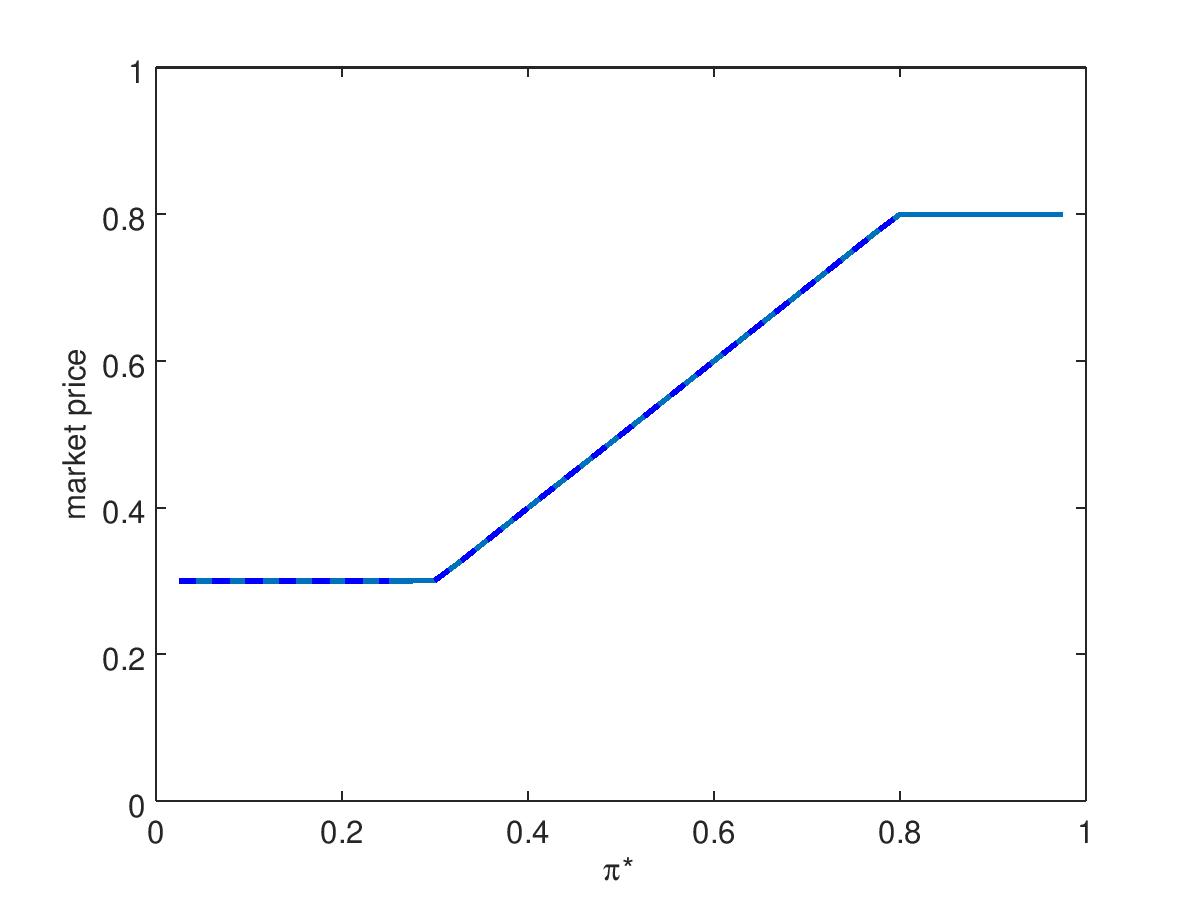}
	\caption{\small Steady-state analysis of expected agents' wealth and market price according to the manual warmup and simulation length settings of \cite{kets2014}. We obtain these results by using \manualrd setting the end of the warmup periods to \np{90000}, and the time horizons to \np{100000}. By setting both $\mathit{bl}$ (the number of simulations in a block) and the maximum number of simulations performed to \np{1000}, we use precisely \np{1000} simulations to estimate each property, perfectly matching the setting used in \cite{kets2014}. 
We consider 39 equally spaced values for $\pi^*$, from $0.025$ to $0.975$, each requiring a separate \mv{} analysis on a correspondingly parametrized instance of the model (we automated this process using an external Octave script).
Confidence intervals computed by \mv for agents' wealth, not reported in the left panel, are such that the maximum recorded width for a statistical confidence of 90\% is below 0.0025. Confidence intervals for the market price are reported in the right panel, with maximum recorded width below 0.00065 for statistical confidence of 90\%.}
	\label{fig:wrong}
\end{figure}

\paragraph{Wealth of the agents}
In Figure \ref{fig:wrong} we report the outcomes of the exercise replicating those from 
Figures 3(c) and 3(d) in \cite{kets2014} considering model variants for 39 different values of $\pi^*$. 
Looking at the left panel one should conclude that there exist model configurations  in which all agents have strictly positive expected wealth share in steady state.
This is, however, in contrast with the analytical analysis from Proposition 4.1 of \cite{bottazzigiachini2019b}, which proves that no more than two agents can have asymptotic positive wealth share.
Thus, the fact that \cite{kets2014} incorrectly suggest that more than two traders can have positive expected wealth in steady state is an artefact of the initial condition bias that affects their analysis. 
Notice that the convergence to zero of the wealth share of at least one trader is asymptotic, thus wealth shares show a bias for any $t$. However, such bias decreases with $t$ and can be made negligible choosing a sufficiently long warmup and simulation length. 
What we observe is that discarding the first \np{90000} observation of every run is simply not enough.

\begin{figure}[t] \centering 
	\includegraphics[width=0.49\linewidth]{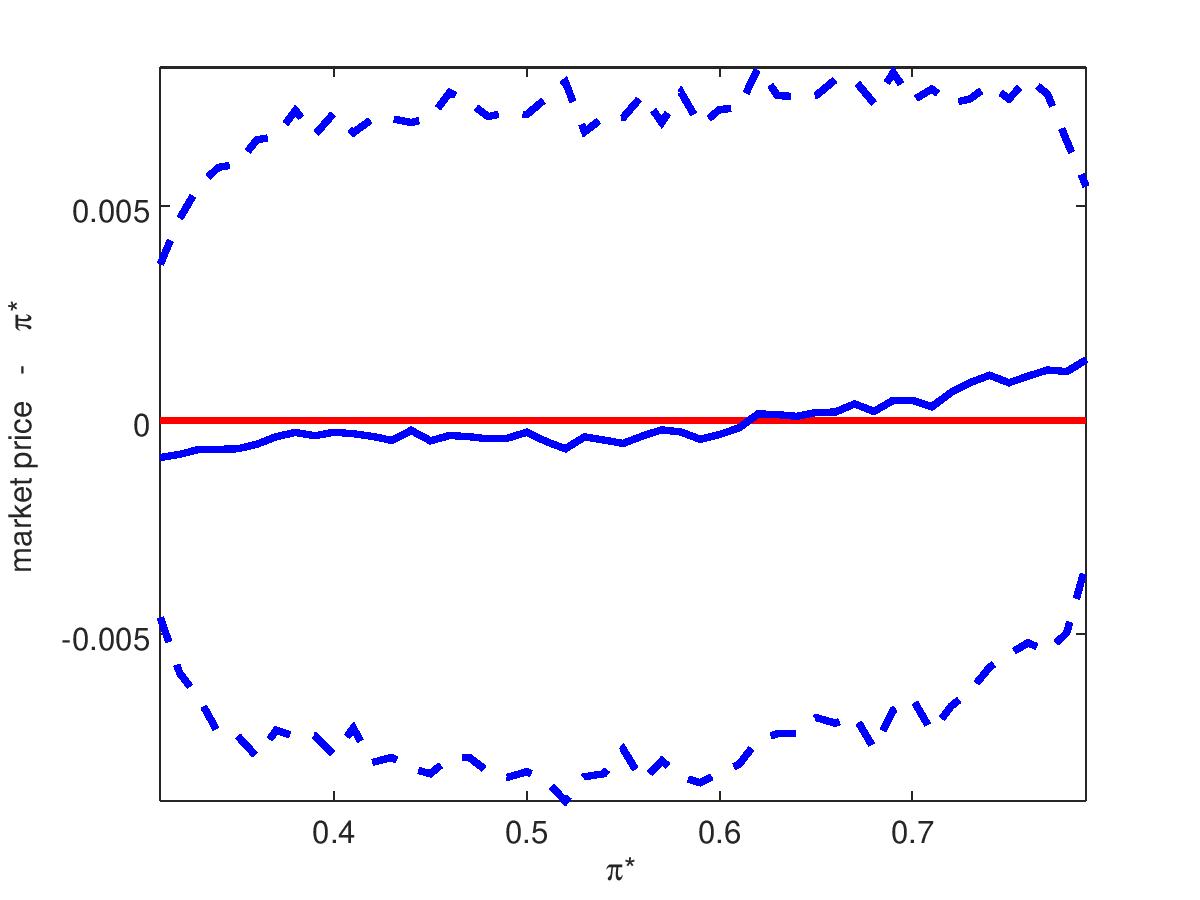}
\includegraphics[width=0.49\linewidth]{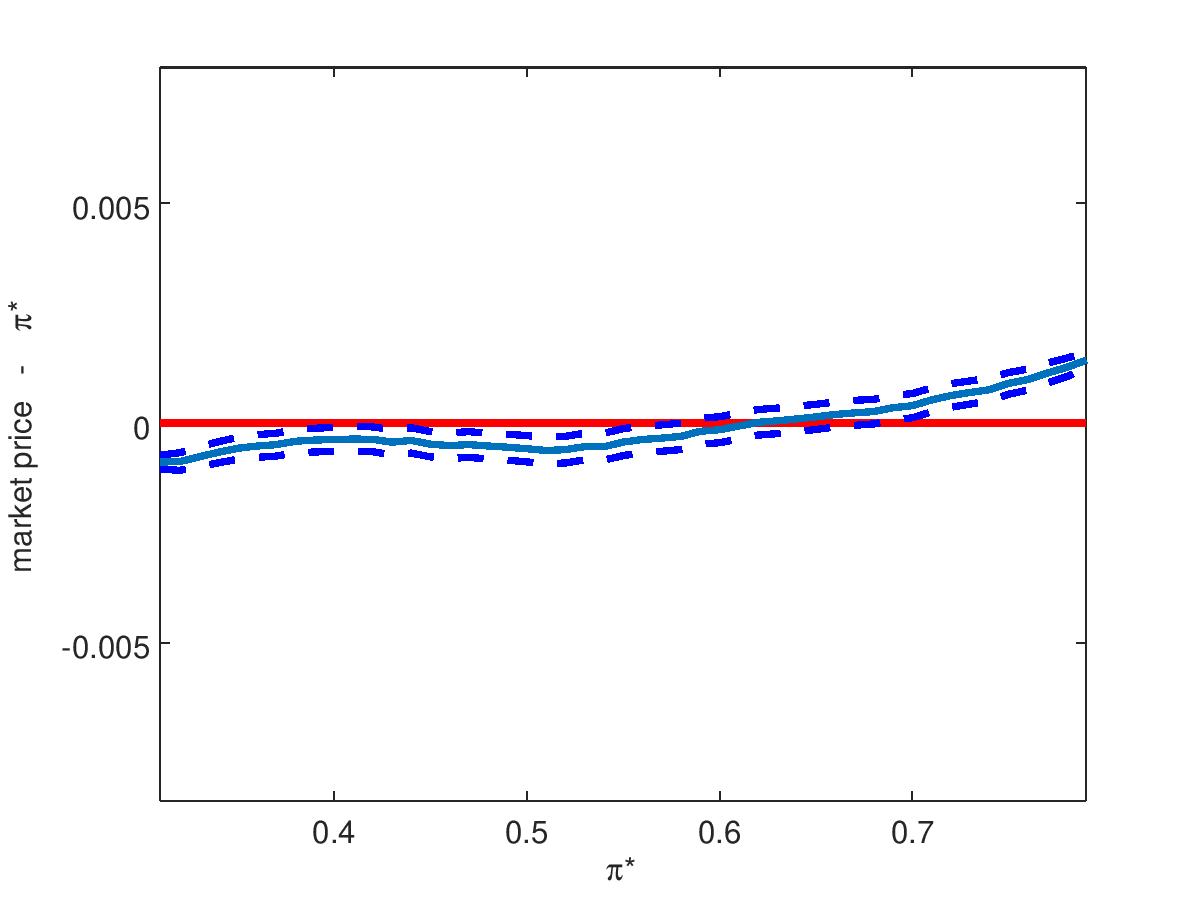}
\caption{\small Estimates of steady-state difference between the expected price and $\pi^*$ using the settings from Figure~\ref{fig:wrong}. We consider 49 equally spaced values for $\pi^*$, from $0.31$ to $0.79$. The dashed lines are CIs built \rev{\sout{on the same samples}} using two different procedures. Left: CIs are erroneously computed \rev{\sout{(using an external Octave script)}} according to the procedure in \cite{kets2014}. 
Right: CIs as computed by \mv using the approach based on the central limit theorem. 
\rev{\mv does not allow for the procedure of \cite{kets2014}, hence the left panel has been produced by an external Octave script. The different pseudo-random number generator 
is responsible for the small discrepancies in estimated steady-state values with respect to the right panel.}
}
	\label{fig:wrongpriceCI}
\end{figure}

\paragraph{Market price}
The right panel of Figure \ref{fig:wrong} shows the average price and should support one of the main results of \cite{kets2014}: the expected market  price  matches $\pi^*$ when $c=0.01$ and $\pi^*$ is  strictly  between  the  lowest  and  the  highest \rev{beliefs of the agents}.
\cite{bottazzigiachini2019b} suggest that such a conclusion is not correct and the source of inaccuracy should be found in the way in which CIs are built by \cite{kets2014}. 
In order to better understand this aspect, we create a new plot in Figure~\ref{fig:wrongpriceCI} focusing on the difference between market price and $\pi^*$. 
In the left panel of the figure we report the CIs (dashed lines) obtained by applying the procedure of \cite{kets2014}, while in the right panel we show those obtained by \rev{following our approach.} 
As one can notice, the procedure of \cite{kets2014} produces large CIs, backing the claim of the authors. Instead, the CIs obtained by \rev{our approach} 
based on the central limit theorem are much tighter and disprove the claim of the authors.
To understand the source of disagreement between the two approaches, consider the following argument: if the \np{10000} observations of $p_t$ from each simulation used to compute the average price of every replication were independent and at steady state, then we would not have spotted any significant difference. Indeed, according to the Central Limit Theorem, we have that each time average is (approximately) distributed as a normal random variable with mean the steady-state expectation of the price and variance the steady-state variance of price over $\sqrt{\np{10000}}$. The initial condition bias lets the expected time average be different from the steady-state expectation. The strong autocorrelation in the price process \citep{bottazzigiachini2019b} 
lets confidence intervals be too wide. While the \manualrd{} procedure used here can do nothing about the initial condition bias, with respect to confidence intervals \rev{our approach} 
does not assume anything about the distribution of time averages, simply relies on the Central Limit Theorem. Indeed, the average across \np{1000} independent replications of the \np{10000}-period time averages is (approximately) distributed as a normal random variable with variance the \rev{one of the time averages} over $\sqrt{\np{1000}}$.

This exercise shows the importance of correctly building confidence intervals when testing hypothesis on steady-state quantities from simulated models. 
We proceed showing that setting the required statistical significance ($\alpha$) and confidence interval width ($\delta$) instead of the total number of independent replicas is a much more 
reliable and efficient procedure to test hypotheses on steady-state expectations. 

\begin{figure}[t] \centering 
\includegraphics[width=0.33\linewidth]{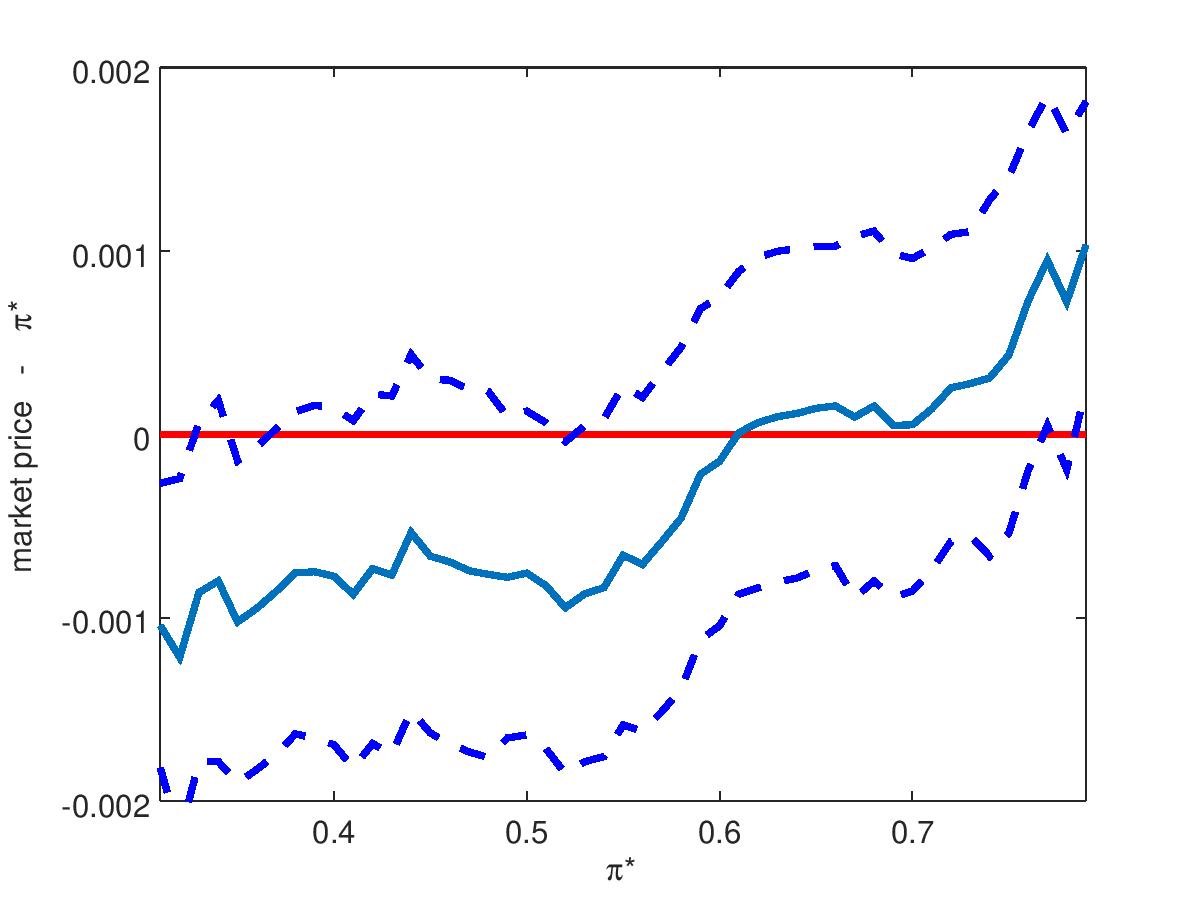}
\includegraphics[width=0.33\linewidth]{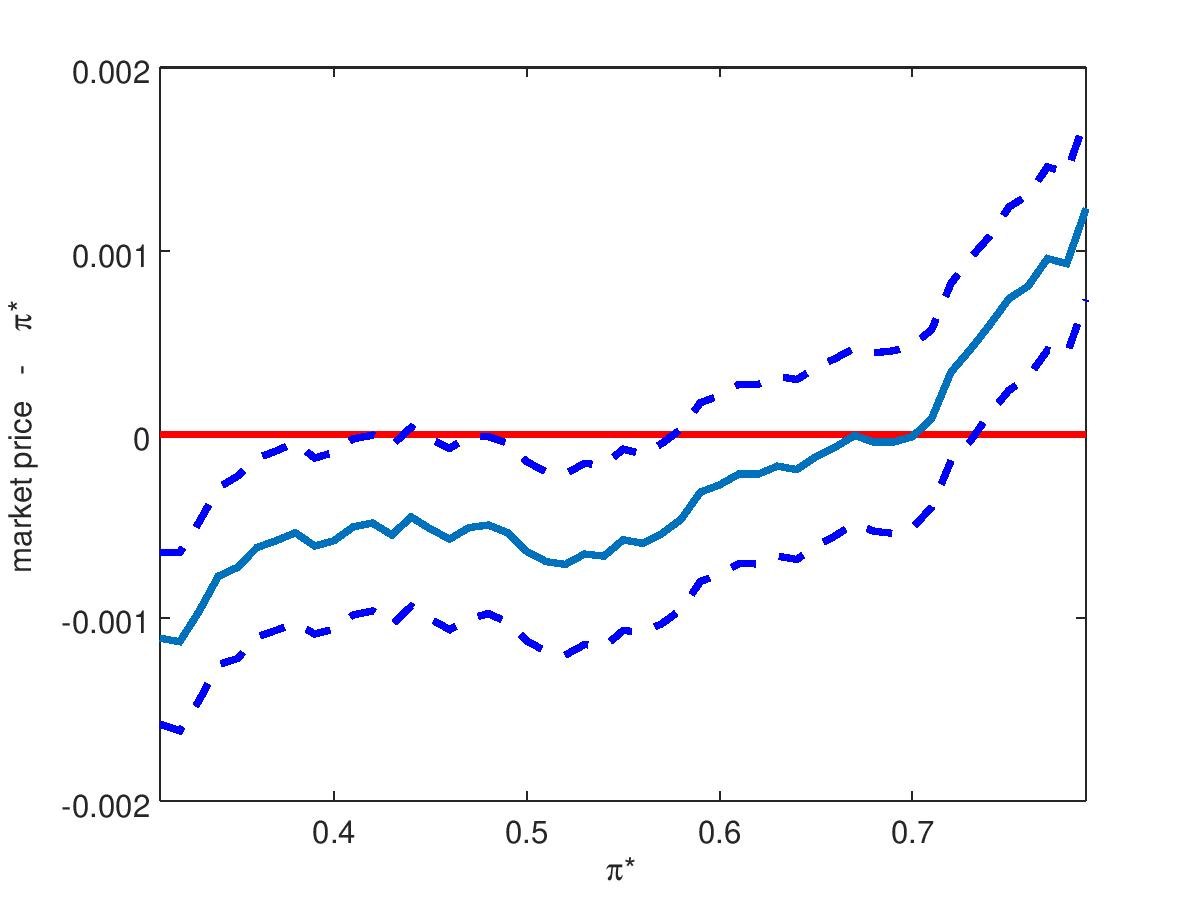}
\includegraphics[width=0.33\linewidth]{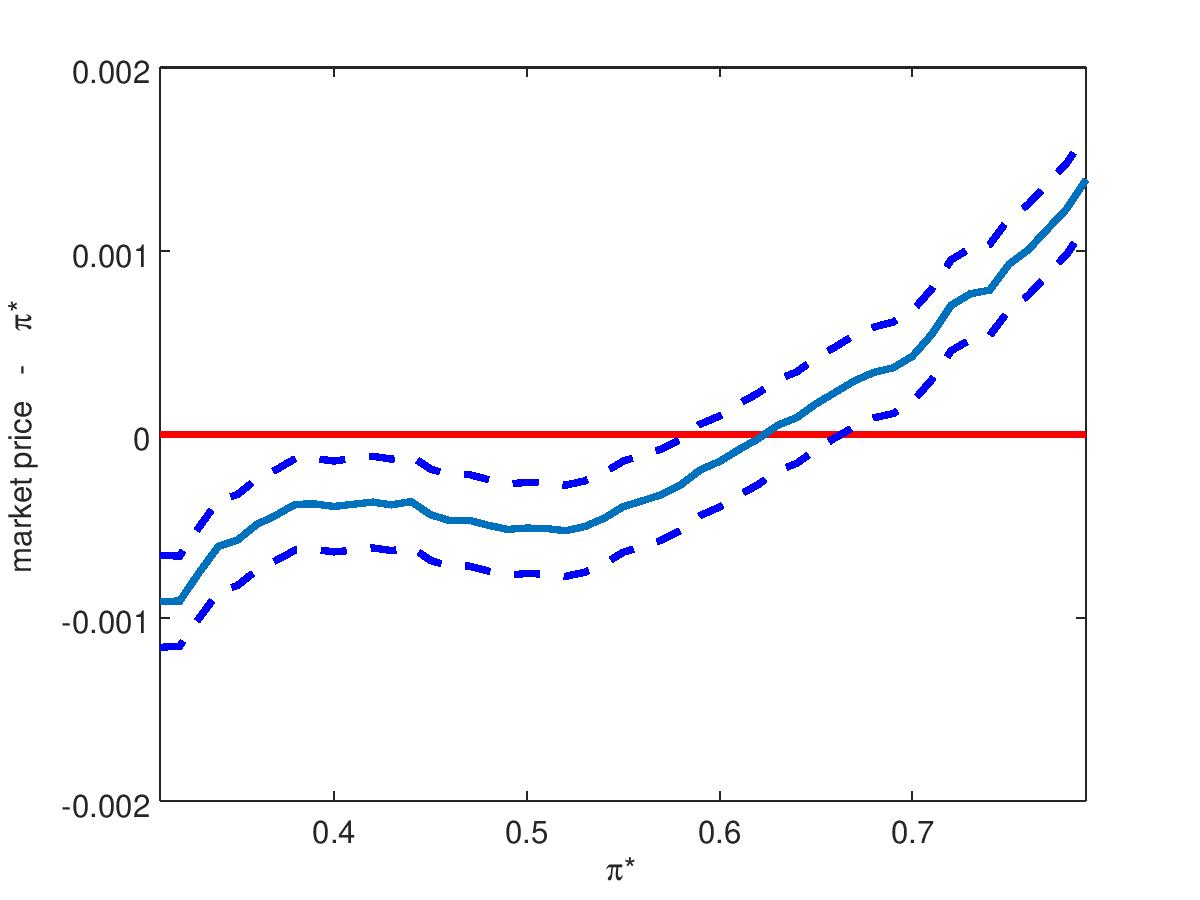}	\caption{\small
Estimates of steady-state difference between the expected price and $\pi^*$ using the settings from Figure~\ref{fig:wrong} for warmup estimation and time horizon. The number of simulations is automatically chosen by \mv{} according to the used values of $\delta$, for $\alpha=0.025$. As in Figure~\ref{fig:wrongpriceCI}, we consider 49 equally spaced values for $\pi^*$, from $0.31$ to $0.79$. 
Left: $\delta=0.002$, the number of simulations varies between \np{60} and \np{120}. Center: $\delta=0.001$, the number of simulations varies between \np{120} and \np{360}. Right: $\delta=0.0005$, the number of simulations varies between \np{420} and \np{1440}.}
	\label{fig:wrongprice}
\end{figure}

\paragraph{Market price for different $\alpha$-$\delta$}
In Figure \ref{fig:wrongprice}, we test the hypothesis from \cite{kets2014} that no difference between the average price  and $\pi^*$ exists under the parametrization in Table \ref{tab:params}. We use again \manualrd keeping the same\rev{, erroneous,} settings for warmup estimation and time horizon discussed in advance\rev{, while we do not impose a maximum number of simulation, which is therefore} 
 automatically chosen \rev{according to Equation \eqref{eq:ci} for the} 
  different values of $\delta$ ({0.002}, {0.001}, and {0.0005}), for $\alpha=0.025$ (i.e., a statistical confidence of 97.5\%). 
If the hypothesis from \cite{kets2014} were correct, the difference should be almost never significantly different from zero for any $\delta$ considered. 
Instead, we notice that $\delta$ plays an important role in assessing the hypothesis testing outcome. 
Indeed, while with $\delta=0.002$ the computed CIs for the difference among market price includes $0$ for almost all $\pi^*$, with $\delta=0.001$ the CIs almost never includes $0$. 
This confirms the point of \cite{bottazzigiachini2019b} and the results we have obtained in Figure \ref{fig:wrongpriceCI} right panel: the hypothesis of no difference between the average price and $\pi^*$ is generically rejected. 
Focusing on $\delta=0.0005$ and looking at the number of required simulations, one notices that there exist cases in which 
the hypothesis that no difference between the average price  and $\pi^*$ exist can be rejected
with less than \np{1000} simulations.\footnote{This is typically the case at the extrema, indeed {0.31} and {0.79} require, respectively, 420 and 480 replicas.} In other cases, instead, \np{1000} independent replications are not enough and one may risk 
to get to the wrong conclusion
simply because of an insufficient number of replicas.\footnote{This may occur for $\pi^*$ around {0.55}, where \rev{our algorithms} 
need \np{1440} simulations to reach the required interval width.} 

Notice, however, that due to the arbitrary choice of the end of the warmup period, all the estimates are biased by the initial conditions. 
We next use \rev{our} 
automated steady-state analysis \rev{techniques} (\autord and \autobm) to accurately estimate steady-state expectations and to finally assess on such obtained results the hypothesis of \cite{kets2014}.

\subsection{Steady-state analysis with \emph{\autoRD} and \emph{\autoBM} using automatic warmup estimation}\label{sec:automatic}

\begin{figure}[t] \centering 
	\includegraphics[width=0.49\linewidth]{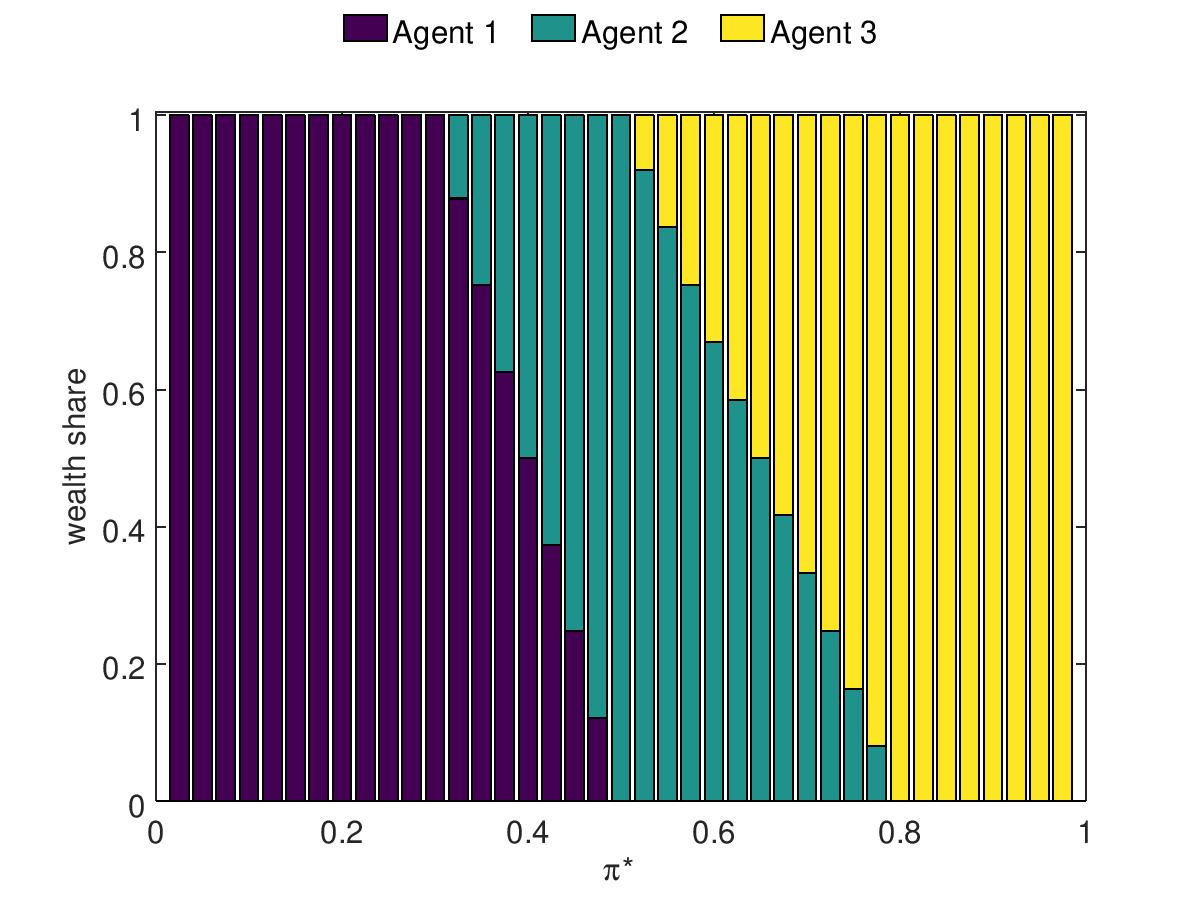}~
	\includegraphics[width=0.49\linewidth]{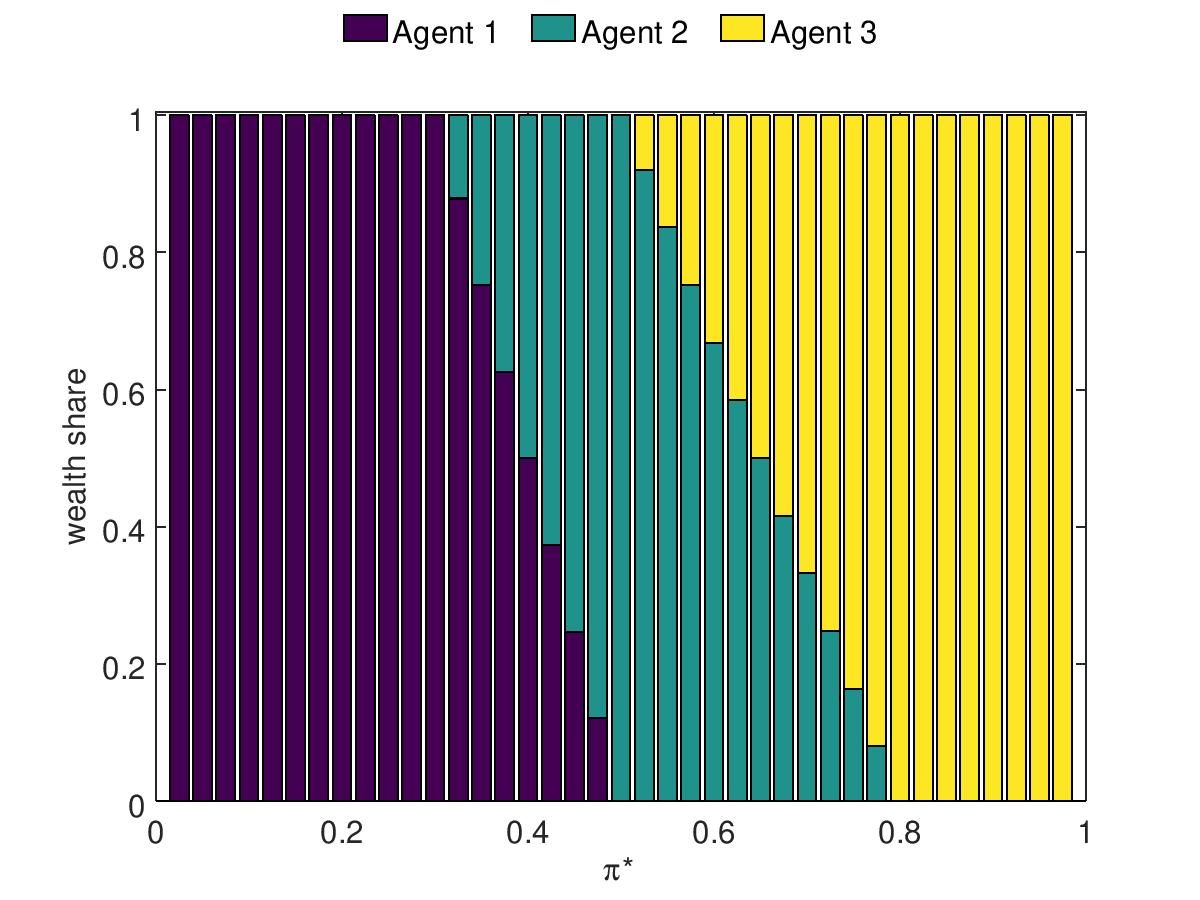}
	\caption{\small Steady-state levels of average wealth shares. Left: \autord{}. Right: \autobm{}. We set $\alpha=0.025$ and $\delta=0.001$, while we consider 39 equally spaced values for $\pi^*$, from $0.025$ to $0.975$, each one of these points requires a separate \mv{} \rev{analysis} 
		that has been invoked and aggregated with the others by means of an external Octave script.}
	\label{fig:wealth}
\end{figure}

Now we repeat the exercises using \rev{our proposed} automated algorithms for steady-state analysis.  
\footnote{\rev{In \ref{appendix:market} we compare the results showed here with those obtained replacing the default Anderson-Darling normality test with the Cramer-Von Mises test. No remarkable differences are spotted.}}

\paragraph{Wealth of the agents} Our results are displayed in Figure \ref{fig:wealth}. 
In the left panel we use the RD approach (\autord) while on the right we use the BM one (\autobm).
As one can notice: $i)$ our results comply with the theoretical and numerical ones of \cite[][cf. Figure 7]{bottazzigiachini2019b} and $ii)$ no significant difference can be spotted between the two pictures. 
Hence, our automated procedures 
allow one to avoid (or, at least, reduce) biases generated by initial conditions. 
As a practical example, let us consider the statistical analysis of steady-state expectations for $\pi^*=0.6$. The \manualrd procedure, using the settings from Section~\ref{manual}, estimates the expected wealth share of agents 1, 2, and 3 to, respectively, \np{0.089}, \np{0.519}, and \np{0.392}. 
Instead, \autoRD and \autoBM estimate the expected wealth shares to, respectively, 0, \np{0.668}, and \np{0.332}, in agreement with the results from~\cite{bottazzigiachini2019b}. 
Looking at the estimated warmup end, \rev{our algorithm proposes values much higher than the threshold set by \cite{kets2014}, see Figure \ref{fig:warmupend} in \ref{appendix:market}}. 
This confirms how manually setting the warmup end to \np{90000} generates a large initial condition bias in the estimation of steady-state expectations of \rev{the wealth of the agents}. 
Our analysis, other than correctly estimating steady-state expected wealth shares, clearly highlights the source of inaccuracy in the exercise of \cite{kets2014}, and stresses the importance of using a reliable automated procedure to pursue steady-state analyses. 

\begin{figure}[t] \centering 
	\includegraphics[width=0.49\linewidth]{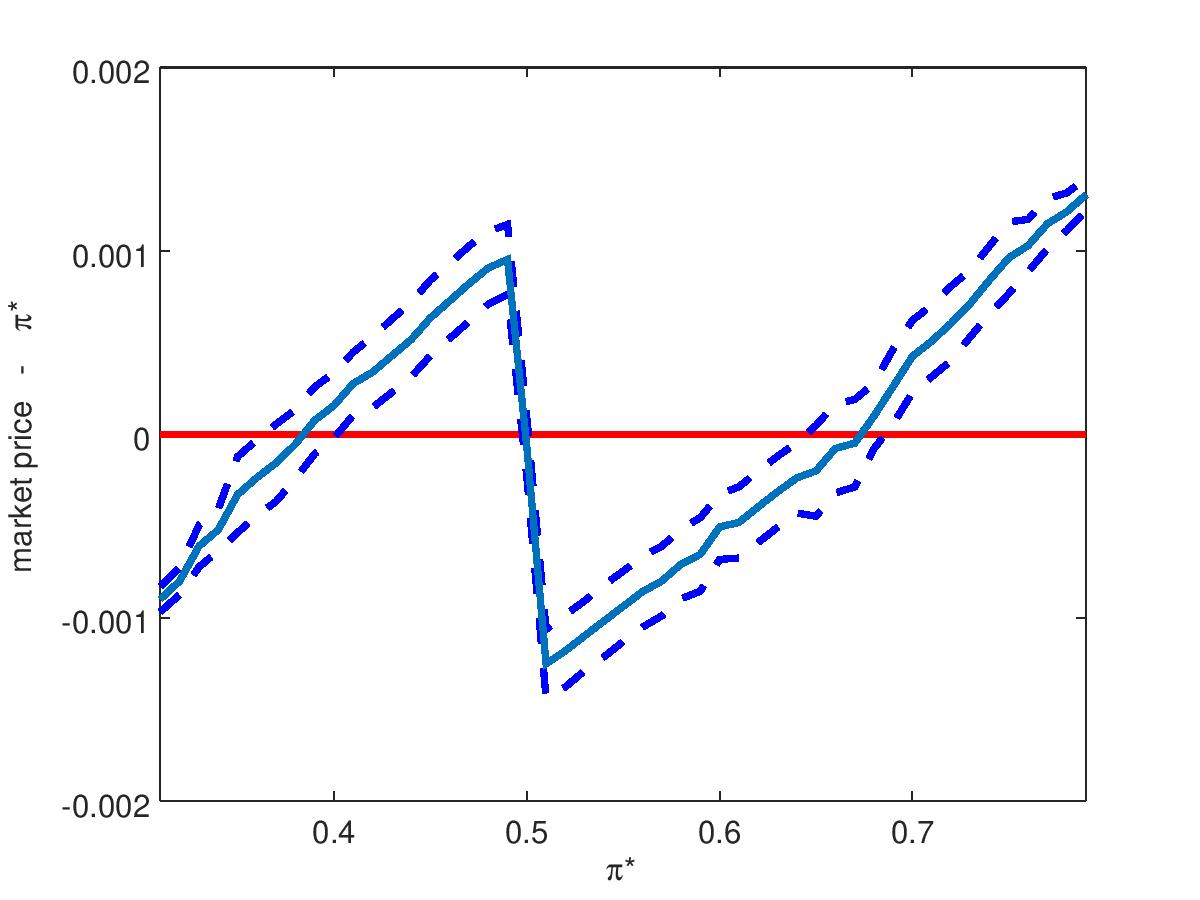}~
	\includegraphics[width=0.49\linewidth]{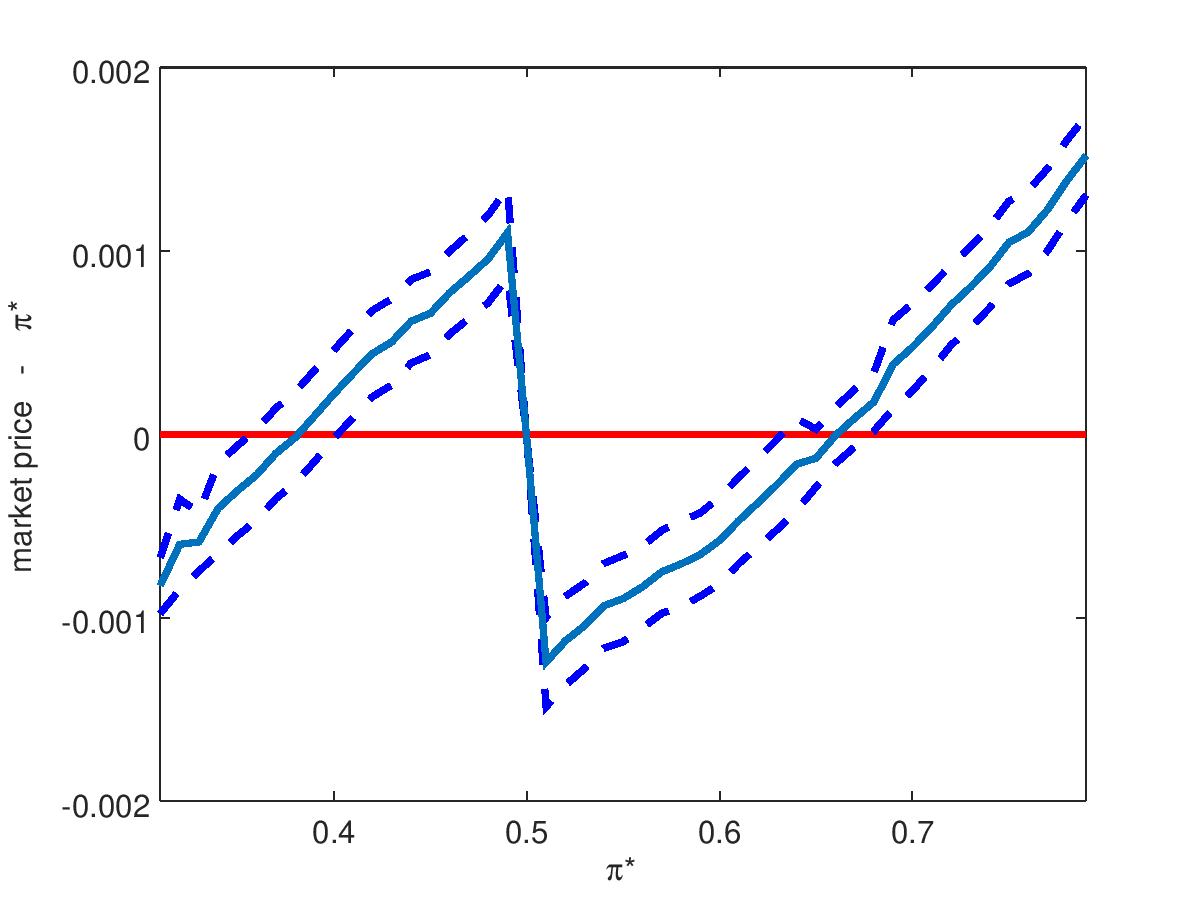}
	\caption{\small Steady-state levels of average price. Left: \autoRD{}. Right: \autobm. We set $\alpha=0.025$ and $\delta=0.0005$, while we consider 49 equally spaced values for $\pi^*$, from $0.31$ to $0.79$, each one of these points requires a separate \mv{} \rev{analysis} that has been invoked and aggregated with the others by means of an external Octave script.}
	\label{fig:price}
\end{figure}

While there are no significant differences in 
the estimates generated by \autoRD and \autoBM, the time required for producing them changes. 
Indeed, the analysis runtime of \autoRD (using parallelism degree 3) is about 3 times 
that of \autoBM. 
As discussed in Section~\ref{sec:bmvsrd}, cases like this with large warmup periods tend to favour \autobm. 

\paragraph{Market price}
Next, we estimate the expected value in steady state of the market price. 
As we did in Figures~\ref{fig:wrongpriceCI} and~\ref{fig:wrongprice}, in order to magnify confidence bands in Figure~\ref{fig:price} we show our estimates of the difference between the expected price and $\pi^*$ for different values of $\pi^*\in(0.3,0.8)$. 
In the left and right panels we show the \autoRD and \autoBM results, respectively. 
We notice that, for any value of $\pi^*$ considered, the estimated difference between the expected price and $\pi^*$ does not change in a significant manner between the two plots.
The emerging expected difference presents a clear pattern: it is larger (in absolute value) when $\pi^*$ is close to the belief of one of the agents.
Moreover, the expected difference appears to be negative when $\pi^*$ is closer to the belief of the agent whose belief is, relatively, the smallest (i.e., among the surviving ones) while it tends to be positive when it is the other way round.
These features are in line with the results obtained by  \cite{bottazzigiachini2019b} in Figure 4.
Hence, the reliability of the steady-state analysis performed by \rev{our techniques} 
 is confirmed.
Moreover, we can conclude that, contrary to what \cite{kets2014} argue, the steady-state value of the average price does not generally match $\pi^*$ when $c=0.01$. 


The analysis presented in Section~\ref{sec:automatic} satisfies all tests of the methodology for ergodicity diagnosis from Section~\ref{sec:methodologyergodicity}, confirming the reliability of the analyses. We show in the next section examples of analysis where this does not hold.

\section{Ergodicity diagnosis in a CRRA prediction market model with noise}
\label{sec:nonergodic}

We apply our methodology for ergodicity diagnosis to variants of the prediction market model. For all analyses we set $\alpha=0.05$ and $\delta=0.01$.

\subsection{Three variants of the prediction market with 2 CRRA traders: IID noise, AR noise, ergodic}
Here we modify the model studied in the previous section to allow violations of the ergodicity assumption.
Following \cite{BottazziGiachini2019a}, it is enough to assume that in the market there are $N=2$ traders who bet  maximizing their next-period CRRA utility to obtain non-ergodic price and wealth dynamics. 
Such a different behavioural assumption changes the betting rules. Indeed, we keep the assumption that agents 1 and 2 have heterogeneous beliefs ($\pi^1$ and $\pi^2$, respectively, with $\pi^1<\pi^2$) and we add risk preferences, assuming that the relative risk aversion coefficient of agent $i$ is $\gamma^i>0$, with $i=1,2$. 
Thus, we replace Equation \eqref{eq:alpha} with
\begin{equation}
	\alpha^1_t= (1-b^1_{t})p_t\quad\text{and}\quad\alpha^2_t=(1-b^2_{t})p_t+b^2_t\,,\quad\text{where}
	\label{eq:alpha1}
\end{equation}
\begin{equation}
 b^1_t=\dfrac{\left(p_t(1-\pi^1)\right)^{\frac{1}{\gamma^1}}-\left(\pi^1(1-p_t)\right)^{\frac{1}{\gamma^1}}} {\left(p_t(1-\pi^1)\right)^{\frac{1}{\gamma^1}}+p_t\left(\pi^1\right)^{\frac{1}{\gamma^1}}(1-p_t)^{\frac{1-\gamma^1}{\gamma^1}}} \quad\text{and}\quad
 b^2_t=\dfrac{\left(\pi^2(1-p_t)\right)^{\frac{1}{\gamma^2}}-\left(p_t(1-\pi^2\right)^{\frac{1}{\gamma^2}}} {\left(\pi^2(1-p_t)\right)^{\frac{1}{\gamma^2}}+(1-p_t)\left(1-\pi^2\right)^{\frac{1}{\gamma^2}}(p_t)^{\frac{1-\gamma^2}{\gamma^2}}}\,.
\label{eq:b}
\end{equation}

\cite{BottazziGiachini2019a} show that, depending on the \rev{values of the parameters} -- in particular $\gamma^1$ and $\gamma^2$ -- several long-run selection scenarios are possible. 
Indeed, one can generically have that: $i)$ one of the two agent has asymptotic unitary wealth share, $ii)$ both agents maintain positive wealth share asymptotically, $iii)$ path dependent scenarios in which either agent 1 obtains unitary wealth share asymptotically while agent 2 loses everything or vice-versa. 
Focusing on the market price $p_t$, in case $i)$  $p_t$ converges to $\pi^i$  (with $i$ the dominating agent), in case $ii)$ $p_t$ fluctuates in the interval $(\pi^1,\pi^2)$, and in case $iii)$ $p_t$ either converges to $\pi^1$ or to $\pi^2$ depending on the particular sequence of events realized.
Case $iii)$ is the one we are interested in: in such a case the ergodicity assumption is violated. 
However, the asymptotic convergence of the price to one out of 
two points makes quite easy to spot the lack of ergodicity and the presence of the two possible long-run price values. 
Hence, we complicate the setting assuming that there exists a third agent in the model who does not trade nor interacts in any way with agents 1 and 2. He simply observes the price and reports it. 
Such report is, however, noisy.
Defining $\tilde{p}_t$ the price such external agent reports, we assume 
	$\tilde{p}_t=p_t+v_t$ 
with $v_t=\theta v_{t-1}+u_t$ and $u_t$ a uniformly distributed random variable: $u_t\sim \mathcal{U}(-\eta,\eta)$, $\eta>0$. 
Such price reports are not taken into account by agents 1 and 2, hence all the properties of $p_t$ deriving from the analysis of \cite{BottazziGiachini2019a} remain unaffected.
Moreover, $v_t$ is an autoregressive process of order 1 with zero mean.
Hence, assuming $|\theta|<1$, we have that in the long-run $\tilde{p}_t$ fluctuates around either $\pi^1$ or $\pi^2$ depending on the sequence of realized events. 
Thus, the lack of ergodicity $\tilde{p}_t$ shows is somehow ``well-behaved''. 
That is, if one isolates the sequences in which $p_t$ converges to a given $\pi^i$, one will obtain that the time averages (i.e., horizontal means) of the relative observations of $\tilde{p}_t$, for $t$ large enough, are approximately normally distributed with mean $\pi^i$. Hence, we can say that $\tilde{p}_t$ presents two stationary points. 
At the same time, studying ergodicity 
of $\tilde{p}_t$ is much more complicated than performing the same tasks on $p_t$.
In what follows, we set $\eta=0.5$ and consider two scenarios for $\theta$.
In the first one, we consider $\theta=0$.
We refer to it as ``IID noise'' scenario, since we have that, in the long-run, the fluctuation described by $\tilde{p}_t$ around either $\pi^1$ or $\pi^2$ are IID. 
In the second scenario, instead, we consider the opposite case: setting $\theta=0.9$ we analyse the performance of our methodology when the noise is highly autocorrelated. We refer to it as the ``AR noise'' scenario.
Finally, as a robustness check, we apply our methodology to a case in which ergodicity should be ensured. We choose a scenario belonging to case $ii)$: long-run survival of both agents. This makes $p_t$ fluctuate in the interval $(\pi^1,\pi^2)$ indefinitely. Moreover, we set $\theta=0.9$ as in AR noise. These two assumptions, even if not affecting the ergodic properties of $\tilde{p}_t$, should make relatively harder for our methodology to work. We refer to this case as ``Ergodic''.
Table~\ref{tab:paramsCRRA} summarizes the parametrization used in our analyses. While the setting for \textit{IID noise} and \textit{AR noise} scenarios ensure the emergence of multiple stationary points, leading to a non-ergodic scenario, the assumptions for the \textit{Ergodic} scenario guarantee the persistent fluctuation of $p_t$ \citep{BottazziGiachini2019a}.

\begin{table}[t]
\center
%
\begin{tabular}{c c c cc cccc cc cc} 
\toprule
 \emph{Scenario} 
& 
&
&\multicolumn{2}{c}{\emph{Beliefs}}
&\multicolumn{4}{c}{\emph{Risk Aversion}} & \multicolumn{2}{c}{\emph{Wealth}}& \multicolumn{2}{c}{\emph{Noise}}
\\
\cmidrule(r){4-5} \cmidrule(r){6-9}  \cmidrule(r){10-11} \cmidrule(r){12-13} 
 & \ $N$ &  \ $\pi^*$ &  \ $\pi^1$ &  $\pi^2$ & \ & $\gamma^1$ & \  $\gamma^2$ & \ & $w^1_0$ & $w^2_0$ & \ $\eta$ & $\theta$ \\
\cmidrule(r){1-13}
IID noise & $2$ & $0.45$ & $0.2$ & $0.5$ & & $2$ & $0.5$ & & $0.5$ & $0.5$ & $0.5$ & $0$
\\
AR noise & $2$ & $0.45$ & $0.2$ & $0.5$ & & $2$ & $0.5$ & & $0.5$ & $0.5$ & $0.5$ & $0.9$
\\
Ergodic & $2$ & $0.45$ & $0.2$ & $0.8$ & & $2$ & $2$ & & $0.5$ & $0.5$ & $0.5$ & $0.9$
\\
\bottomrule
\end{tabular}
\caption{Parameters used for the prediction market model with CRRA traders and noisy price reporting. 
}
\label{tab:paramsCRRA}
\end{table}

\subsection{Application of the methodology for ergodicity analysis}

\paragraph{IID noise}
We start our analysis by applying \autoBM and \autoRD 
to the IID noise case. The former requires $\np{33792}$ steps of simulation. It signals that the warmup ends after the first batch of \np{1024} steps, and estimates the steady-state mean as 
$0.498$. 
Instead, \autoRD signals that the warmup ends after \np{1032} steps. 
After \np{2604} independent replications, it estimates the steady-state mean as $0.426$.  
With reference to our methodology for ergodicity analysis in Figure~\ref{fig:ergo}, we performed step 1, and passed the termination check of step 2. After that, step 4 requires to compare the results of \autobm and \autord. 
The difference among the two results is larger than $\delta$, suggesting an ergodicity problem. 
According to our method, we already have an indication of non-ergodicity. However, for illustrative reasons we also performed the Anderson-Darling normality test on the horizontal means computed by \autord (step 5), obtaining a p-value equal to 6.092E-251,  
which allows us to reject the null hypothesis that the horizontal means are normally distributed. 
Hence, our methodology is able to correctly spot that the IID noisy price lacks ergodicity.

\paragraph{AR noise}
We consider now the AR noise case starting with \autobm. 
The algorithm estimates the warmup to end in \np{1024} steps,
and 
the steady-state mean as $0.499$.
Instead,
by performing \autoRD 
we obtain that the warmup is estimated to end after \np{1032} steps. 
The total number of independent replications  needed by \autoRD to reach the IC width is \np{2709}, obtaining as result $0.426$.  
Therefore, the two algorithms provide significantly different results for the used $\delta$, suggesting an ergodicity problem (step 4).
This is confirmed by the Anderson-Darling normality test of step 5 which 
computes a p-value of $1.458$E-136, rejecting the normality assumption.  
Therefore, our methodology is able to correctly spot that also the AR noisy price lacks ergodicity.

\paragraph{Ergodic}
Finally, we perform a robustness check on our methodology by applying it to the Ergodic scenario.
Using \autobm, the warmup is estimated to end after \np{1024} steps producing as result $0.4027$.
Using \autoRD, 
one gets that the warmup is estimated to end after \np{1032} steps.\footnote{\rev{As one can notice, the estimated warmup is the same for every model specification we considered. This is due to the shortness of the model's warmup phase. Thus, \rev{our techniques} signal that the warmup is over after the first check (which, for implementation reasons, is 1024 steps for \autobm and 1032 for \autord}).}
The number of independent replications needed for reaching CIs of width $\delta$ is \np{210}, obtaining as result $0.4035$.  
Thus, the two algorithms give results within the tolerance of $\delta$ (step 4). 
The normality test from step 5 
cannot reject the null hypothesis of normality, as we get a p-value of $0.691$. 
Therefore, our methodology correctly suggests that no violation of ergodicity is observed.


\section{Conclusion}\label{sec:conclusions}
In this article we presented a fully automated framework \rev{of techniques and algorithms} for the statistical analysis of simulation models and, in particular, agent-based models (ABM). The framework, \rev{that we have } implemented 
in 
the statistical analyser \mv, 
provides a novel \rev{methodological and practical}
toolkit to the ABM community. 
These \rev{techniques} 
range from transient analysis, with statistical tests to compare results for different model configurations, to warmup estimation and the exploration of steady-state properties, including a
procedure for diagnosing ergodicity 
-- and 
hence 
the reliability of any steady-state analysis.

Our approach can be easily applied to simulators written in Java, Python, R or C++\rev{. We also support JMAB,} 
a framework for building macro stock-flow consistent ABMs.
Our \rev{techniques} 
allow modellers to automate 
\rev{the exploration of the models,} 
save time and 
avoid mistakes originating from semi-automated and error-prone 
tasks. 
Importantly, this facilitates 
reproducibility of experiments and promotes the use of a minimal set of \emph{default} 
analyses that should be performed when proposing or studying a model.

We 
validated our approach on two \rev{ABMs widely studied in} 
the literature: a large-scale macro financial ABM and a small scale prediction market ABM (and variants thereof). We obtained new insights on these models, 
\rev{and we identified and fixed} erroneous 
results from 
prior analyses.
Our framework also allows one to easily parallelize 
simulations 
within the cores of a machine or in a computer network. 
For instance, 
we reduced the analysis runtime for the macro ABM 
from 15 days to 16 hours on a machine with a CPU with 20 cores. Indeed, our 
toolkit enables 
modellers to run extensive tests in a unique environment (i.e., without the need of exporting data) and optimizing computational time (which is often precious; see also the discussion in \citealp{lamperti2018agent}).

Our approach is rooted in results from the simulation, computer science and operations research communities, which we aim to 
make available to the ABM community. 
Connecting these communities is 
critical to leverage the most effective techniques and approaches across fields. For example, the stationarity analysis proposed by \cite{grazzini2012analysis} mentioned in Section~\ref{sec:outputanalysis}
can be viewed as  a
non-automated version of the batch means approach by~\cite{Conway1963} and \cite{doi:10.1287/opre.27.5.1011}.


In the near future, we plan to integrate \rev{our tool-implementation of the approach, \mv{},} with other popular platforms used to 
build and analyse simulation models -- including the LSD environment for 
ABMs~\citep{valente2008},  the JASMINE environment for discrete-event simulations \citep{richiardi2017jas} \rev{and  NetLogo~\citep{netlogo}.}
We see this 
article as a first step in bringing practices from the statistical model checking (SMC) tool-set \rev{from computer science} to  
the ABM computational economics community. 
Of particular interest in this respect are
SMC techniques
developed to mitigate two classic problems of Monte Carlo methods: 
dealing with models 
that present rare events~\citep{DBLP:conf/rp/LegayST16}, 
and
using machine learning 
techniques 
to reduce the number of simulations~\citep{DBLP:conf/rv/BortolussiMS15}.
%
Finally, we will expand the family of \rev{proposed} automated analysis techniques.  
For instance, we 
will extend and refine our 
ergodicity diagnostics procedure, e.g., tackling the problem of identifying multiple stationary points (assuming they are finitely many) by means of clustering algorithms.
\rev{Of course, the analysis of an ABM typically goes beyond estimation of average behaviours as we consider here, which however is often considered as a necessary first step in ABM analysis. For this reason, we} 
also plan to further improve our proposals for 
the analysis of simulation output, 
e.g., by 
introducing corrections for multiple testing across the time domain, and move beyond it, 
e.g., by considering \rev{bifurcation analysis,} sensitivity analysis and parameter calibration, which are prominent in the ABM community. 
\rev{For instance, our methodology for estimating the steady-state average offers a reasonable solution for those who would like to perform sensitivity analyses on models whose target outcome is not (or cannot be) well-specified.\footnote{\rev{The reference framework of sensitivity analysis, as in \cite{saltelli1999quantitative}, assumes that a model is a function $f$ that links the vector of inputs $x$ with the scalar outcome variable $y$: $y=f(x)$. In the models we consider here, selecting the outcome $y$ is not straightforward since we obtain  time series as output instead of single values. One may choose a given time horizon and record the value of the variable of interest at that horizon, but this solution appears rather arbitrary.}}   Indeed, under the assumptions we make, any parameter setting is uniquely related to a steady-state average of the variable of interest. This connection allows one to perform sensitivity analysis exercises in a well-defined manner.\footnote{Notice that the exercises we perform in Figures \ref{fig:wrong}, \ref{fig:wrongpriceCI},  \ref{fig:wrongprice}, \ref{fig:wealth}, \ref{fig:price}  can be considered as simple local sensitivity analyses in which we observe how the average wealth shares or price change as we increase $\pi^*$.} Thus, our methods should be considered instrumental and complementary with respect to other types of analysis, like the global sensitivity analysis of \cite{saltelli1999quantitative}.}

\newpage
\singlespacing

\bibliographystyle{chicago}
\bibliography{refs}

\appendix

\rev{
\section{Multiple hypothesis problem, a critical discussion}
\label{appendix:MHT}
Here we provide a critical discussion on the problem of multiple hypothesis testing in our framework and propose a solution to such an issue when it may negatively affect our results.\footnote{\rev{See \cite{austin2014multiple} for a review of the multiple hypothesis testing problem.}} First of all, notice that the multiple comparison problem may occur in three points: when conducting counterfactual analysis in the macro ABM model, when testing for the warmup end, when testing for significant differences between estimated average price at steady state and true probability in the market model.

Concerning the tests we performed to check for significant changes in average dynamics produced by the macro ABM under different parametrizations, we assumed a nominal significance value of $\alpha$ for each single test. In particular, for each experiment $T=400$ statistical tests are conducted.
Under independence of tests, the family-wise (series-wise in our case) type-I error is $1-(1-\alpha)^T$.  The correlation in time series data may generate dependencies between the test we perform at a given time step $t$ and the tests we perform for subsequent time steps, invalidating the independence of tests assumption. Standard procedures may still control the family-wise error under positive dependence \citep{sarkar1997simes}, but more accurate procedures require some information on the underlying data-generating process \citep[see, e.g.,][]{sun2009large}. Alternative methods are based on the false discovery rate, but they still require independence of test. However, as shown by \cite{benjamini2001control}  and \cite{sarkar2002some}, the algorithms of \cite{benjamini1995controlling} and of  \cite{benjamini1999step} for controlling false discovery rate still work well under positive dependence of tests. Hence, addressing the multiple hypothesis testing problem in a consistent way should require some form of assessment of the dependence structure in synthetic data. A promising approach for our case may be the factor-analysis-based approach of \cite{friguet2009factor}. We have to notice that a very simple, but rather coarse and a bit time consuming, procedure one can immediately implement in our framework to limit the consequences of multiple testing is controlling if the test results are stable as  $\delta$ decreases. Indeed, decreasing $\delta$ standard errors become lower and estimates more precise. Thus, the series-wise probability of committing a type-I error in the comparison between the two different settings naturally decreases. Hence, if decreasing the value of $\delta$ the results provided by the tests remain rather stable, we can provide reasonable conclusions about the comparison without explicitly controlling for the multiple hypothesis testing problem.

The second case in which the multiple comparison problem is potentially affecting our results is when we use statistical tests to assess whether the warmup has ended or not.
We argue that such a problem does not invalidate our procedures, it actually makes them more conservative. Indeed, following the previous approaches of 
\cite{steiger2005asap3} and \cite{GilmoreRV17}, we use statistical tests to check whether the distribution of batch means is significantly different from a normal distribution or not. The inflation of type-I error, in this case, implies that we might reject the null of the distribution being normal when it actually is. However, this has the only effect of letting us increase the length of the simulation and, thus, of estimating a larger warmup period. Even if potentially costly in terms of computational time, a longer warmup horizon does not negatively affect the estimation of steady-state values. 

Finally, the  multiple comparison problem may be an issue when we perform  significance tests for the difference between average market price and true probability over several different values of $\pi^*$. Here the same caveats discussed in advance concerning the counterfactual analysis apply and observing how the results change as $\delta$ decreases may help to control the multiple testing problem. 
}

\rev{\section{Application: Transient analysis of a large macro ABM - more on counterfactual analysis}\label{appendix:macro}}
\rev{We present further experiments related to counterfactual analysis for the macro ABM by \cite{caiani2016agent}.}

\begin{figure}[t]
	\centering
	\subfloat[ \rev{CIs width for $\alpha=0.025$ and $N=100$ simulations.T-tests ``\emph{are means point-wise equal?}'' not rejected for significance $a_w\!=\!0.025$}]{\includegraphics[width=0.47\linewidth]{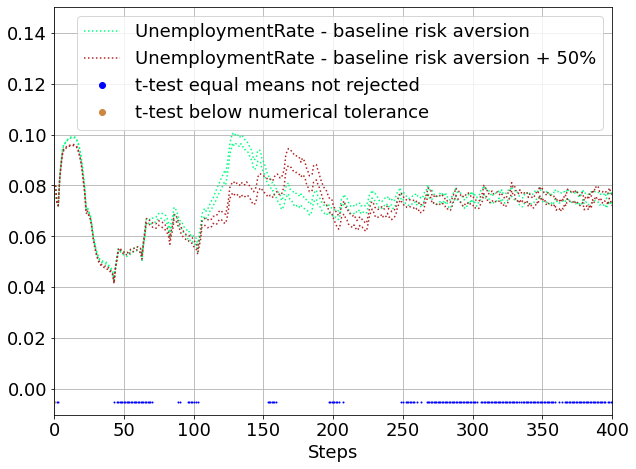}
	}
	\hfill
	\subfloat[\rev{CIs width for $\alpha=0.025$ and $\delta=0.005$. T-tests ``\emph{are means point-wise equal?}'' not rejected for significance $a_w\!=\!0.025$}]{\includegraphics[width=0.47\linewidth]{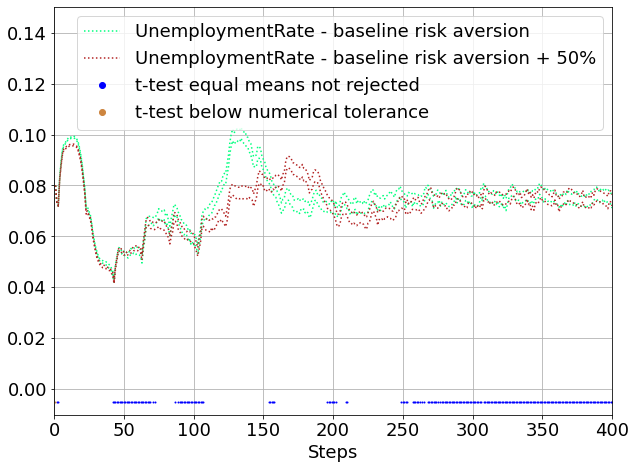}
	}
	\\
	\subfloat[Power of t-test in (a) for difference $\varepsilon=0.005$]{\includegraphics[width=0.47\linewidth]{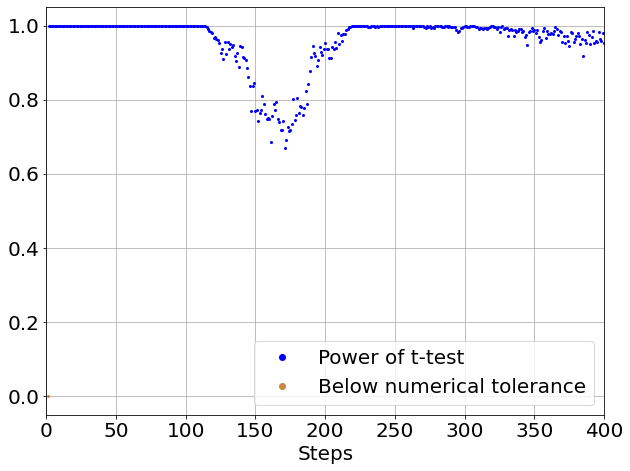}}
	\hfill
	\subfloat[Power of t-test in (b) for difference $\varepsilon=0.005$]{\includegraphics[width=0.47\linewidth]{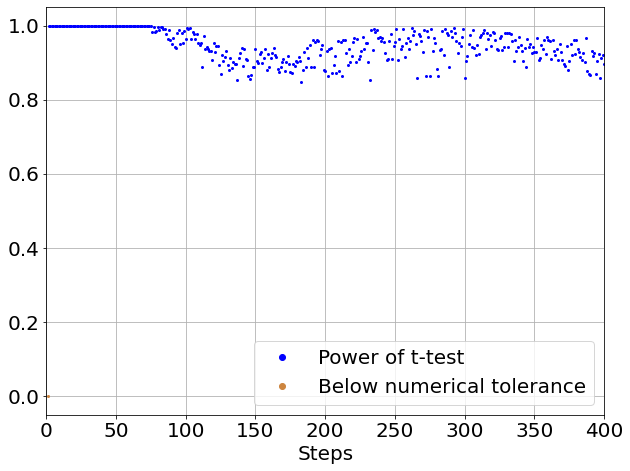}}
	\caption{
		Evolution of unemployment rate for \rev{two} different risk aversions for consumption firms: are they point-wise equal? \rev{This figure has same structure as Figure~\ref{fig:behaviouralWithPower}.} 
	}
	\label{fig:behaviouralWithPower_unemplRate}
\end{figure}

\subsection{\rev{Automatic experiment comparison and statistical testing: unemployment rate}}\label{sec:counterfactual_unempl}
\rev{In this section we extend the analysis performed on the macro ABM by \cite{caiani2016agent}. In particular, we extend the counterfactual analysis done in Section~\ref{sec:counterfactual} for bankruptcies by considering the unemployment rate. 

The results are shown in Figure~\ref{fig:behaviouralWithPower_unemplRate}, which has the same structure of Figure~\ref{fig:behaviouralWithPower} from the main text.
The figure provides the same study done for bankruptcies for the unemployment rate. The results are confirmed, even though the discrepancy among the dynamics of the two model variants is less marked.
}

\subsection{\rev{Counterfactual analysis with u-test}}\label{sec:utest}

\begin{figure}[t]
\centering
\subfloat[\rev{Not rejected t- and u-tests  for bankruptcies for the setting as in Figure~\ref{fig:behaviouralWithPower} (a) 
}]{\includegraphics[width=0.47\linewidth]{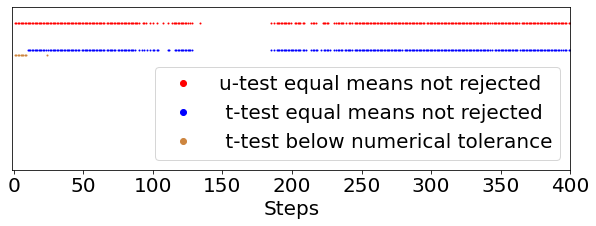}
 }
\hfill
\subfloat[\rev{
Not rejected t- and u-tests for bankruptcies for the setting as in Figure~\ref{fig:behaviouralWithPower} (b) 
}]{\includegraphics[width=0.47\linewidth]{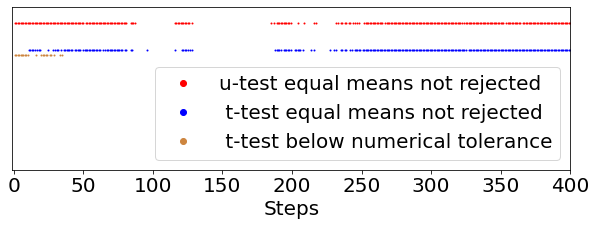}
}
\\
\subfloat[\rev{Not rejected t- and u-tests for unemployment rate for the setting as in Figure~\ref{fig:behaviouralWithPower_unemplRate} (a)} ]{\includegraphics[width=0.47\linewidth]{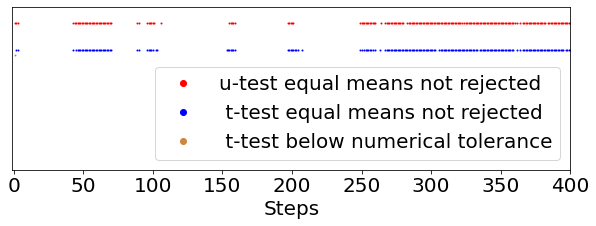}}
\hfill
\subfloat[\rev{Not rejected t- and u-tests for unemployment rate for the setting as in Figure~\ref{fig:behaviouralWithPower_unemplRate} (b)} ]{\includegraphics[width=0.47\linewidth]{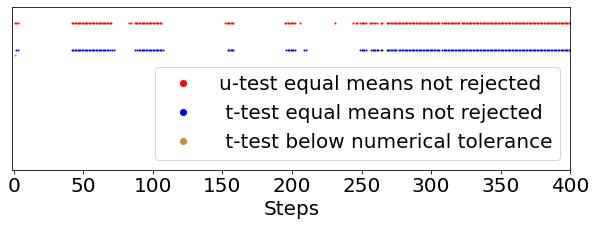}}
\caption{\rev{Each plot provides t- and u-tests ``\emph{are means point-wise equal?}'' not rejected for significance $a_w\!=\!0.025$ for bankruptcies (a,b) and unemployment rate (c,d). As in Figures~\ref{fig:behaviouralWithPower} and~\ref{fig:behaviouralWithPower_unemplRate}, we compare analysis results for two different risk aversions $C$ for consumption firms. As in Figures~\ref{fig:behaviouralWithPower} and~\ref{fig:behaviouralWithPower_unemplRate}, the left-column considers an analysis setup involving $n=100$ simulations for each time point, while in the right-column we let our algorithms find automatically the correct $n$ for each time point. Yellow dots denote initial steps with variances so small to get intermediate results below the numerical tolerance of our implementation of the t-test (1E-15).
}
}
\label{fig:utest}
\end{figure}

\rev{As discussed in Section~\ref{sssec:eqtest_power}, our framework allows for counterfactual analysis on results obtained from two different parametrizations of a model, to decide whether the changes in the parameters led to significant changes in the average dynamics. In particular, we do this for transient analysis by comparing the obtained point-wise average behaviours using Welch's t-test~\citep{welch1947}. In the main text, we opted for such test because, as further demonstrated in Section~\ref{sec:macro}, it is possible to compute its \emph{power} as in  \cite{chow2002}. However, there exist further tests for this type of analysis which make weaker assumptions than Welch's t-test. An example is the so-called \emph{Wilcoxon-Mann-Whitney} test, or just \emph{u-test}~\citep{10.1214/aoms/1177730491}. Differently from Welch's t-test, the u-test does not assume that the two populations being compared are normally distributed, however, to the best of our knowledge, no closed-formula exists for estimating its power.

Despite the asymptotic normality of our random variables (batch means) lets us deem the assumptions underlying the t-test  as not too strict in our framework, we also support the u-test. This test can be used for performing counterfactual analyses as those performed in Section~\ref{sec:counterfactual}. We hereby reproduce the experiments from  Section~\ref{sec:counterfactual} using the u-test rather than Welch's t-test. The results are depicted in Figure~\ref{fig:utest} for bankruptcies (first row) and unemployment rate (second row). The figure considers the settings from Figure~\ref{fig:behaviouralWithPower} (a) and (b) for the first row, and those from Figure~\ref{fig:behaviouralWithPower_unemplRate} (a) and (b) for the second row. The provided t-tests are those presented in the corresponding original figures. 

From the figure we can see that, for the considered macro ABM and analysis of interest, the results of the two tests are very similar.
}

\subsection{\rev{Automatic experiment comparison and statistical testing: Experiment on policy tax rate - Power}}\label{sec:counterfactual_tax_power}
\begin{figure}[t]
\centering
\subfloat[\rev{Power for t-tests in Figure~\ref{fig:policyTaxRate} (a) 18 vs 15 for $\varepsilon=0.5$.}]{\includegraphics[height=0.17\linewidth]{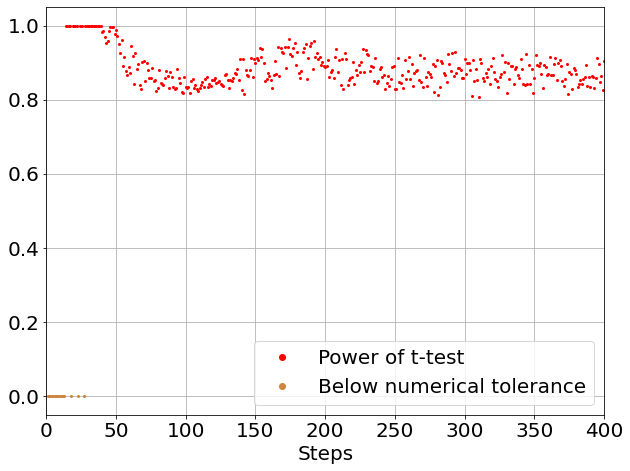}
 }
\quad
\subfloat[\rev{Power for t-tests in Figure~\ref{fig:policyTaxRate} (a) 18 vs 21 for $\varepsilon=0.5$.}]{\includegraphics[height=0.17\linewidth]{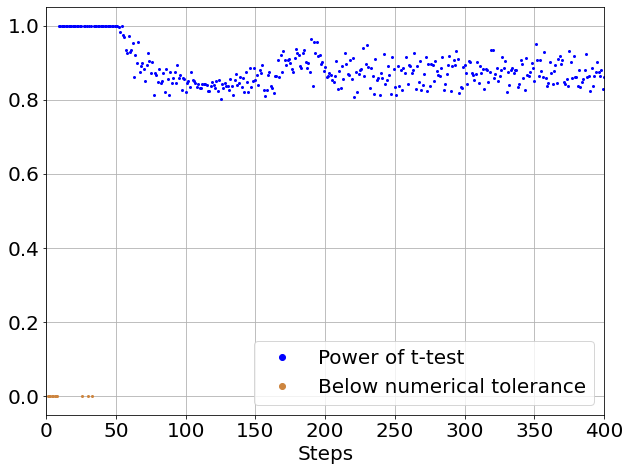}
}
\quad
\subfloat[\rev{Power for t-tests in Figure~\ref{fig:policyTaxRate} (b) 18 vs 15 for $\varepsilon=0.01$.}]{\includegraphics[height=0.17\linewidth]{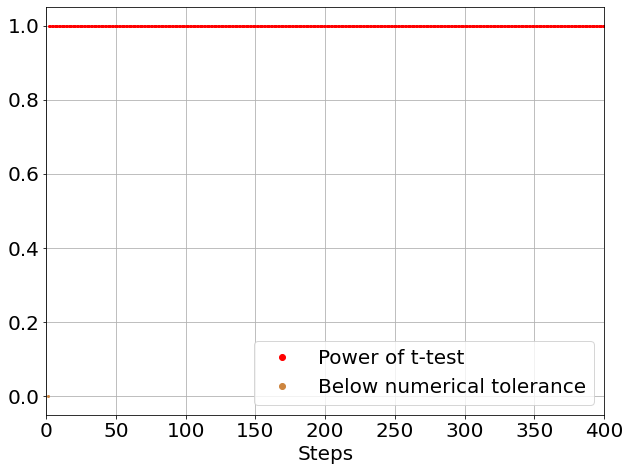}}
\quad
\subfloat[\rev{Power for t-tests in Figure~\ref{fig:policyTaxRate} (b) 18 vs 21 for $\varepsilon=0.01$.}]{\includegraphics[height=0.17\linewidth]{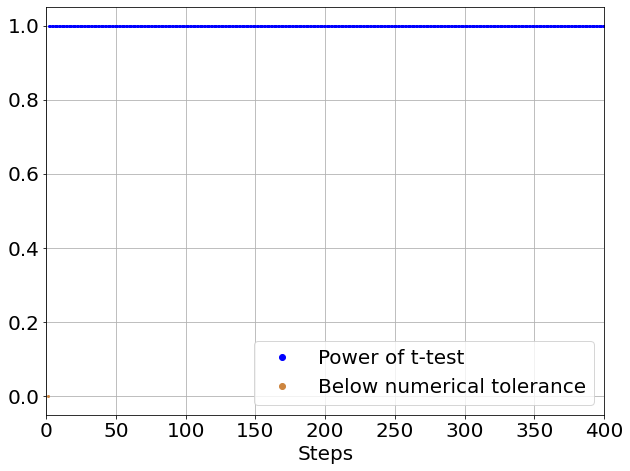}}
\\
\subfloat[\rev{Power for t-tests in Figure~\ref{fig:policyTaxRate} (c) 18 vs 15 for $\varepsilon=500$.}]{\includegraphics[height=0.17\linewidth]{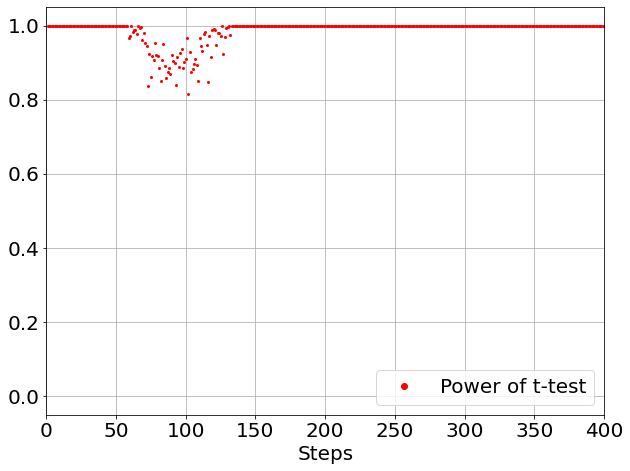}}
\quad
\subfloat[\rev{Power for t-tests in Figure~\ref{fig:policyTaxRate} (c) 18 vs 21 for $\varepsilon=500$.}]{\includegraphics[height=0.17\linewidth]{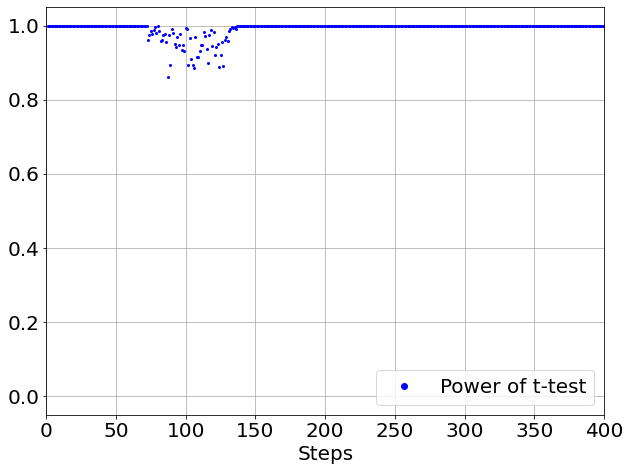}}
\quad
\subfloat[\rev{Power for t-tests in Figure~\ref{fig:policyTaxRate} (d) 18 vs 15 for $\varepsilon=300$.}]{\includegraphics[height=0.17\linewidth]{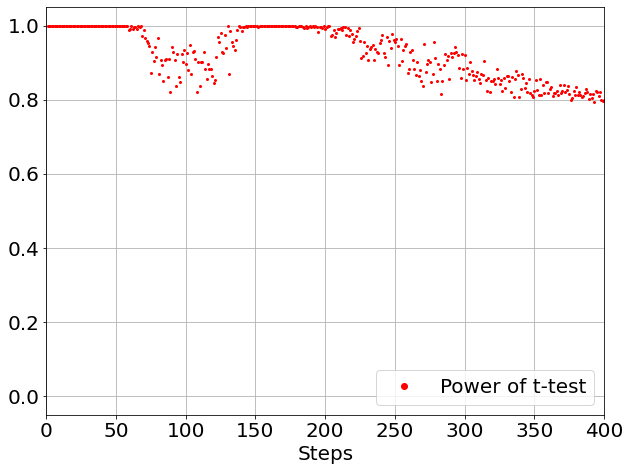}}
\quad
\subfloat[\rev{Power for t-tests in Figure~\ref{fig:policyTaxRate} (d) 18 vs 21 for $\varepsilon=300$.}]{\includegraphics[height=0.17\linewidth]{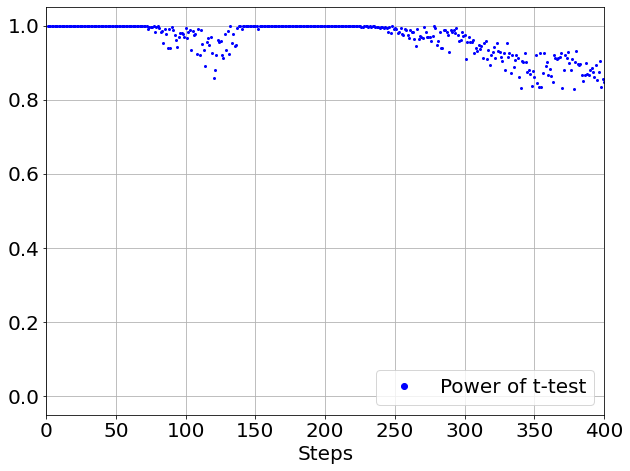}}
\caption{
\rev{Powers for the t-tests presented in Figure~\ref{fig:policyTaxRate}. Yellow dots denote initial steps with variances so small to get results below the numerical tolerance of our implementation, 1E-15.}
}
\label{fig:policyTaxRate_power}
\end{figure}

\rev{
In this section we provide the power for the t-tests computed in Section~\ref{sec:counterfactual_tax}, and in particular in Figure~\ref{fig:policyTaxRate}. 
The power is of the t-tests is shown in Figure~\ref{fig:policyTaxRate_power}. In all cases, the power is high, an in particular higher than the threshold of 0.8 mentioned in the main text.
}

\rev{
\section{Detailed description of the prediction market model}
\label{appendix:ketsetal}
}

The model is a pure exchange economy in discrete time, indexed by $t \in \mathbb{N}$, where $N$ agents repeatedly bet on the occurrence of a binary event. 
That is, in every $t$ two contracts are available for wagering: the first pays 1 dollar if the event occurs and zero otherwise, while the second pays 1 dollar if the event does not occur and zero otherwise. 
We model the event by means of a Bernoulli random variable $s_t$, such that $s_t=1$ means that the event at time $t$ has occurred, 
and $s_t=0$ otherwise. 
The probability of observing $s_t=1$ is a constant $\pi^*\in(0,1)$. 
Every agent $i\in\{1,2,\ldots,N\}$ assigns a subjective probability $\pi^i$ to the realization of the event at any time $t$. 
Agent $i$ has initial wealth equal to $w^i_0$ and at the end of every betting round it evolves in $w^i_t$ depending on the results of her betting. 
The total initial wealth in the market is normalized to 1, such that, since wealth is only redistributed by the betting system, it is $\sum_{i=1}^N w^i_t=1$ for all $t$.\footnote{Hence, one can indifferently refer to $w^i_t$ as both the wealth and the wealth share of agent $i$ at time $t$.}
In every period, the agents trade in the competitive market according to rules as in Equation \eqref{eq:alpha} and contracts' prices are fixed by means of market clearing conditions.
Without loss of generality, we assume that contracts are in unitary supply. Hence, calling $p_{1,t}$ and $p_{2,t}$ the price of the first and second contract, respectively, we have $\forall t$
\begin{equation}
	1=\sum\limits_{i=1}^N \dfrac{\alpha^i_t}{p_{1,t}}w^i_{t-1}\quad\text{and}\quad 1=\sum\limits_{i=1}^N \dfrac{1-\alpha^i_t}{p_{2,t}}w^i_{t-1}\,.
	\nonumber
\end{equation}
Since wealth sums up to 1 in every period, one has $p_{1,t}+p_{2,t}=1$, hence we call $p_{1,t}=p_t$ and $p_{2,t}=1-p_t$. 
Substituting with Equation \eqref{eq:alpha} and applying simple algebraic manipulations, one obtains
\begin{equation}
	p_t=\sum\limits_{i=1}^N\pi^i w^i_{t-1}\quad\forall t\,.
	\label{eq:price}
\end{equation}
After the market round, the outcome of the binary event is revealed and the wealth of agent $i$ evolves according to 
\begin{equation}
	w^i_{t}=
	\begin{cases}
		\dfrac{\alpha^i_t}{p_t}w^i_{t-1}\,=\,\left(1-c+c\,\dfrac{\pi^i}{p_t}\right)w^i_{t-1} & \text{if }s_t=1\,,\\[0.35cm]
		\dfrac{1-\alpha^i_t}{1-p_t}w^i_{t-1}\,=\,\left(1-c+c\,\dfrac{1-\pi^i}{1-p_t}\right)w^i_{t-1} & \text{if }s_t=0\,.
	\end{cases}
	\label{eq:wealth}
\end{equation}

\rev{\section{Application: steady-state analysis in a model of market selection using Cramer Von-Mises normality test}\label{appendix:market}}
\rev{
In this section we further discuss the normality tests supported in our algorithms for steady-state analysis, namely the one by Cramer Von-Mises, and replicate the analysis performed in Sections~\ref{sec:predmarket} and~\ref{sec:nonergodic} replacing the Anderson-Darling test with the Cramer Von-Mises one.

Notice that, when we test for normality of batch means inside \autow{}, 
we are taking as input of the test variables that are only approximately normal. 
One can reasonably assume that these random variables are closer to normality in the centre of the distribution rather than in the tails. This may create problems with the Anderson-Darling test we provide as default. Thus, we provide the option of using the Cramer-Von Mises normality test, that, weighting the tails less than the Anderson-Darling, should be less affected by the approximated normality of batch means. 

Here we report the results of the steady-state analysis performed on the market model replacing the Anderson-Darling normality test with the Cramer-Von Mises test. As one can notice in Figures \ref{fig:wealth_CvM}-\ref{fig:price_CvM}, the results are very close to the ones reported in Figures \ref{fig:wealth}-\ref{fig:price}.  

\begin{figure}[t] \centering 
	\includegraphics[width=0.49\linewidth]{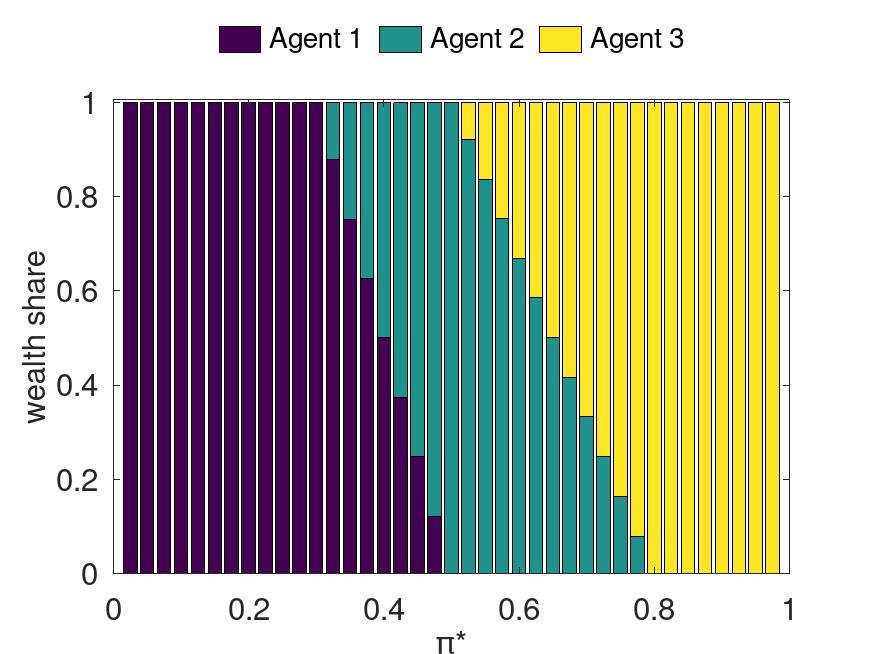}~
	\includegraphics[width=0.49\linewidth]{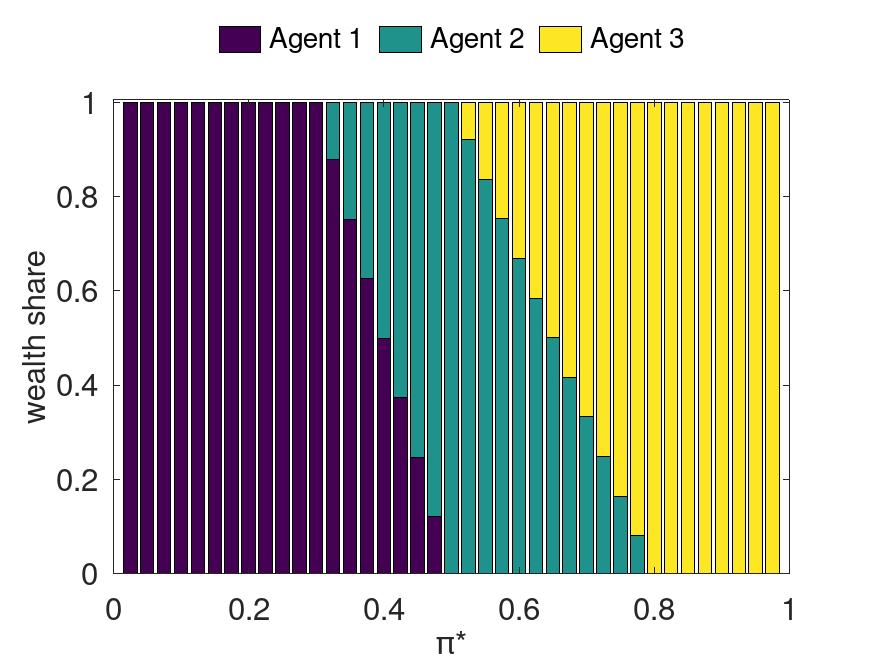}
	\caption{\small \rev{Steady-state levels of average wealth shares. Left: \autord{}. Right: \autobm{}. Same settings of Figure \ref{fig:wealth}, the only difference is the use of the Cramer-Von Mises test to check for the normality of batch means.}}
	\label{fig:wealth_CvM}
\end{figure}

\begin{figure}[t] \centering 
	\includegraphics[width=0.49\linewidth]{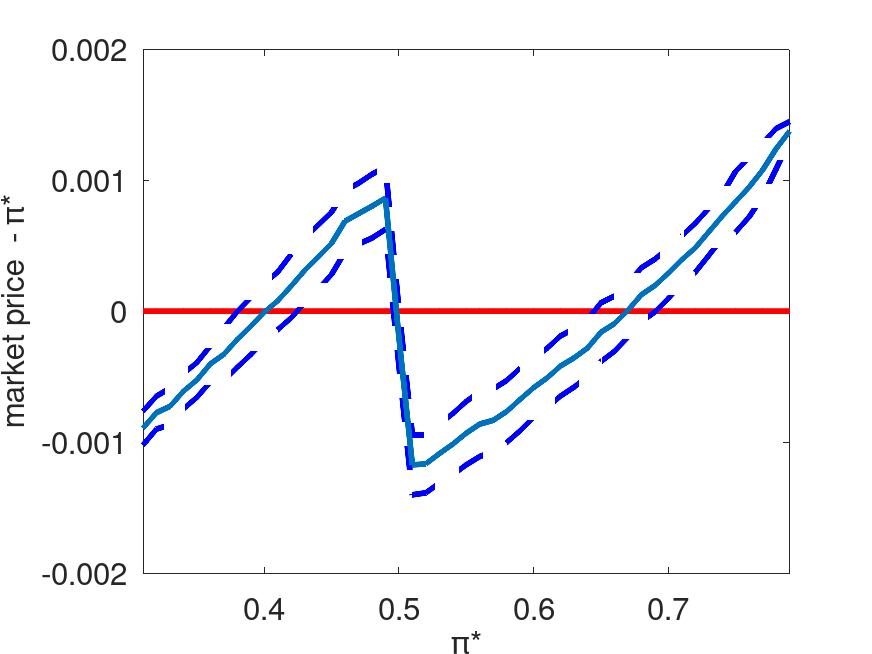}~
	\includegraphics[width=0.49\linewidth]{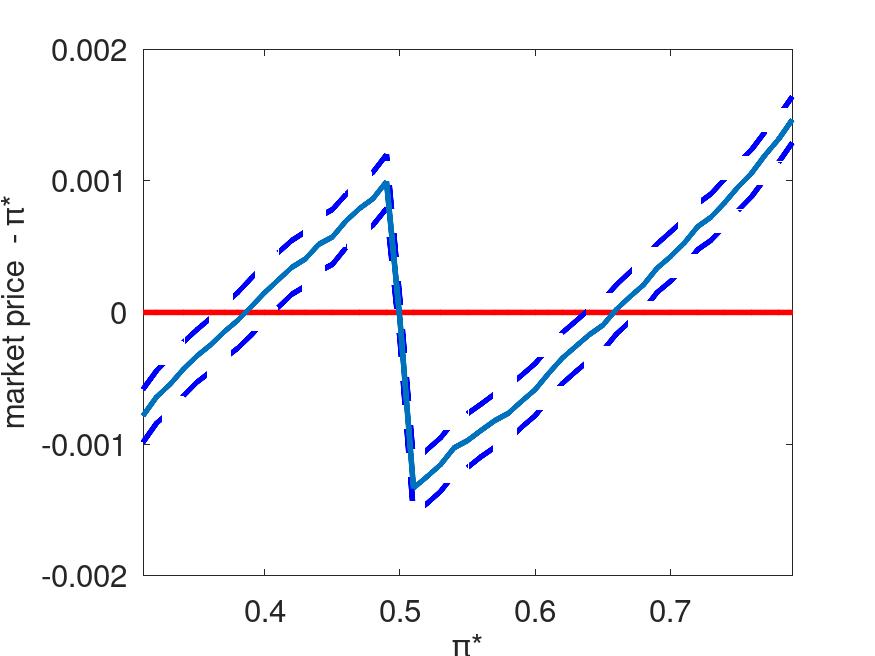}
	\caption{\rev{\small Steady-state levels of average price. Left: \autord{}. Right: \autord{}. Same settings of Figure \ref{fig:price}, the only difference is the use of the Cramer-Von Mises test to check for the normality of batch means.}}
	\label{fig:price_CvM}
\end{figure}

We control whether the two tests produce large differences in estimated warmup ends in Figure \ref{fig:warmupend}. As one can notice, the two tests generically yield coherent estimates, with the Anderson-Darling test generating more conservative estimates when differences are observed.

\begin{figure}[t] \centering 
	\includegraphics[width=0.49\linewidth]{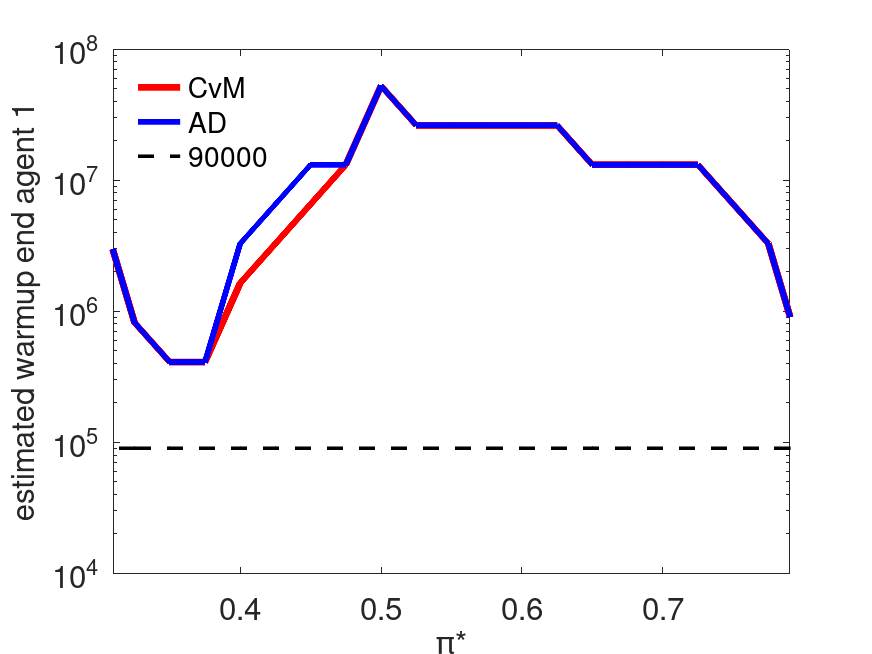}~
	\includegraphics[width=0.49\linewidth]{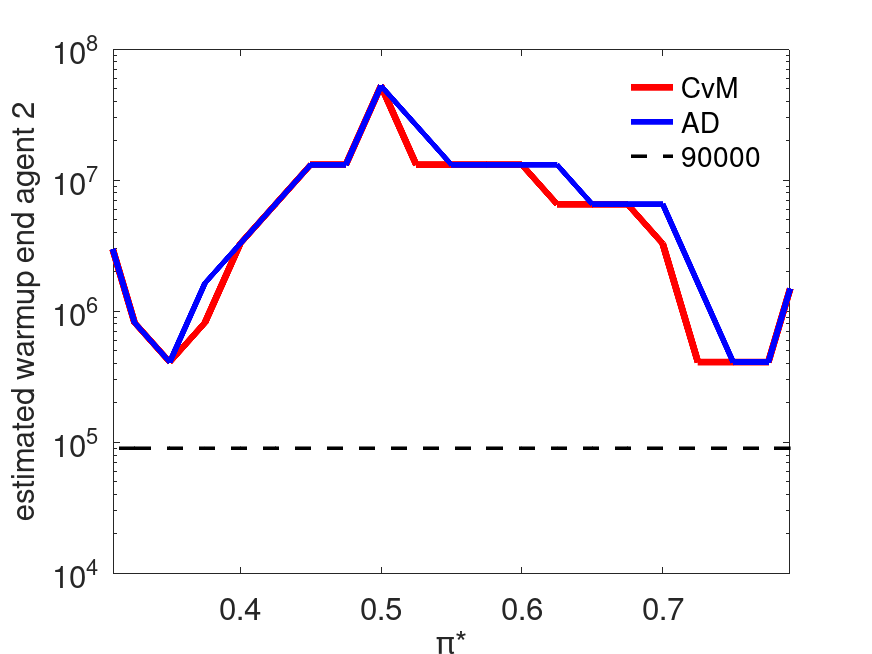}
	\includegraphics[width=0.49\linewidth]{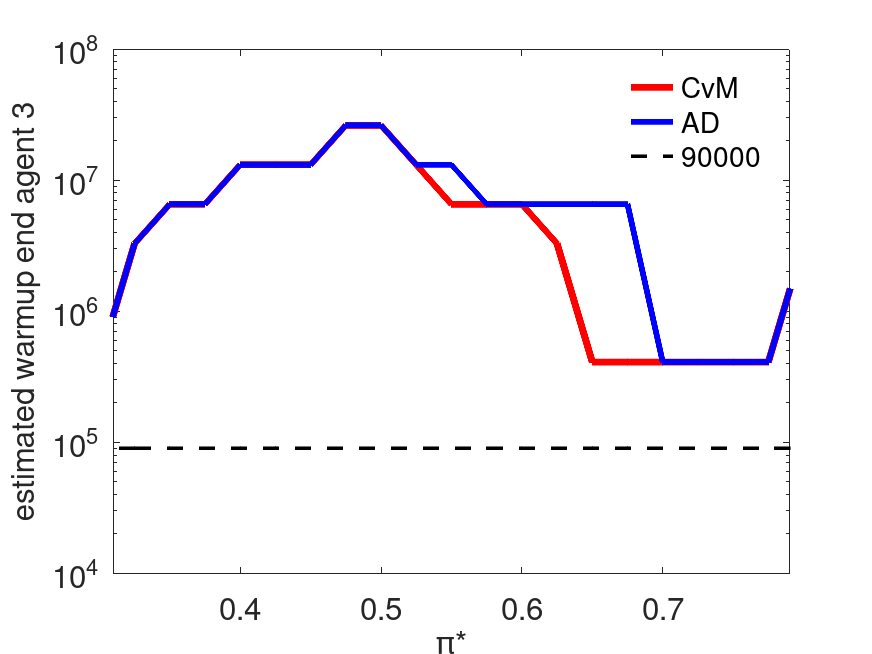}~
	\includegraphics[width=0.49\linewidth]{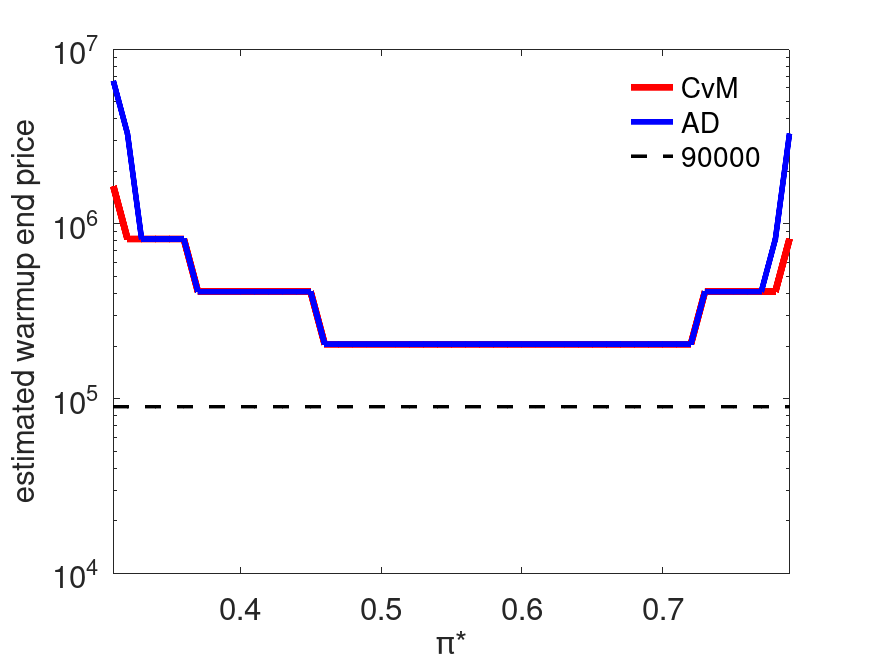}
	\caption{\rev{\small Comparison between warmup end estimated using the Cramer-Von Mises test (CvM) and using the Anderson-Darling test (AD). The value 90000 set by \cite{kets2014} has been added for reference.}}
	\label{fig:warmupend}
\end{figure}
}

\rev{\section{Operationalizing the framework: Statistical Model Checking and \mv} 

\label{sec:mv}}
This section discusses how we operationalise our approach. 
In particular, we frame our approach to ABM analysis in the context of Statistical Model Checking and show how we integrate it into \mv, a model-agnostic statistical model checker that can be integrated with existing simulators. 
\subsection{Statistical Model Checking}
Statistical Model Checking (SMC)~\citep{Agha18,LLTYSG19} is a successful simulation-based verification approach from 
computer science.  SMC allows to study quantitative properties of large-scale models through completely automated analysis procedures equipped with statistical guarantees. 
Following the principle of \emph{separation of concerns}, the idea is to offer a simple external language to express properties of interest that can be \emph{queried} on the model using predefined analysis procedures. The goal of SMC is therefore that of offering a \emph{one-click-analysis} experience to the modeller which is freed from the burden of modifying the model to generate large CSV files every time a new analysis is required, and then analysing such CSV files in an error-prone semi-automated manner. 
This guarantees that the analysis procedures are written once and then extensively tested, decreasing the possibility of errors. 
Making a parallel with databases, we do not have to explicitly manipulate the internal representation of the data every time a new query is needed, rather we define the data to be selected using compact languages (e.g., SQL). 
%

Several statistical model checkers exist, 
most of which require to implement models 
into proprietary languages. We consider \emph{black-box} SMC~\citep{sen2004statistical,younes2005probabilistic}, where the idea is to offer a 
model-independent analysis framework that can be easily attached to existing simulation models, effectively enriching them with automated statistical analysis techniques. \rev{In particular, we use 
\mv~\citep{SebastioV13,GilmoreRV17}, maintained by one of the authors, redesigned and extended here with the techniques presented in this paper to tailor it for the ABM community.} 

%
%

\subsection{Simulator integration}\label{sec:integration}
\mv{} only needs to interact with a simulator by triggering 3 basic actions:
$(i)$ \texttt{reset(seed)}, to reset the simulator to its ``initial state'', and update the random seed used to generate pseudo-random numbers. This is necessary to reset the model before performing a new simulation\rev{. \mv{} takes care of random-seed generation, meaning that it generates adequate sequences of seeds for the necessary simulations. \mv{} allows for \emph{random-seed control}, i.e., the user can fix such sequence across different experiments by providing a \emph{seed-of-the-seeds}, a parameter used to univocally generate all necessary seeds;}
$(ii)$ \texttt{next}, to perform one step of simulation;
$(iii)$ \texttt{eval(obs)}, to evaluate an observation in the current simulation state, 
where an observation (\texttt{obs}) can be any feature of the aggregate model or of any group of agents.
A new model can be integrated with \mv by implementing an \emph{adaptor} between \mv{} and the considered simulator, obtained by instantiating \mv's (Java) interface. 
As a consequence, it natively  supports Java-based simulators, but it has been also integrated with C- and Python-based simulators,  and it has been recently extended to support R-based ones. 
%
\rev{We remark here that our algorithms assume that the model at hand always computes well-defined numeric observations. By providing adequate implementation of the \texttt{eval(obs)}, one can provide a model-specific handling for unexpected/special cases like \emph{infinity} or \emph{not-a-number}.}

For the ABM macro model from Section~\ref{sec:macro} we are interested in two aggregate features of the model: the number of bankruptcies and the unemployment rate in a given step. Therefore, the model has been integrated such that these can be  obtained using  \texttt{eval("bankruptcy")}, and \texttt{eval("unemploymentRate")}, respectively. 
Instead, the prediction market models from Sections~\ref{sec:predmarket} and~\ref{sec:nonergodic} have been integrated such that \texttt{eval(i)} gives a particular feature of agent $i$ (its current wealth), and \texttt{eval("price")} gives a certain aggregate feature of the  model (the prevailing price).

\subsection{\mv query language and supported analysis}\label{sec:queries}
\mv offers a powerful and flexible \emph{property specification language}, MultiQuaTEx, which allows to express transient and steady-state properties, including warmup analysis. 

\paragraph{Transient properties}
Intuitively, a MultiQuaTEx query might describe a random variable (e.g., the number of bankruptcies in an ABM macro model at a certain point in time during a simulation). 
Following the discussion in Section~\ref{sec:outputanalysis}, the expected value of a MultiQuaTEx query is estimated as the mean $\overline{x}$ of~$n$ samples (taken from $n$ simulations), with $n$ large enough (but minimal) to guarantee that the  $(1-\alpha) \cdot 100\%$ CI centred on $\overline{x}$ has size at most $\delta$, for given $\alpha$ and $\delta$. 

MultiQuaTEx actually allows to express more random variables in one query, all analysed independently reusing the same simulations. 
Listing~\ref{ls:examplePropertyTr} depicts a \mq{} query used in Section~\ref{sec:macro} to study the evolution of the number of bankruptcies and of the unemployment rate in an ABM macro model.

\lstset{caption=A transient MultiQuaTEx query, label=ls:examplePropertyTr}
\begin{lstlisting}[float=t,mathescape,morekeywords={next,autoIR,parametric,eval,E,if,fi,then,else,s,eval,evalME,evalOnceME},numbers=left,belowskip=-8pt]
obsAtStep(t,obs) = if (s.eval("steps") == t)!\label{ls:if}! !\label{ls:examplePropertyTrif}!!\label{ls:rval}!
			then s.eval(obs) !\label{ls:rval2}!
			else next(obsAtStep(t,obs))!\label{ls:examplePropertyTrNext}!
		   fi ;!\label{ls:examplePropertyTrfi}!
eval autoIR(E[ obsAtStep(t,"bankruptcy") ],E[ obsAtStep(t,"unemploymentRate") ],t,1,1,400) ;!\label{ls:examplePropertyTrParametric}!
\end{lstlisting}
Coming to the structure of a MultiQuaTEx query, it contains a list of \emph{parametric operators} 
that can be used in an \texttt{eval} \transient command to specify the properties to be estimated. 
\llines{ls:examplePropertyTrif}{ls:examplePropertyTrfi} of Listing~\ref{ls:examplePropertyTr} define the parametric operator \texttt{obsAtStep} having two parameters, \texttt{t}  and \texttt{obs}, respectively the step and observation of interest. Such operator is evaluated, in every simulation, as the value of \texttt{obs} at time point \texttt{t}. 
Before discussing the \emph{body} of the operator, we note that 
\lline{ls:examplePropertyTrParametric} uses it twice for observations the number of bankruptcies and the unemployment rate for each step from 1 to 400 (with increment 1). Therefore 800 properties will be studied (400 for each observation), all evaluated using the same simulations and with their own CI. 
%
%
The body of an operator (\llines{ls:rval}{ls:examplePropertyTrfi}) might contain:
\begin{enumerate}
\item conditional statements (the \texttt{if}-\texttt{then}-\texttt{else}-\texttt{fi}); 
\item real-valued observations on the current simulation state (the \texttt{s.eval} in \lline{ls:rval} and \lline{ls:rval2}); 
\item a \texttt{next} operator that triggers the execution of a  simulation step (\lline{ls:examplePropertyTrNext}); 
\item recursion, used in \lline{ls:examplePropertyTrNext} to evaluate \texttt{obsAtStep(t,obs)} in the next simulation step;
\item arithmetic expressions.
\end{enumerate}
%
This is general enough to express a wide family of properties at varying of time.
In the case of Listing~\ref{ls:examplePropertyTr}, we check whether we have reached the step of interest (\lline{ls:if}), in which case we return the required observation (\lline{ls:rval2}). Otherwise,   we perform a step of simulation (\lline{ls:examplePropertyTrNext}), and evaluate recursively the operator in the next simulation state. 
\rev{\mv has been extended to support counterfactual analysis as discussed in previous sections.}

\paragraph{Steady-state properties and warmup analysis}
\mq{} has been extended to support \mv{}'s extension with steady-state and warmup analysis capabilities discussed in Section~\ref{subsec:steadystate}. 
Listing~\ref{ls:examplePropertySS} provides a \emph{steady-state} MultiQuaTEx query used in Section~\ref{sec:predmarket} to study the average value at steady state of the wealth of three agents (0, 1, and 2), and of the price in our testbed market selection model. 
The query is simple, as in this case the operator \texttt{obs} just returns the observation of interest, while \llines{ls:ssan1}{ls:ssan3} show how to run the three types of supported analysis.
\lstset{caption=A steady-state MultiQuaTEx query. Only one of the three eval commands should be used at a time., label=ls:examplePropertySS}
\begin{lstlisting}[float=t,mathescape,morekeywords={warmup,autoRD,autoBM,parametric,eval,E,if,fi,then,else,s,rval,evalME,evalOnceME},numbers=left,belowskip=-8pt]
obs(o) = s.eval(o) ;
eval warmup(E[ obs(0) ],E[ obs(1) ],E[ obs(2) ],E[ obs("price") ]) ;!\label{ls:ssan1}!
eval autoBM(E[ obs(0) ],E[ obs(1) ],E[ obs(2) ],E[ obs("price") ]) ;!\label{ls:ssan2}!
eval autoRD(E[ obs(0) ],E[ obs(1) ],E[ obs(2) ],E[ obs("price") ]) ;!\label{ls:ssan3}!
\end{lstlisting}
In particular, a steady-state query is composed of two parts:
A list of \texttt{next}-free operators,  
and one of the three \texttt{eval} commands in Listing~\ref{ls:examplePropertySS}, provided with a list of operators to study. 

Intuitively, a steady-state \mq query defines observations on single simulation states, implicitly studied at steady state. 
In particular, \texttt{warmup} performs the warmup estimation procedure (Section~\ref{sec:steadystatewu}) for each of the listed properties. Indeed, every random variable defined on a process might have a different warmup period. We \rev{have seen} examples of this in Section~\ref{sec:predmarket}. 
Instead, \texttt{autoBM} performs a warmup estimation on each property, and begins computing the batch means procedure (Section~\ref{sec:steadystatebm}) on each of them as soon as the property completes its warmup period. 
The command \autord is similar, but it first completes the warmup analysis for all considered properties, and then feeds this information to the replication deletion procedure from Section~\ref{sec:steadystaterd}.  
In all cases, the default values described in Section~\ref{sec:algorithms} will be used if not otherwise specified by the user when running the analysis.

\mq supports two further \texttt{eval} commands: \manualbm and \manualrd. These behave the same as \autobm and \autord, respectively, but 
skip the warmup analysis phase and required as input an estimation of the warmup period.
These might be useful in case one has this information due to previous analyses. In Section~\ref{sec:predmarket} we use them to replicate erroneous steady-state analyses from the literature based on a wrong estimation of the warmup period.

\begin{figure}[t] \centering 
	\includegraphics[width=0.55\linewidth]{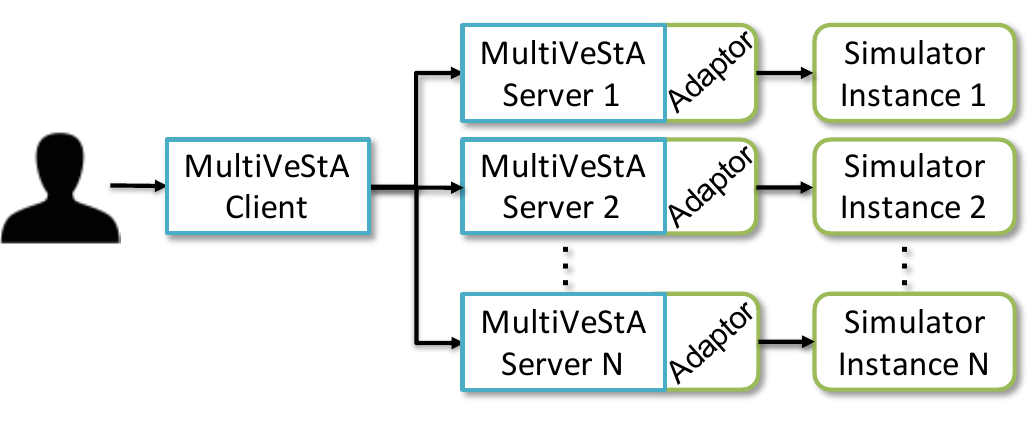}
	\vspace{-0.3cm}
	\caption{\mv's client-server architecture enabling parallelization of simulations. 
	}
	\label{fig:cs}
\end{figure}

\subsection{\mv's distributed architecture}\label{mvarch}
\mv{} has a client-server architecture as sketched in Figure~\ref{fig:cs}. This is a classic software architecture for 
distributing tasks in the cores of a machine or in the nodes of a network. We distribute the simulations of \transient and \autord.  
In the figure, arrows denote visibility/control/activation of the source component on the target one: 
\begin{itemize}
\item A user runs the 
client specifying the model, query, CI, and the parallelism degree $N$. Transparently to the user, the client will trigger, distribute, and handle the necessary simulations providing to the user the results.

\item The 
client  creates $N$ 
servers among whom distributes the analysis tasks. 

\item Each 
server runs independently, therefore in parallel, the required simulations. Each 
server creates its own instance of the simulator, and controls it through the adaptor to perform the simulations. 
\end{itemize}

\begin{figure}[h!]
\centering
\includegraphics[width=0.6\linewidth]{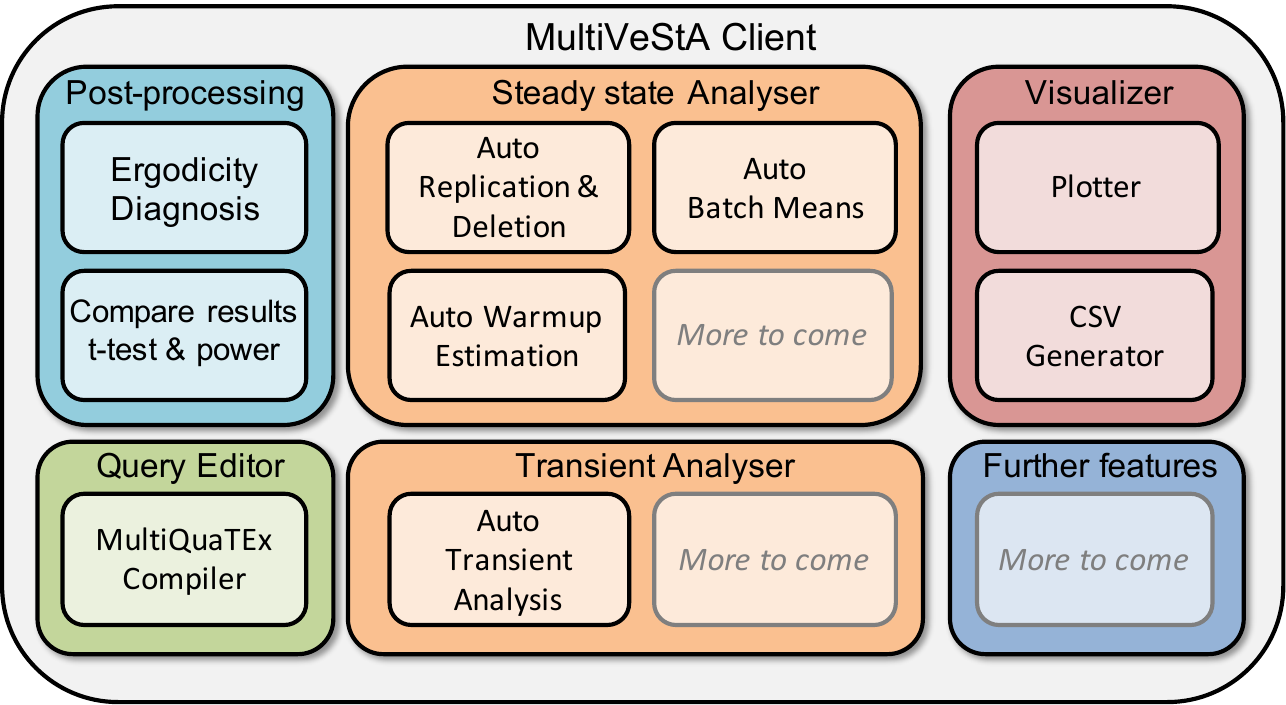}
\caption{\label{fig:clientarchitecture}The novel architecture of the \mv{} client}
\end{figure}

As discussed, we extended \mv{} with a number of analysis techniques. In particular, we mainly extended the client, where the analysis logic is localized. 
The new architecture of the client is depicted in Figure~\ref{fig:clientarchitecture}.
%
It consists of a number of modules, the central ones regarding steady-state and transient analysis. Further modules regard: 
post-processing of analysis computed by \mv like
t-tests and power computation to compare results obtained for different model configurations (Section~\ref{sssec:eqtest_power}), or the methodology for ergodicity analysis (Section~\ref{sec:methodologyergodicity}); 
%
support for the creation and parsing of \mq{} queries, offered by a novel compiler for \mq{} queries; 
visualization of the analysis results through a plotter and of a CSV file creator. 
%

\rev{\section{Parallelization study}\label{sec:parallelizationstudy}}
\begin{figure}[t] \centering 
	\includegraphics[scale=0.35]{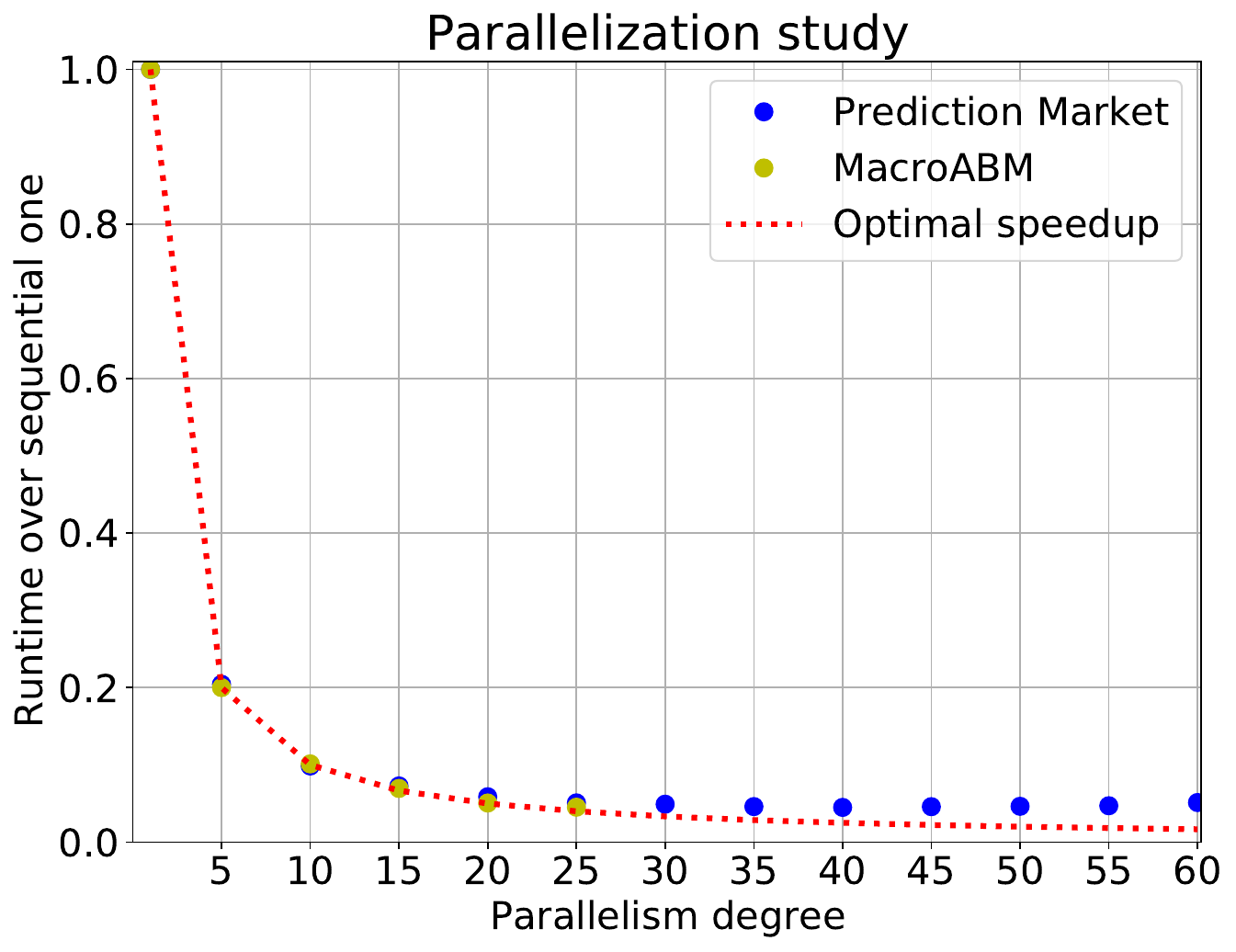}
\hfill
	\includegraphics[scale=0.35]{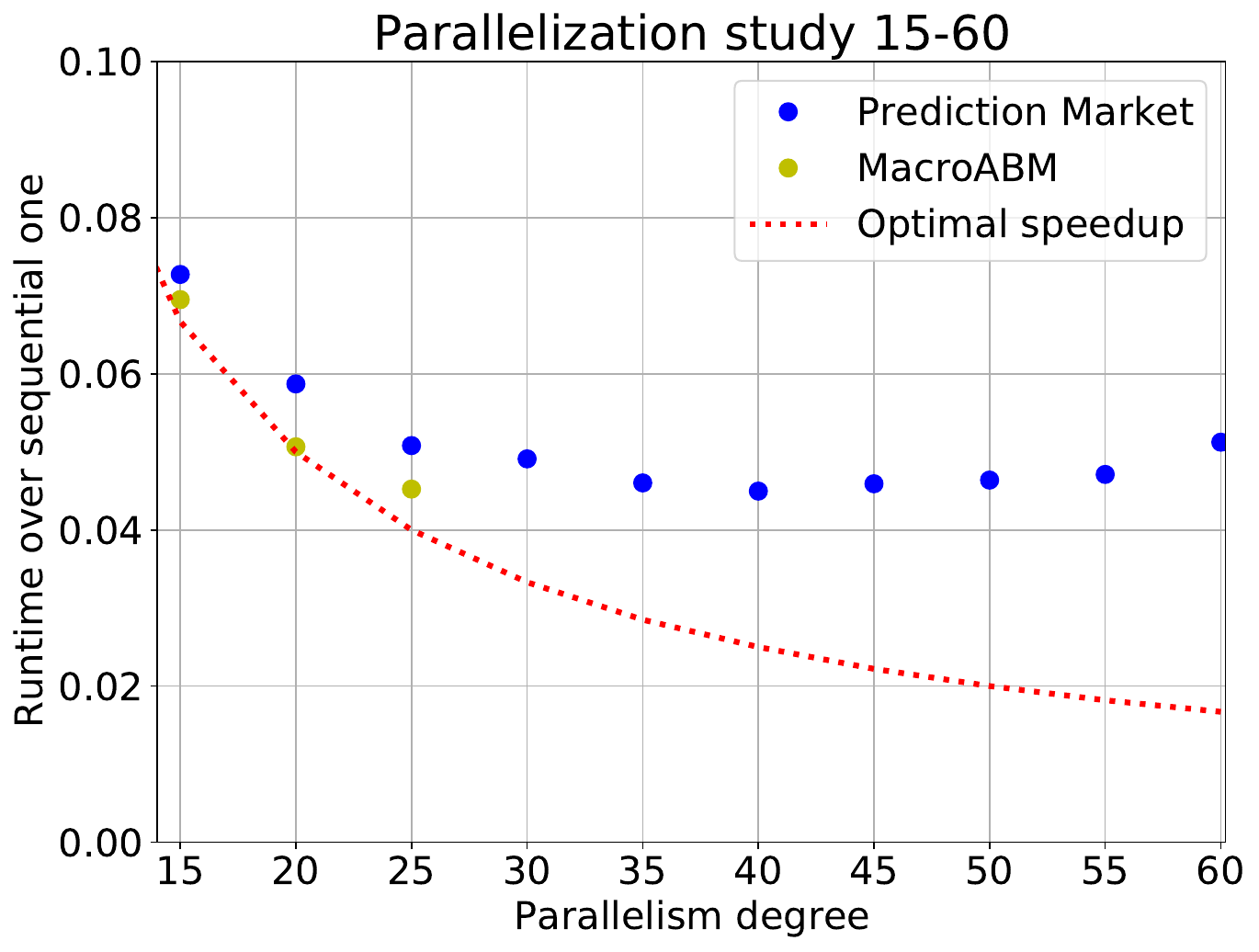}
	\caption{Analysis of the runtime speed-ups on a machine with 20 physical cores.}
	\label{fig:speedup}
\end{figure}
One of the key issues in the analysis of ABMs in social sciences concerns computational time; while some approaches have recently proposed to take advantage of machine learning surrogates \citep{lamperti2018agent, van2019}, the most direct approach to speed-up simulation is an efficient parallelisation of the experiments.
In this section we discuss how \mv can efficiently and automatically parallelize the various runs. Notably, we demonstrate the potential analysis speed-ups showing 
an analysis that requires about 15 days when performed in sequential, and about 16 hours when parallelizing it on a machine with 20 cores. 
%
In particular, we show the actual runtime gains obtained on the analysis of our case studies when using different degrees of parallelism on a machine with 1 CPU Intel Xeon Gold 6252 (20 physical cores) and 94GB of RAM. This machine allows to perform up to 20 processes in parallel, but 
\emph{hyperthreading} 
further allows for limited speed-ups also with parallelism degrees higher than 20. 

Figure~\ref{fig:speedup} (left) shows the results of our study considering the sequential case and parallelism degrees $N$ multiple of $5$ up to $60$.
Intuitively, in the ideal case an analysis using parallelism degree $N$ should take $\frac{1}{N}$ of the time required by a sequential analysis (i.e., with $N=1$). For this reason, 
the red dashed line provides the optimal obtainable speed-ups: 1 (no speed up) for the sequential case, and $\frac{1}{N}$ for all considered $N$.
The blue and yellow dots, instead, show the actual speed-ups obtained for our two case studies. In particular, in order to compare with the optimal speed-up, for each value of $N$ we provide the ratio among the runtime obtained with parallelism degree $N$ over the one of the sequential case.
For the prediction market model, we consider the \autord analysis from Figure~\ref{fig:wealth} (left) for $\pi^*=0.45$, while for the macro model we consider the analysis from Figure~\ref{fig:analysisMacroABM}.
Notably, the analysis of the macro model took about 15 days when executed sequentially, while it goes down to about 18 hours for $N=20$, and 16 hours for $N=25$. The analysis failed for higher values of $N$ due to the high memory requirements of the model. 
Instead, the analysis of the prediction market model requires about 14 minutes in sequential and about $50$ seconds for $N=20$. The analysis could be performed for all considered $N$, with a minimum runtime of about $38$ seconds for $N=40$.

Overall, for both case studies we note speed-ups very close to the optimal ones up to $N=20$, while they tend to deteriorate for higher values of $N$. 
Figure~\ref{fig:speedup} (right) focuses on the values of $N$ from $15$ onwards. We see that the speed-ups obtained for the macro model tend to be closer to the optimal ones. This is because simulations are computationally intensive, taking more than 1 hour. Therefore, the \emph{overhead} (i.e., the extra computations) introduced by the communications among the \mv{} client and servers has almost no impact on the overall runtime. Instead, the prediction market model is not particularly computationally expensive, making the extra communications influence more the overall runtime.
In particular, the figure shows that relatively limited speed-ups are obtained for $N$ greater than $25$. This is expected, as discussed. Interestingly, increasing $N$ further than $40$ actually worsens the performances,  as the processor is not anyway able to perform more than 20 processes in parallel while the overhead costs increase. 

\end{document}